\pdfoutput=1
\documentclass[12pt]{article}
\usepackage{graphicx}
\usepackage{amssymb,amsfonts,amsmath}

\def\lsim{\mathrel{\rlap{\lower3pt\hbox{\hskip1pt$\sim$}}
     \raise1pt\hbox{$<$}}} 
\def\gsim{\mathrel{\rlap{\lower3pt\hbox{\hskip1pt$\sim$}}
     \raise1pt\hbox{$>$}}} 
\newcommand{\be}{\begin{equation}}
\newcommand{\ee}{\end{equation}}
\newcommand{\ba}{\begin{eqnarray}}
\newcommand{\ea}{\end{eqnarray}}

\newcommand{\A}{\mathcal{A}}
\begin{document}
\title{The role of monopoles in a Gluon Plasma}
\author{Claudia Ratti and Edward Shuryak\\
\small{\it Department of Physics and Astronomy, State University of New York}\\ 
\small{\it Stony Brook, NY 11794 USA} \\
\small{\it shuryak@tonic.physics.sunysb.edu}}
\maketitle
\begin{abstract}
We study the role of magnetic monopoles  at high 
enough temperature $T>2T_c$,
when they can be considered heavy, rare objects embedded into
matter consisting mostly of the usual ``electric" quasiparticles, quarks and gluons.
We review available lattice results on monopoles at finite temperatures.
Then we proceed
to classical and quantum charge-monopole  scattering,
solving the problem of gluon-monopole
scattering for the first time. We find that, while this process hardly
influences thermodynamic quantities, it does produce a
large transport cross section, significantly exceeding that
for pQCD gluon-gluon scattering up to quite high $T$. Thus,
in spite of their relatively small density at high $T$,
monopoles are extremely important for QGP  transport properties,
keeping viscosity small enough for hydrodynamics to work at LHC.
\end{abstract}
\section{ Introduction}
\subsection{Overview}
As it is known from 1970's, QCD at
 high temperature $T$ is weakly coupled \cite{Collins:1974ky}
and provides perturbative screening of the charge \cite{Shuryak:1977ut},
thus being called Quark-Gluon Plasma (QGP). 
Creating and studying this phase of matter in the laboratory
has been the goal of experiments at CERN SPS and recently
 at the Relativistic Heavy Ion Collider (RHIC)
facility in Brookhaven National Laboratory, soon to be
continued by the ALICE collaboration at the Large Hadron Collider (LHC).
  RHIC experiments have revealed 
robust collective phenomena in the form of radial and elliptic flows,
which turned out to be quite accurately described by near-ideal
 hydrodynamics. QGP thus seems to be the most perfect
liquid known, with the smallest viscosity-to-entropy 
ratio $\eta/s$.

 The theory of QGP has shifted from the 
perturbative-based one, appropriate for a weakly coupled (gas)
regime, to the non-perturbative methods needed to
address the strongly coupled QGP (sQGP for short) regime.
This ``paradigm shift", documented in Refs.
\cite{Gyulassy:2004zy,Shuryak:2004cy}, is still profoundly affecting
the developments. The methods addressing strongly coupled gauge theories
include in particular the so-called AdS/CFT correspondence,
relating strongly coupled gauge theory to weakly coupled
string theory in a particular setting. We will not discuss
it in this paper, for a recent review see
e.g. \cite{Shuryak:2008eq}. On pure phenomenological grounds, it has been argued
 that, since many substances exhibit a
minimum of the viscosity at some phase transitions, perhaps
QGP is the ``best liquid''  at the QCD phase transition as well, namely at 
$T=T_c$ \cite{Csernai:2006zz}.

  Another duality which has been used to explain 
unusual properties of
the sQGP is the $electric-magnetic$ duality.
Liao and one of us have proposed the so-called
``magnetic scenario"
\cite{Liao:2006ry}, according to which  the near-$T_c$ region 
is dominated by
 magnetic monopoles. This is not surprising, if the
 deconfinement phase transition 
is basically interpreted as their Bose condensation. 
Based on molecular dynamics
of classical plasmas with both electric and magnetic
 quasiparticles, it has been further argued in that work
  that the minimal
viscosity/entropy ratio (the ``best liquid'')
does not correspond 
 to the phase transition point $T=T_c$, but rather to 
the ``electric-magnetic equilibrium'',  at $T\approx 1.4 T_c$,  
where both components of QGP contribute about equally 
to transport coefficients. We will review more recent works
on the subject below.

   One of the central questions
 is how  sQGP with ``perfect fluidity'' will change into a weakly
coupled wQGP with increasing $T$. In view of the
  next  round
of heavy ion experiments at LHC, a quite urgent question
is what transport properties
 are expected to be
  observed there,
at temperatures reaching about twice those reached at RHIC.
In order to answer this question, one of course has to understand
where the ``perfect fluidity'' property of QGP comes from. 
As an important example of perturbative point of view,
we mention the work by
Xu, C. Greiner and St\"ocker \cite{Xu:2007jv} who
argued that the QGP is only moderately coupled,  
with $\alpha_s=0.3..0.6$,  explaining the small viscosity
 by inclusion of the next-order radiative processes,
 $gg\leftrightarrow ggg$. We will discuss this issue
partially in the next section, dealing with parametric
dependences of densities and scattering rates, and also
 at the end of the
paper in the discussion section. Here we only
notice that, if this should be the explanation,
one would expect a very slow transition to weakly coupled QGP,
induced by the logarithmic running of the coupling.

In this paper we address the issue of QGP 
transport properties in the ``magnetic scenario'' framework, 
moving  away from the phase transition region to higher temperatures,
where QGP is still dominated by the usual
 electric quasiparticles  -- quarks and gluons -- and the coupling is
 moderately small.
Our goal is to
study the interaction between electric and magnetic sectors.
The main difference, with respect to the Liao-Shuryak paper \cite{Liao:2006ry},
is that their classical (manybody) treatment of this interaction
(via Lorentz force) is replaced
by our (two-body) quantum-mechanical calculation.
Our main result is the explicit solution of the rather 
difficult problem of quantum 
gluon-monopole scattering, from which we calculate the
 corresponding transport cross sections.
 
Two types of ``enhancement mechanisms'' are found: one is related to 
the couplings and the other one (of much more nontrivial
origin) is due to enhanced large-angle scattering.
The former is rather easy to explain: while ``electric'' $qq,qg,gg$
scattering has cross sections proportional to small
$\alpha_{electric}^2\ll1$,
the scattering of electric-magnetic type contains the product
$\alpha_{electric}\alpha_{magnetic}$ which, according to the
celebrated Dirac condition, cannot run and is
equal to an integer (just one for the 't Hooft-Polyakov
monopole we will consider). 
Therefore, in this problem there is no small parameter  
at any temperature. For many other details the reader should read
the paper further; nevertheless, we emphasize here our main result,
namely that the gluon-monopole scattering rates are very large.
Parametrically they are  $\dot{w}_{gm}/T\sim \log(T)^{-3}$, just one
power of the log down compared to the gluon-gluon 
pQCD scattering rates  
$\dot{w}_{gg}/T\sim \log(T)^{-2}$. Thus, if the gluon-monopole scattering
is the dominant one, one should expect stronger
decrease of scattering  
($\eta/s$ growth) at LHC, compared to the perturbative scenario
\cite{Xu:2007jv}.

 Below we include a rather extensive introduction
 to the electric-magnetic duality in gauge theories
in general, and to the magnetic sector of
QCD at high $T$ in particular. This choice, which we hope 
will be useful to many readers, can be justified
 by the somewhat intermittent nature of this field, 
with many important results obtained and then
 semi-forgotten outside a relatively small group of experts.
 In particular, the
 scattering of an electric particle on a magnetic monopole is
a century-old problem.
It is hard to explain
the quantum scattering without introducing classical notations and
results first. Thus, although this
 part of the story is rather old, 
 we include some of this material for completeness as
an extended introduction. We will then proceed to the extension of 
this theory to gauge particles.

\subsection{Scattering on monopole in classical approximation}
\label{sec_classical}
 Let us start with a very old (19th century) problem in classical electrodynamics:
 an electrically charged
  particle with charge $e$ is moving  in the magnetic field of 
 a static monopole with magnetic coupling $g$.
Classically, one does not need
the vector potential, thus
many subtleties are absent. The interaction
is simply given by the Lorentz force
\be m\ddot{\vec r} = -eg { {\dot{\vec r}\times \hat{r}} \over r^2}  \ee
where $\times$ indicates the
 vector product and $ \hat{r}=\vec r/r$ is the unit vector
along the line connecting both charges.
It is worth noticing that only the product of the couplings ($eg$)  appears.

Furthermore, the cross product of the magnetic field of the monopole and
electric field of the charge leads to a nonzero Poynting vector,
which rotates
around $\vec r$: thus, the field 
itself has a nonzero angular momentum. The total conserved
angular momentum  for this problem has two parts
\be 
\vec J= m \vec r  \times\dot{\vec{r}}+eg\hat{r}.
\ee
The traditional potential scattering only has the first part: therefore,
in that case, 
the motion entirely takes place in the so-called
 ``reaction plane'' normal to $ \vec J$. In the 
charge-monopole problem,
the second term restricts the motion to the so-called
 Poincar\'e cone \cite{Poincare}: its
half-opening angle $\pi/2-\xi$ being
\be 
\sin(\xi)={eg\over J} \label{eqn_cone}. 
\ee
Only at large $J$ (large impact parameter scattering) the angle $\xi$
is small and thus the cone opens up, approaching the scattering plane.

Following Ref.~\cite{Boulware:1976tv}, one can project
the motion on the cone to a planar motion, by introducing
\be 
\vec R= {1\over \cos(\xi)} [\vec r - \hat{J} (\vec r\cdot \hat{J})]
\ee 
where the first scale factor is introduced to keep 
the same length for both vectors $\vec R^2=\vec r^{~2}$. 
Now,  
two integrals of motion are
\ba 
\vec J&=& m \vec R \times \dot{\vec{R}}\\
E&=&{m\dot{\vec R}^2 \over 2}- {(eg)^2\over 2m R^2}
\ea
and the problem seems to be reduced to the motion of a particle of mass $m$
in an inverse-square potential.
The scattering angle $\Delta\psi$ for this {\em planar} problem 
can be readily found: it is the variation of $\psi$
as $R$ goes from $\infty$ to its minimum $b$ and back to $\infty$
\be 
\Delta\psi =\pi\left({1\over \cos{\xi}}-1\right) =\pi\left(\sqrt{1+\left({eg\over mvb}\right)^2}-1\right).
\ee 
Note that at large $b$ (small $\xi$) we have $\Delta\psi\sim 1/b^2$,
 as expected for
the inverse-square potential.
Yet  this is $not$ the scattering angle of the original problem,
 because
one has to project the motion back to the Poincar\'e cone.  The
 true scattering angle
-- namely the angle between the initial and final velocities
-- is $\cos{\theta}=-(\hat{v}_i\cdot
\hat{v}_f)$. By relating velocities on the plane and on the cone one can find
it to be
\be 
\left(\cos{\frac{\theta}{2}}\right)^2=(\cos{\xi})^2 \left(\sin{{\pi \over 2 \cos{\xi}}}\right)^2.
\ee
Thus for distant scattering -- small $\xi$ -- one gets 
$\theta\approx 2\xi=2eg/(mvb)$, which is much larger than $\Delta \psi\sim 1/b^2$.
The important lesson that we learn from these formulae is that
the small scattering angle is given by the opening angle
of the cone, rather than the scattering angle in the planar,
 inverse-square effective potential.
Calculating the cross section by $d\sigma=2\pi b db$ one finds that,
at small angles, it is 
\be 
{d\sigma \over d\Omega}=\left({2eg \over mv}\right)^2{1\over  \theta^4}, 
\ee
similar to the Rutherford scattering of two charges.
The difference (apart from different charges) is also the additional
second power of velocity, originating from the Lorentz force.

\subsection{ Quantum charge-monopole scattering problem} 
\label{sec_quantum}
Jumping from 1900's to 1930's and from classical to
quantum mechanics, one first has to deal with
the necessity of introducing the vector potential $A_\mu(x)$.
If this is done, the divergence of the magnetic field is zero\footnote{
One can introduce the so-called dual vector potential related to $\vec
B,\vec E$ in the opposite way: then it is the electric field which
would be forced to have zero divergence and Dirac strings.
}. Dirac \cite{Dirac} brilliantly solved this difficulty
by introducing the celebrated Dirac string: he then proved
it to be an unphysical gauge artefact, provided the charge
quantization condition is fulfilled
\be {eg \over 4\pi}= integer. \ee
The same condition appears, if we recognize the necessity of quantizing 
the field angular momentum to (semi-integer) multiples of $\hbar$.

Now we jump to mid-1970's, when the 
 quantum scattering problem  was  solved 
in Refs. \cite{Boulware:1976tv,Schwinger:1976fr}
for the scattering of a scalar particle on a monopole, and
in Ref. \cite{Kazama:1976fm} for a monopole-spin 1/2 particle
 scattering. 

The  wave function 
 in spherical coordinates is as usual
 a sum of products of certain $r$-
dependent radial functions, times the angular functions. The former
basically  follow from the
 inverse-square law potential and thus are easily solved
in Bessel functions: they
correspond to  the auxiliary planar projection of the
classical problem of the previous subsection. The nontrivial
part happens to be
 in the unusual angular functions. Before introducing those,
let us hint why the usual set of angular harmonics
$Y_{lm}(\theta,\varphi)$ should $not$ to be used.
 The reason is that their classical limit -- for large
values of the indices  $m\approx l \gg 1$ -- corresponds to flat
planar motion near the $z=0$ plane. Nevertheless, we
have already learned  from the classical limit
 to expect the motion 
to be concetrated around the Poincar\'e
cone instead! 

The functions that we need (for example in the scalar sector)
must satisfy the following set of conditions
  \be
 \left\{ 
 \begin{array}{l}  
 \vec{T}^2 \\  T_3 \\  \vec{I}^{~2} \\  \left(\hat{r}\cdot \vec{I} \right) \end{array}
 \right\}  
 \phi^{mn}_{ti}(\theta,\varphi)=
  \left\{  \begin{array}{l}  
 t(t+1) \\  m \\  i(i+1) \\  n \end{array}
 \right\}  
 \phi^{mn}_{ti}(\theta,\varphi).  
 \label{eigenvalues}
  \ee
  where $\vec{T}$ is the total angular momentum $\vec{T}=\vec{L}+\vec{I}$, $\vec{L}$ is
  the orbital angular momentum and $\vec{I}$ the isospin.
  The unusual condition in the above set is the last one, since the vector $\vec{I}$ must
  be projected
  to the (space-dependent) radial unit vector. The functions satisfying
  this requirement \cite{Boulware:1976tv,Schwinger:1976fr} will be introduced in
  Sec.~\ref{scalarfluctuations} in the case of a scalar particle, and in Sec. \ref{vectorfluctuations}
  in the case of a vector one. 
  Here we can anticipate that, in order to check if the hint we provided above
  is satisfied, it is enough to explore  
the large $l,n$ limit of the $D$-functions involved. The result
\be D^l_{nl}\sim e^{i(l-n)\varphi} \exp[-l (\theta-\theta^*)^2/2]\ee
where $\cos(\theta^*)=n/l$, shows that they indeed correspond to the Poincar\'e cone.

\subsection{Brief review on QCD monopoles}
\label{sec_mono_review}
The discussion in the previous two subsections was restricted to
electrodynamics, in which there is neither an explicit
explanation of the monopole structure nor a real necessity
to have them\footnote{The famous Dirac's quote is that he would be
  surprised
that Nature would not make use of them.}.
However, non-Abelian gauge theories do have monopole-like
solitons, as demonstrated by 't Hooft and Polyakov \cite{thooft,polyakov}
in the Georgi-Glashow model framework. Three decades later,
Seiberg and Witten \cite{sw} have discovered beautiful
ways in which monopoles and dyons can get alive and
even replace the usual electric particles (gluons/gluinos)
in $\cal N$=2 Supersymmetric-Yang-Mills theory. 

Since those developments are very well documented in
books and reviews, we directly jump to finite-$T$ QCD.
 Explicit calculations show
that the static magnetic field 
is not screened perturbatively in QGP  \cite{Shuryak:1977ut}:
yet  Polyakov  conjectured \cite{Polyakov:1978vu}  that magnetic
screening should appear
non-perturbatively, at the so-called ``magnetic scale''\footnote{
In order to keep with the notations used in monopole problems,
in this work we use $e$ to indicate the usual gauge coupling,
reserving
$g$ for magnetic coupling. We hope $e$ will not be confused with
QED charge, which is not used in this work.
} 
 $ E_M\sim e^2 T$. Linde \cite{Linde:1980ts}
 related this magnetic scale and screening to monopole solutions,
which appear at finite $T$ when the Higgs fundamental
scalar of  the Georgi-Glashow model is replaced by
$A_0$, the 0 component of the gauge field. Shortly afterwards, the
monopole Bose-condensation was the central
idea of the so called ``color superconductor''
picture of confinement, proposed by Mandelstam and 't Hooft
\cite{'tHooft-Mandelstamm}. The abelian Higgs model (an incarnation of
Landau-Ginzburg action) was worked out in detail and shown to be
in excellent agreement with lattice results on the structure of confining flux tubes
at $T=0$ by a large number of people. The profiles of the
flux tubes nicely follow the celebrated Abrikosov solution,
and lattice monopoles do rotate around them, providing
a ``coil'' needed to contain the electric flux.

Lattice monopoles are defined by the procedure 
\cite{DeGrand:1980eq} which basically locates the ends
of singular Dirac strings by calculating the total magnetic flux through
the boundary of  elementary 3-d boxes.
Since the strings and thus the procedure depend on a certain gauge,
for decades sceptics kept the viewpoint that  
those objects are just unphysical  UV gauge  noise.
Yet, many specific questions  -- e.g. monopole density and correlations \cite{D'Alessandro:2007su} -- produced very
reasonable and consistent answers, which are apparently 
independent of the particular lattice parameters.  It is hard to imagine that lattice gauge 
artefact have consistent correlations at relatively large distances, accurately
reproducing those in a Coulomb plasma.  We thus
assume instead, that these monopoles are indeed
meaningful physical degrees of freedom,
present in QGP as quasiparticles and being the source of a Coulomb-like magnetic field.

The picture of the (uncondensed)
 monopole gas  at  $T>T_c$ was discussed in particular
in the important paper by
Smit {\it et al.}
\cite{Smit:1989vg}, in which the
 relation between the monopole density and such observables as the spatial 
string tension were considered. One question (especially
important for our work) is whether monopoles form
 a gas or a liquid. To that purpose, the authors of 
 Ref.~\cite{Smit:1989vg} 
have pointed out that the combination $n_{m} /M_m^3\sim .03 \ll 1$ 
is small\footnote{
For more recent discussion of this feature see 
Korthals Altes \cite{Korthals Altes:2006gx}.}, while in a
weakly coupled (Debye) plasma it should be large.
(Indeed, in high-$T$ electric plasma $n\sim T^3, E_E\sim eT$, and thus  
$n /E_E^3\sim 1/e^3 \gg 1$).
 Yet such ratio can be small
in a liquid; in fact the ensemble of instantons in the QCD
vacuum -- the ``instanton
liquid" extensively studied by one of us \cite{Shuryak:1981ff,Schafer:1996wv} --
 has the same property.
The $T=0$ instanton density
$n_{inst}\sim 1\,$ fm$^{-4}$ while the screening mass of the topological charge
is $m_{\eta'}\approx .94\,$ GeV, thus $n_{inst}/M_{\eta'}^4\sim .002 \ll 1$
as well. Taking the corresponding roots for comparison,
one even finds  the small parameters in both cases  to be similar:
$n_{m}^{1/3}/M_m\approx .3,$
 $n_{inst}^{1/4}/M_{\eta'}\approx .2$.

Recent interest in
QGP  monopoles (at $T>T_c$) started with the ``magnetic scenario" 
\cite{Liao:2006ry}  which basically suggested the magnetic sector
to be a  strongly coupled (magnetic) Coulomb
plasma of monopoles, in its liquid form.  
Another line of work based 
on lattice monopoles has led  Chernodub and Zakharov
  \cite{Chernodub:2006gu} at the same time to 
a very similar  conclusion. An
important feature of this scenario \cite{Liao:2006ry} is 
 the opposite running of the electric
coupling $e$ and the
magnetic one $g$, induced by the Dirac condition $e g=const$.
 As recently shown in \cite{Liao:2008jg}, this feature has been
dramatically confirmed
by the behavior of the lattice correlation functions 
\cite{D'Alessandro:2007su}, which indeed display 
monopole-monopole and antimonopole-monopole correlations
$increasing$ with $T$. As shown in \cite{Liao:2008jg},
those correlations  are well described by a picture
of classical Coulomb gas. The main input in those calculations
is the magnetic Coulomb coupling $g$  $growing$ with $T$.
Furthermore, it  was shown to be
simply the inverse of the gauge coupling $e$,
as Dirac predicted.
 We consider the correlations observed in \cite{D'Alessandro:2007su}
  to be a decisive
confirmation of the existence of the
 long-distance magnetic Coulomb
field of the monopoles.
If so, the scattering
of the dominant ``electric'' particles on this
magnetic field is an important component of transport, and this
is what we are going to study below.

  In order to complete this brief overview, let us comment on 
the issue of ``Higgsing'' in hot gauge theories. It is well known
 that, at high $T$, one can consider a 
dimensionally reduced effective 3d theory, 
which is basically a Georgi-Glashow model
with $A_0$ being the adjoint scalar.
Its lattice-derived phase diagram, as well as  
the effective lines of the high-$T$ 4-d gauge theories
derived to 1 and 2 loops
can be found in Ref.~\cite{Kajantie:1997tt}. If taken
literally, the  line for QCD 
 crosses into the broken phase, except at extremely high $T$:
but the authors themselves argued 
that this line should in fact remain in the symmetric
phase. Thus, even the asymptotically high-$T$ behavior is
under debate.
In the temperature range of 
few-$T_c$ we know that the expectation value
of the Polyakov line changes from zero to unit value, and that
its effect on quarks \cite{PNJL} and gluons \cite{Meisinger:2003id}
is large, producing significant suppression. This approach
in particular gives a very good description of the 
quark number susceptibilities \cite{Ratti:2007jf}.
  
One more practical aspect of the issue 
comes from heavy-ion phenomenology. Dumitru and
collaborators \cite{Bazavov:2008qh} have used this form of the effective Lagrangian
to study the real-time evolution of $A_0$.
The main conclusion from their work is that
$\langle A_0\rangle$  belongs to the class of so-called  {\em slow variables},
and its evolution in heavy ion collisions has to be treated
separately from the overall equilibration.  They have numerically solved
the EOM for $A_0$, starting from the ``suddenly quenched" value corresponding
to its vacuum form, moving toward its
 minimum at the deconfined phase at $T=2T_c$.
 The main finding of this work
is that  the relaxation of this variable is very slow, taking approximately 40 fm/c.
This time significantly
exceeds the QGP lifetime at RHIC,  which is only about 5 fm/c,
which suggests that in real collisions we should treat $A_0$ essentially as a
random variable frozen at some value and color direction
during hydro evolution. This means that there is a chaotic out-of-equilibrium Higgsing,
slowly rolling down, like in cosmological inflationary models: thus
one would like to know as much as possible about phase transitions
and EoS for $all$ values of the Higgs VEV.

\begin{figure}
\begin{center}
\includegraphics{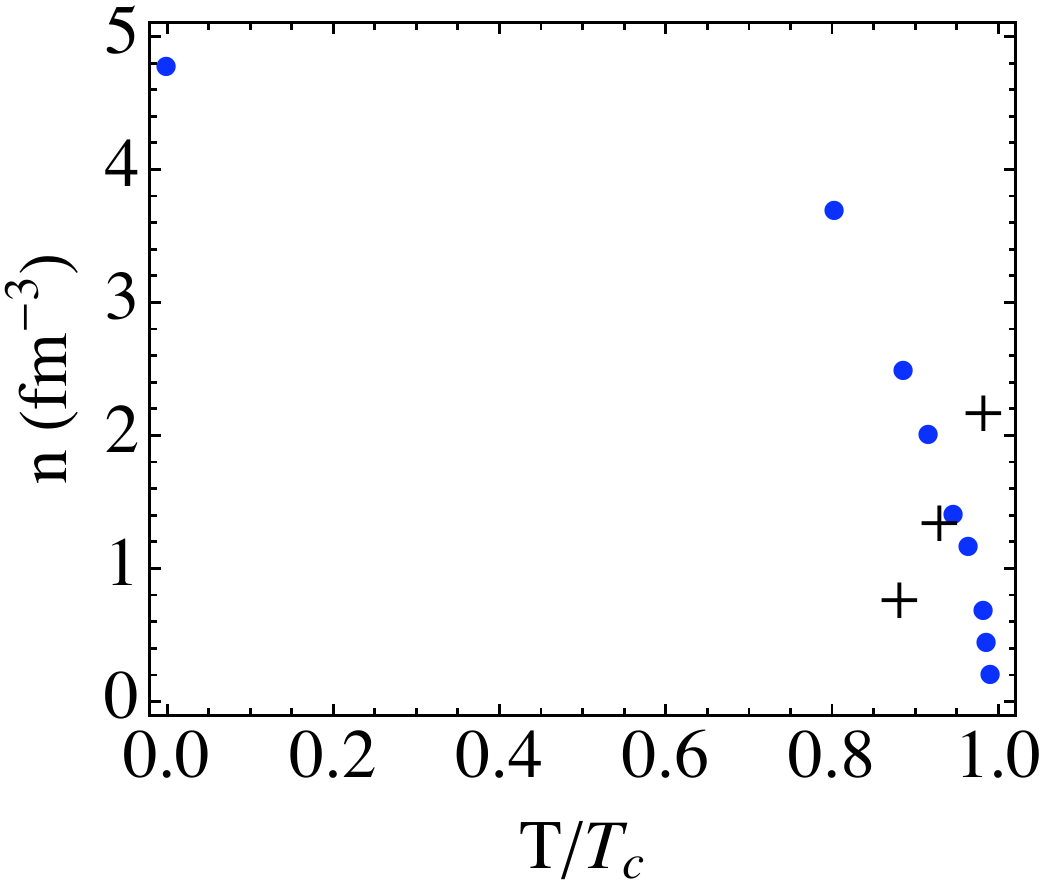}
\vspace{.5cm}\\
\includegraphics{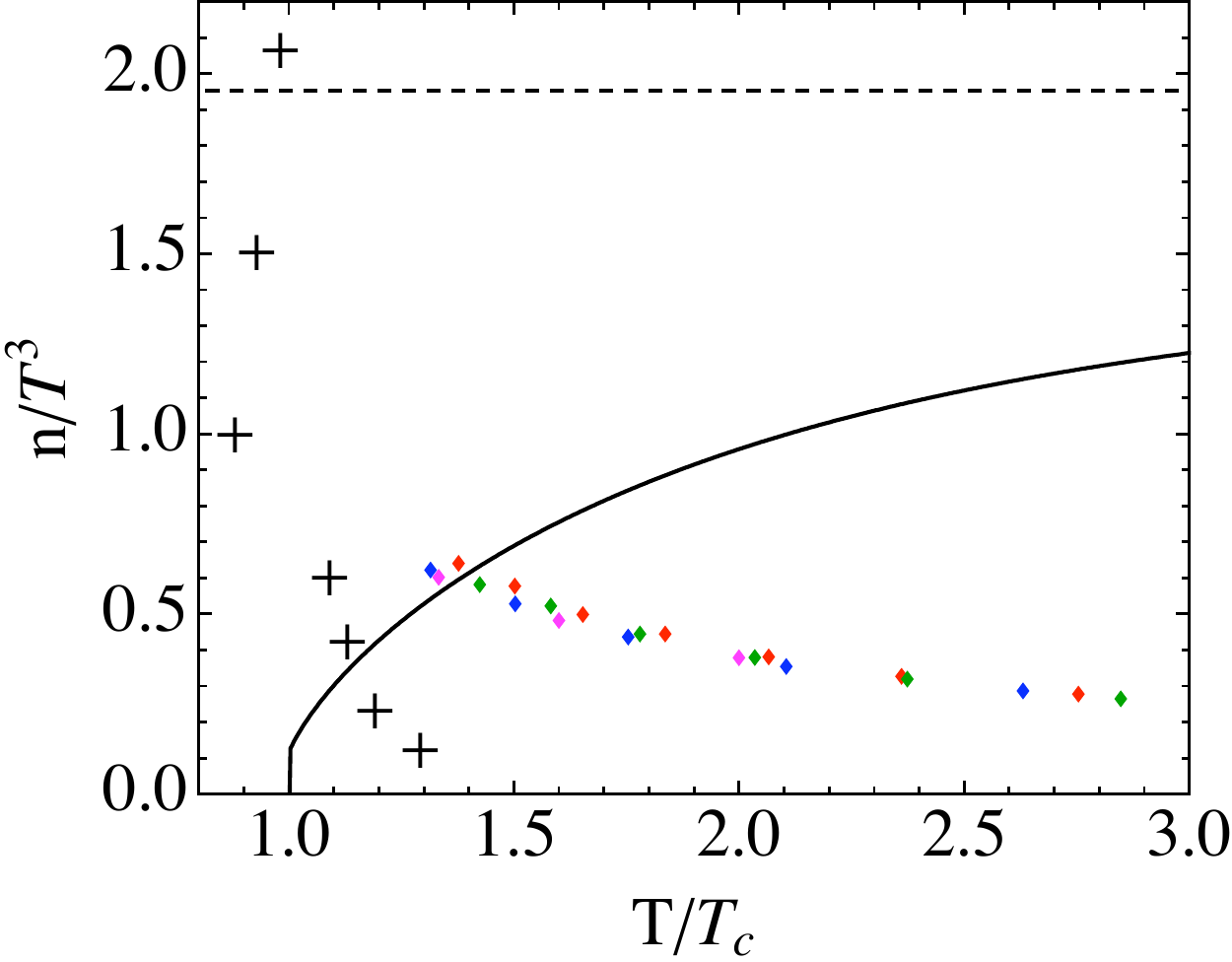}
\caption{(a) The monopole density $n_m$ in units of $[$fm$^{-3}]$  
versus $T/T_c$ for the confined phase $T<T_c$.
(b) The normalized density $n/T^3$ versus $T/T_c$ for the 
deconfined phase $T>T_c$.}
\label{fig_density}
\end{center}
\end{figure}

\section{Monopoles in gauge theories at high $T$}
\subsection{The monopole density and properties} 

In section \ref{sec_mono_review} we have already
briefly reviewed the main issues and ideas related with
magnetic monopoles in hot QCD.
Now we begin a much more quantitative discussion of
the main parameters and lattice results involved,
which will be needed for our discussion of transport 
cross sections.

In Fig. \ref{fig_density} we compile the lattice-based
data on the monopole density. 
Since we compare various theories, SU(2) and SU(3) pure 
gauge theories as well as those with quarks, we need to explain the units.
Following the lattice tradition, physical units
are defined by insisting
that the string tension is the same in all of them,
$\sqrt{\sigma}=426$ MeV. We will always show the
temperatures in units of the corresponding deconfinement transition 
temperature $T_c$. 

In Fig. \ref{fig_density}(a) the vacuum value of the density
is taken from Bornyakov {\it et al.} \cite{Bornyakov}, 
$(n_m)_{T=0}/\sigma^{3/2} = 0.5$. The dots and crosses
are from Liao and Shuryak  \cite{Liao:2008vj}: 
they correspond to ``condensed'' and ``decondensed'' monopoles,
respectively\footnote{The origin of those points resides in the lattice results
on string tensions corresponding to free energy and potential energy,
respectively.}. Note that  these two
add together to an approximate constant  total monopole density,
for all $T<T_c$. 

In Fig. \ref{fig_density}(b) we summarize what is known about monopoles
in the deconfined phase. The diamonds of different colors show the
direct lattice observation of monopole density by 
D'Alessandro and D'Elia
\cite{D'Alessandro:2007su}, scaled up by the factor 2 which
accounts for the transition from SU(2) to SU(3) gauge group. 
The best  fit to the data of
D'Alessandro and D'Elia  (not shown) is
\be 
n_m/T^3 = A/\log(T/\Lambda_{\mathrm{eff}})^\alpha 
\ee
with $A = 0.48, \alpha = 1.89$ and $T_c/\Lambda_{\mathrm{eff}} = 2.48$:
we discuss the expected parametric dependence at high $T$ below.

The crosses in Fig. \ref{fig_density}(b) are from 
Ref. \cite{Liao:2008vj},  as those in Fig. \ref{fig_density}(a),
but for temperatures larger than $T_c$. They have been obtained
from (metastable) flux tubes in the plasma phase. 
The latter should be compared to the solid line,
which  represents gluons, affected by the nonzero VEV of 
the Polyakov line effects\footnote{The reader is reminded that it reflects 
the nonzero VEV
of $A_0$, which acts on electrically charged particles as
an imaginary chemical potential. }
\cite{Meisinger:2003id}. (At high $T$ the normalized gluon density 
becomes a constant corresponding to Stefan-Boltzmann
 ideal gas of massless
SU(3) gluons.) 
Note that the densities
of gluons and monopoles cross each other at  $T\simeq1.3T_c$.

\begin{figure}[t]
\begin{center}
\includegraphics[width=16cm]{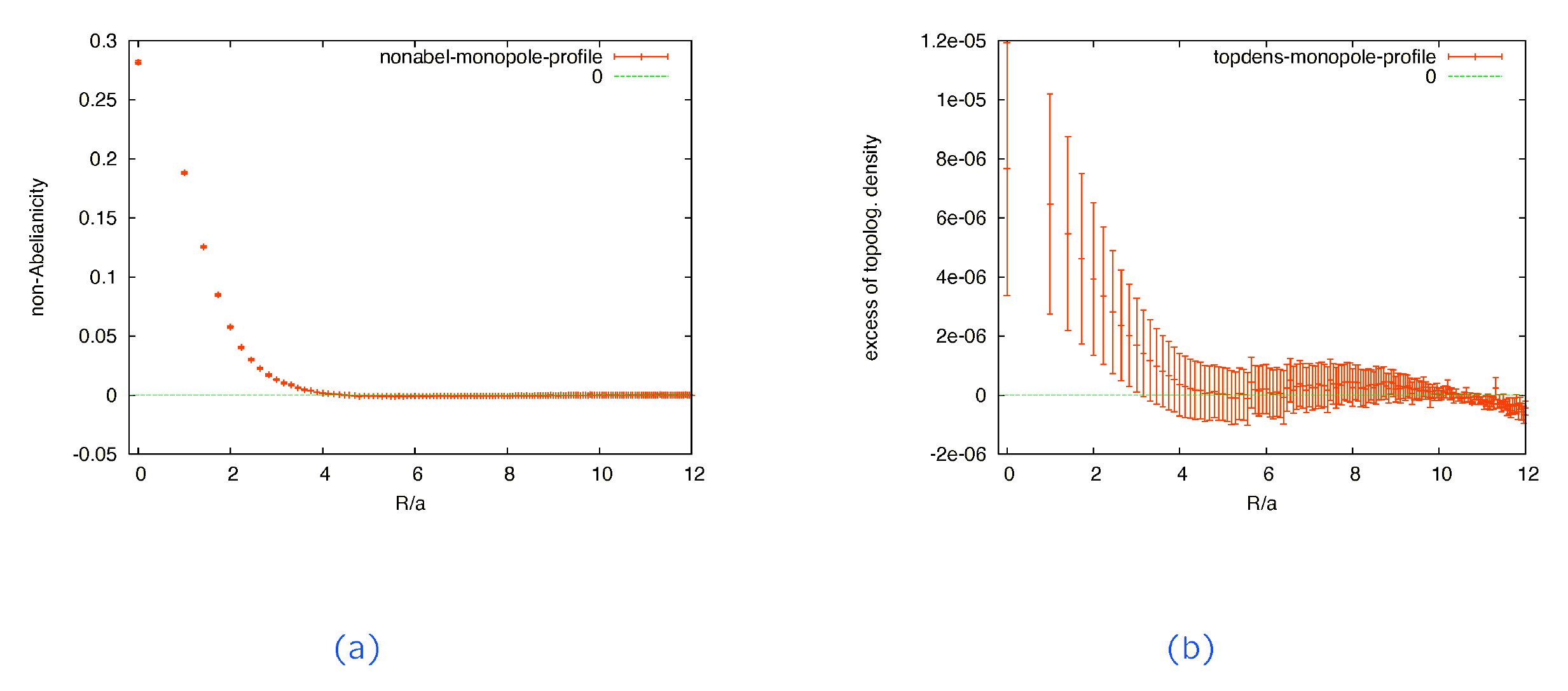}
\caption{
The square of non-Abelian components of the gauge potentials (a) and the topological charge 
(b) correlated with the monopole path (taken from Ref.~\cite{Ilgenfritz:2007ua}).
}
\label{fig_Ilgen}
\end{center}
\end{figure}

The available lattice results on
the internal monopole structure are not too detailed. Fig. \ref{fig_Ilgen}
is taken from the talk given by Ilgenfritz at Lattice 2007 \cite{Ilgenfritz:2007ua}.
Panel (a) and panel (b) show the structure, transversal to a monopole
world line, of the ``non-Abelianicity'' (the gauge potential
 not in the Abelian direction)
and of the modulus of the topological 
density\footnote{Provided by the unfiltered
overlap-defined topological density.}, respectively. 
These two quantities are definitely correlated with
the monopole current, however they display
different sizes. The
 non-Abelianicity shows a narrower peak, with a small width of
only about 
\be r_m\approx 1.5a \approx 0.15 \, \mathrm{fm} \label{eqn_size}\ee
(The lattice spacing was $a = 0.105$ fm in this simulation).
It is hardly surprising that we don't have
any real insights about the monopole structure, as their size
is so close to $a$ (the UV cutoff) of the best lattice calculations.

From Fig.(b) it is evident that the 
topological charge shows a larger radius:
the interpretation is that this is actually the size of
dyons\footnote{The reader is reminded that the lattice definition
of monopoles, based on the end of the Dirac strings, includes
all magnetically charged objects. The density that we showed in
Fig. \ref{fig_density}
also includes all  monopoles and dyons of all kinds.
},
having both $\vec E,\,\vec B$ along the radius.
Although we will not go into this subject, let us note that
selfdual dyons, possessing the topological charge and zero
modes, have been studied
via fermionic methods $without$ any gauge fixing: nice 
correspondence to
Kraan-vanBaal solution \cite{Kraan:1998sn} was found.

The issue of monopole mass is currently under intense study.
We have been informed by D'Elia and D'Alessandro about their
derivation of the mass from monopole paths. We will only mention
that the mass grows with $T$ substantially, justifying the
static monopole approximation adopted in this work. Only  close
to $T_c$ the monopole mass becomes lighter than the gluonic one,
reaching about 300 MeV at $T_c$, as anticipated from
Bose condensation arguments \cite{Shuryak:2008eq}. This
region is excluded from discussion in the present work.

\subsection{Parametric dependence at high $T$}

  Let us now switch to the domain of asymptotically high $T$
and discuss what should happen when the electric sector
gets really weakly coupled, namely when $e^2(T)/4\pi\ll 1$.
  At high $T\gg T_c$ we are in weakly coupled QGP, with ``electric''
degrees of freedom -- quarks and gluons\footnote{One may ask
  why we call gluons ``electric'' excitations, as they have $\vec{E}$ and
  $\vec{B}$ of equal strength. A more accurate formulation of what we mean is:
 they are the excitations which can be inferred directly from the Lagrangian
  of the
``electric'' formulation of the gauge theory, in which the magnetic 
objects are solitonic objects. The ``magnetic'' formulation would have a
Lagrangian with monopoles as sources,
 containing of course the gauge fields but no quarks.
}-- dominating. Perturbative scales include (i) the $hard$ scale $T$;
(ii) the $electric$ scale, e.g. that of the electric mass
 found perturbatively \cite{Shuryak:1977ut}, $E_E\sim eT$.
In the magnetic sector, perturbatively there is no screening but
power infrared divergencies
of the diagrams, on which basis  Polyakov \cite{Polyakov:1978vu}
conjectured that magnetic screening will be developed nonperturbatively at the
(iii) $magnetic$ scale  \be E_M\sim e^2T\sim T/\log(T). \ee
This conjecture is supported by the lattice data at high $T$
 (e.g. by Nakamura {\it et al.}
\cite{Nakamura:2003pu}). To give an idea of the magnitude,
 let us mention that
at $T/T_c\approx 10$ the numerical values
of the two screening masses are $M^{screening}_M/T\approx 1,
M^{screening}_E/T\approx 2.3$. The total contribution of the magnetic
sector to thermodynamics (pressure) is of the order of
\be p_m\sim e^6 N_c^5 T^4,\ee similar to what one obtains
from 4 (and more) gluonic
loops. This value is much smaller than the total
pressure $p\sim N_c^2 T^4$.

Perturbative arguments for the magnetic sector
suggest for its entropy, monopole density, magnetic
screening mass and spatial string tension to scale
 with Polyakov's ``magnetic
scale'' $e^2 T$. 
In fact,  lattice results for the monopole density  
\cite{D'Alessandro:2007su} that we discussed above  
can be fitted as $n_m\sim \log(T)^{-3}$
 for $T/T_c>4$ (omitting the near-$T_c$ points
 which are not expected to
obey the asymptotic behavior).

One  general property of the magnetic scenario \cite{Liao:2006ry}
is the expectation of a strong magnetic coupling at high $T$.
More specifically, it is expected  \cite{Liao:2008jg} to be a
monopole plasma in
 liquid form, with the ratio of potential to kinetic energies
(known as plasma parameter $\Gamma$) to be constant
 $\Gamma(T)\rightarrow\Gamma^*(\infty)\approx 5$. 
Yet, there are many more specific questions about the magnetic sector
which remain open. In general, the monopole properties
are formed under the influence of the following  three important
 effects:
(i) interaction with $A_0$, the ``Higgs field'';
(ii) interaction with other monopoles;
(iii) interaction with ``electric'' quasiparticles, quarks and gluons. 

Let us comment on them, subsequently. 

(i) At nonzero $T$ the role of scalar adjoint Higgs
is taken by $A_0$, whose nonzero VEV breaks the color group
into massive non-Abelian and massless Abelian gauge bosons. 
This phenomenon, related to the expectation of the Polyakov loop,
is widely studied, see e.g. \cite{Ratti:2007jf}: it is most important
 at $T\sim(1-4)T_c$. The
 Euclidean-real
$eA_0/T$ produces a O(1) phase factor -- also known as imaginary
chemical potential -- for electric particles, quarks and gluons.
The corresponding extra $i$ in Minkovski VEV does not allow
to use standard 't Hooft-Polyakov monopole solution
of the  Georgi-Glashow model directly: yet one would still
expect that the monopole mass is heavier than that for quarks/gluons
by the factor $1/e^2(T)\sim \log(T)$, thus with $M_m/T$ growing
at high $T$. 

(ii) Monopole-monopole interaction is Coulombic,
and (as shown in \cite{Liao:2008jg}) the
available lattice data on same and opposite sign monopole
correlations are in good correspondence with MD simulations
for such plasma. One important point is that, at high $T$,
the
 stronger magnetic coupling cancels the relevant diluteness
of monopoles, thus the corrections to their mass are finite
 \be \Delta M_m/T\sim g^2n_{m}^{-1/3}/T\sim g^2 e^2 \sim O(1); \ee 
similar implications hold for the magnetic screening mass
\be \left(M^{screening}_{M}\right)^2\sim g^2 n_m/T\sim g^2 e^6 T^2 \sim (e^2 T)^2\ee
which is thus at the expected ``magnetic scale''.
Below we will consider more detailed models of
the monopole correlations and the collective (magnetic or dual)
potential resulting from them.

(iii) Electric-magnetic interaction is what
 this paper is mostly about. Let us only note here that,
since such interaction is based on
the Lorentz force, it  contains
the coupling combination $ge\sim O(1)$ locked by the
Dirac condition, thus here there is no small
parameter involved.

\subsection{Parametric estimates of the scattering rates}

We define the dimensionless scattering rate of a particle of type 
$i$ on all particles of type $j$ as
\be 
{\dot w_{ij} \over T}={\langle v_{ij} n_j\sigma_{ij}\rangle\over T} 
\ee
where $n_j$ is the density of ``scatterers'' and $v_{ij}\sigma_{ij}$
is the relative velocity and corresponding (transport) cross section.

We start by counting the powers of $\log(T)$
in
the well-known results for $gg\rightarrow gg,\,ggg$ processes,
to be compared with the monopole-related ones. 

We have
\be {\dot w_{gg} \over T}={\langle n_g\sigma_{gg}\rangle\over T}\sim \alpha_s^2
\sim \left({1\over \log(T)}\right)^2 \ee
Although  the radiative process $gg\rightarrow ggg$ is a higher-order process
and has extra coupling, it also has one ``soft''
propagator carrying momentum at the ``electric scale" $eT$,
which puts it at the same level as the elastic process.

 As we argued above, the monopole density scales 
with the ``magnetic'' scale:
 $n_m\sim (e^2 T)^3$ and thus it is small; nevertheless, there is no coupling
constant in the cross section. Thus
\be {\dot w_{gm} \over T}={\langle n_m\sigma_{gm}\rangle\over T}
\sim {\log(\log(T))\over \log^3(T)} \ee
where the factor $\log(\log(T))$ comes from divergent transport cross section
(to be derived below).
Thus, the process that we consider is 
parametrically subleading at high $T$, but not by much.

There are many other details to be included below -- such as numerical
constants in all densities and cross sections -- but at this point
we only address the dependence on one more
general parameter, the number of colors $N_c$. 
The number of gluons scales as $N_c^2$, while that of monopoles
scales as $N_c^1$.
The perturbative $gg$ cross section contains the
 't Hooft coupling squared,
$\lambda^2=(e^2N_c)^2$ with extra powers of $N_c$, but
the result only makes sense as an expansion in $\lambda\ll1$,
so  the limit of large $N_c$, irrespective of $e$, cannot be taken.
The non-perturbative gluon-monopole amplitude has partial 
scattering  phases which are
pure numbers $O(1)$, thus one can rather write that the ratio
of interest is
$\dot w_{gm}/\dot w_{gg}\sim 1/\lambda^2 N_c$,
with small $\lambda$ but large $N_c$.

\subsection{The monopole mass}
 In the original context of the Georgi-Glashow model, the monopole mass
can be written as a perturbative series
\be  M_{m}={4\pi M_W \over g^2}\left[f_0\left({M_H\over M_W}\right)+ {e^2 \over 4\pi}f_1
\left({M_H\over
    M_W}\right)+O(e^4)\right].
\ee
The first classical term contains the known functions $f_0(0)=1,f_0(1)=1.238$,
obtained from the numerical solution of the classical EOM. As for the second term,
only its limit at small argument has been calculated
\cite{Kiselev:1988gf} 
\be 
f_1(z)=\frac{1}{2\pi}\log(z^2) \hspace{1cm}
\mathrm{for}~~z\rightarrow 0. 
\ee

There was one lattice calculation of the monopole mass, by Rajantie
\cite{Rajantie:2005hi}, 
which was performed at $z=M_H/M_W=1$, with
very weak coupling $e^2=1/5$, at which the deviations from the
classical limit were small.
From these results one can estimate that $f_1(1)\approx -(1.5-2)$.
If so,  extrapolating it to $\alpha_e=e^2/4\pi\sim 0.7$ (which we
have at $T\sim$ a few times $T_c$ in QGP), one finds that  quantum corrections
would be as large as the classical mass, cancelling it to zero. While one
cannot trust this extrapolation quantitatively, it certainly shows
the trend. 

What is the mass which is needed to reproduce the observed monopole
 density?
We write the  expression for the density as 
\be 
n_m= {(\mathrm{dof})\over 2\pi^2} \int { p^2 dp\over  \exp[(\epsilon_p+V(0))/T]-1}
\ee
with $\epsilon_p^2=M_{m}^2+\vec{p}^{~2}$. We included in the above expression
the mass as well as the 
{\em correlation energy}, which is due to a 
(dual Coulomb) potential induced by all other monopoles at the
position of one of them. We will write it as
\be 
{V(0)\over T}=-\left({g^2 \over 4\pi a T}\right)M
\ee 
with $a=(n_m)^{-1/3}$ and $M$  the so called ``Madelung'' dimensionless
constant: its value is the subject of the next subsection.
The effective degrees of freedom are dof=2 for the SU(2)
gauge group and 4 for the SU(3) one, which corresponds to
 (spinless) monopoles and antimonopoles.

If one wants the monopole ensemble to mimic the scaling
relations mentioned above, the mass (minus $V(0)$) should
slowly $grow$ with $T$   
\be 
{M_{m}+V(0) \over T}\approx 3 \log (\log(T)) 
\ee
in contrast to the perturbative masses of quarks and gluons 
which slowly decrease:
$m_g/T\sim e \sim [\log(T)]^{-1/2} $.
This leads to the picture that we consider here, namely 
heavy, rare monopoles
embedded into a dense plasma of light gluons and quarks.
If so, the main interaction to be included is a strong-coupling
mutual magnetic Coulomb one.

\subsection{Monopole  interactions}
Before we get into the details of the possible models describing
the monopole ensemble, we provide the numerical values of
some key variables at two temperatures (we choose them near the ends of the
interval but, prudently, we do not take the endpoints):
\vspace{.2cm}\\
\begin{center}
\begin{tabular}{|r|r|r|r|r|} \hline
$T/T_c$ & $n_m (\mathrm{fm}^{-3})$ & $aT$ & $n_m/T^3$ & $g^2/(4\pi aT)$ \\ \hline
1.424 & 2.834 & 1.5 & 0.29 & 1.3 \\
9.865 & 169.2 & 2.6 & 0.056 & 2.4 \\ \hline 
\end{tabular}
\end{center}
\vspace{.2cm}
Let us start by comparing the dimensionless densities 
$n_m/T^3=1/(aT)^3$ in the table, to that of a massless Bose gas
with two
 degrees of freedom, $n/T^3=.244$ (like that for 
the blackbody photons).
While at the higher $T/T_c=9.865$ the monopole density is several times
smaller than this number
 (ascribed to a mass-induced suppression above),
 at the lower  $T/T_c=1.424$ it is somewhat $above$ it,
which would require ``overcompensation'' $M_{m}+V(0)<0$.
Since in this work we will not discuss the near-$T_c$ region,
with all its complications, we conclude for now that  $T/T_c=1.424$
is a bit too low for our basic picture to apply in full.
The reason why we will still include its discussion below
 is pragmatic:  monopole correlations
are not yet known at really high $T$.

The value of the effective plasma parameter $(g^2/ (4\pi a T))$
is given in the above table, the coupling is from correlations \cite{D'Alessandro:2007su}.
Note that,  if the Madelung
constant is simply=1\footnote{We will discuss its
realistic values below.}, the interaction-induced factor
is about $\exp(g^2/ (4\pi a T))= 3.6, 11$ at the two considered 
temperature values. This cannot be taken literally, but shows
that the interaction effect can  be quite substantial.
 The values of $\Gamma$ itself 
grow from 1 at $T=T_c$, to about 4.5 at $T/T_c=4$, as calculated
in ref. 
\cite{Liao:2008jg}. Its $T$-dependence shows that it approaches a
constant value $\Gamma^*(\infty)\approx 5$ for large $T$. 

In the expression for the density, $V(0)$ implicitly depends
on how the other monopoles are distributed in space, so we 
have to solve a manybody plasma problem. 
The distribution of those monopoles in space can be described by three
 models, correponding to (i) ``weak'', (ii) ``medium''
and (iii) ``strong'' correlations. Physically, they correspond to a 
``gas'', ``liquid''
and ``molten salt'' examples, respectively, 
all well known and studied for classical Coulomb plasmas.

In the first case (i)
 we simply ignore the correlations (perturbative Debye
   theory includes a cloud of correlated charges with total charge
-1, but with large size and small Debye
correlation energy $E=-(g^2/(4\pi R_D))\sim g^3 T$).

The second model \cite{LS_note}
 (ii)  is based on two assumptions/inputs:
that the static $two$-body correlation between particles is
given by the lattice results for monopoles  \cite{D'Alessandro:2007su};
and that $manybody$ correlations are absent. If so,
the correlated charge density 
is spherically symmetric and thus 
\be\Delta Q(r)=(n_m/2)\int_0^r(g_{++}-g_{+-})4\pi r^2 dr.\ee
The interaction field and
the potential induced by the correlated charge 
 can be calculated via Gauss' law.
For one typical case discussed in \cite{LS_note} in detail, namely for
$T/T_c=1.42$ when $\alpha_m=g^2/4\pi\approx 2$ and $a\approx .24$
 fm,
the potential induced by the correlated charge is harmonic at small $r$.
Its value at the origin is
$V(0)\approx 35$ MeV, thus the correction induced by it
is $\exp(-V(0)/T)\approx 1.086$. Such corrections $O(10\%)$
are typical for this model (ii). 

The $third$ model (iii) has two different assumptions: (a) that 
local multi-body correlations are
significant;  (b) that those can be  described by
a cubic lattice of alternating charges, since this is known to be
the ground state for ion lattices such as $NaCl$ (thus the name
 ``molten salt'').
Collective potentials
in solid plasmas are
easily calculable from obvious triple sums over the atoms.
Although it is not difficult to calculate them numerically,
as the sum over expanding cubes converges well,
quite a lot of mathematical efforts have been made over 
nearly a century to get
the best analytic representation of it. The latest one that
  we found is
 ref.~\cite{madelung}, which tells that the so called Madelung
 constant is
\be 
M= -{1\over 8} -{\log 2\over 4\pi} -{4\pi\over 3}
+{1\over 2^{3/2}}+{\Gamma(1/8)\Gamma(3/8)\over\pi^{3/2}2^{1/2} }+S\approx-1.74764594
\ee
where $S$ is explicitly exponentially small  $O(10^{-10})$. 

Physicswise, by elevating the Coulomb energy of a particle at
the exact crystal node to a Coulomb potential, induced by all
charges but one, one can see the directions in which it decreases.
This reflects the obvious fact that classically the ion lattice is
simply unstable against opposite charges falling at each other.
That does not happen quantum-mechanically, of course, because
of the ``localization potential'' $\sim \hbar^2/m r^2$. Furthermore,
both atoms in salts and monopoles in QGP are extended objects,
which only interact via the Coulomb law at large distances $r$, while
there
exist
some repulsive cores at small $r$. For molecular dynamics simulations
\cite{Liao:2006ry} we used a potential of the form
\be 
V(r)={g^2\over 4\pi a}\left[ -{1\over (r/a)}+{1\over p (r/a)^p}\right] 
\ee
where $p$ is some power, chosen to be 9. (The factor $(1/p)$
in the second term is needed to make the force zero at distance $r=a$.)
It leads to a nice stable
in-lattice potential
with a near-isotropic harmonic well for small deviations from the node.
The energy per particle for this potential is 
\be 
V(0)\approx -1.1 {g^2\over
4\pi a} 
\ee
So far we don't have lattice results on $manybody$ correlations
of monopoles and thus we are not sure which of those
is more realistic.  
We have presented those results for future comparison,
and also to emphasize that the
collective magnetic (dual) potential is large and non-negligible
compared to the monopole mass. Both should be included in
the monopole density calculations, which are thus more involved
compared to a weakly coupled gas considered before.

\section{Scalar fluctuations}
\label{scalarfluctuations}
We set up the problem of quantum scattering on monopoles in the framework
of the Georgi-Glashow model \cite{ggm} (see Appendix A for the details on the model and the
notations we use). The generalization to QCD will be discussed in the following.
The Georgi-Glashow model describes the interaction between gauge fields (two
massive, spin-1 $W$s and a massless photon) and a scalar Higgs field. We start
with the scattering of the scalar fluctuations on monopoles.
We introduce a total angular momentum operator $\vec{T}$, which is the sum of the orbital angular momentum
$\vec{L}$ and the isotopic spin $\vec{I}$:
\be
\vec{T}=-i\vec{r}\times\vec{\nabla}+\vec{I}
\ee
with $(I^a)_{bc}=-i\epsilon_{abc}$.
In terms of these operators, the wave equation for the scalar fluctuations
can be written in the form (for a derivation of this equation see Appendix A):
\ba
&&
\!\!\!\!\!\!\!\!\!\!\!\!\!\!\!\!\!\!\!
\left[\frac{\partial^2}{\partial r^2}-\frac{2}{r}\frac{\partial}{\partial r}-\frac{\left(\vec{T}^2
-\left(\hat{r}\cdot\vec{I}\right)^2\right)}{r^2}-
\partial_0^2\right]\vec{\chi}+\frac{2K(\xi)\left(\vec{I}\cdot\vec{T}-\left(\hat{r}\cdot\vec{I}\right)^2
\right)}{r^2}\vec{\chi}
\label{eea}
\\
&-&\frac{K(\xi)^2\left[\vec{I}^2-\left(\hat{r}\cdot\vec{I}\right)^2\right]}{r^2}\vec{\chi}
-\lambda\left[2\frac{\vec{r}\cdot\vec{\chi}}{e^2r^4}H(\xi)^2\vec{r}+\left(
\frac{H(\xi)^2}{e^2r^2}-v^2\right)\vec{\chi}\right]=0.
\nonumber
\ea
The term which is proportional to $K[\xi]$ induces charge-exchange reactions.

We can define a simultaneous eigenfunction $\phi_{ti}^{mn}(\hat{r})_a$ of the commuting operators 
$\vec{T}^2$, $T_3$, $\vec{I}$ and $\hat{r}\cdot\vec{I}$ (see eq. (\ref{eigenvalues})).
This function depends only on the angular variables specified by $\hat{r}$. A solution to the
equation (a specific partial wave) 
can be written as the product of the angular function $\phi_{ti}^{mn}(\hat{r})$
and a radial function $S_t^n(r)$
\be
\chi(\vec{r})_a=\phi_{ti}^{mn}(\hat{r})_aS_{t}^{n}(r).
\ee
The angular function $\phi_{ti}^{mn}(\hat{r})_a$ is peculiar because the operator $\vec{I}$ is projected
along $\hat{r}$.
Therefore, the angular function must be rotated, from the standard cartesian frame, to a ``radial" frame.
This construction can be achieved by making use of a {\it spatially dependent} unitary matrix
which rotates $\hat{r}\cdot \vec{I}$ into $I_3$:
\be
U(-\varphi,-\theta,\varphi)=e^{-i\varphi I_3}e^{-i\theta I_2}e^{i\varphi I_3}.
\ee
We therefore have
\ba
\hat{r}\cdot\vec{I}\,\,U(-\varphi,-\theta,\varphi)&=&U(-\varphi,-\theta,\varphi)I_3
\nonumber\\
\vec{T}\,\,U(-\varphi,-\theta,\varphi)&=&U(-\varphi,-\theta,\varphi)\vec{\mathcal{T}}
\ea
where
\be
\vec{\mathcal{T}}=-\vec{r}\times\left(i\vec{\nabla}+e\vec{\mathcal{A}}I_3\right)+\hat{r}I_3
\label{jj}
\ee
and
\be
e\vec{\mathcal{A}}=\frac{\hat{r}\times\hat{z}}{r+z}.
\ee
Eqs. (\ref{eigenvalues}) are satisfied by the following function
\be
\phi_{ti}^{mn}(\hat{r})_a=(U(-\varphi,-\theta,\varphi)\chi_{i}^{n})_a\mathcal{D}(\hat{r})
\ee
where $(\chi_{i}^{n})_a$ is an eigenvector of $I_3$ in the cartesian basis
\be
I_3 (\chi_{i}^{n})_a=n(\chi_{i}^{n})_a
\ee
and the function $\mathcal{D}(\hat{r})$ obeys
  \be
 \left\{ 
 \begin{array}{l}  
 \vec{\mathcal{T}}^2 \\ \mathcal{ T}_3  \end{array}
 \right\}  
\mathcal{D}(\hat{r})=
  \left\{  \begin{array}{l}  
 t(t+1) \\  m \end{array}
 \right\}  
 \mathcal{D}(\hat{r})  
 \label{eigenvalues2}
  \ee
where $I_3$ in eq. (\ref{jj}) is now replaced by its eigenvalue, $n$. We have
\be
\mathcal{D}(\hat{r})=\mathcal{D}^{(t)}_{nm}(-\varphi,\theta,\varphi)=\langle t,n|e^{-i\varphi T_3}
e^{i\theta T_2} e^{i\varphi T_3}|t,m\rangle.
\ee
We can write the function $\phi_{ti}^{mn}(\hat{r})_a$ by making use of the following expansion
\ba
\phi_{ti}^{mn}(\hat{r})_a&=&\sum_{n'}(\chi_{i}^{n'})_a(-1)^{n-n'}\times
\label{solution}\\
&\times&\sum_{l=|t-i|}^{t+i}\langle i,-n,t,n|l,0\rangle
\mathcal{D}_{0,m-n'}^{(l)}(-\varphi,\theta,\varphi)\langle l,m-n'|i,-n',t,m\rangle
\nonumber
\ea
where
\be
\mathcal{D}_{0,m}^{(l)}(\alpha,\beta,\gamma)=\sqrt{\frac{4\pi}{2l+1}}Y_{l}^{m}(\beta,\gamma).
\ee
These functions are normalized in the following way
\be
\int_{0}^{2\pi}d\varphi\int_{-1}^{1}d\cos\theta\phi_{t_1i}^{m_1n_1}(\theta,\varphi)^\dagger\phi_{t_2i}^{m_2n_2}(\theta,\varphi)=\frac{4\pi}{2t_1+1}\delta_{t_1,t_2}\delta_{m_1,m_2}\delta_{n_1,n_2}.
\ee
Since the above function is a polynomial in $x/r,~y/r,~z/r$, the function in eq. (\ref{solution})
is analytic everywhere. This clearly shows that there are compensating singularities in 
$U(-\varphi,-\theta,\varphi)$ and $\mathcal{D}(\hat{r})$.
Another way of building the spherical harmonics \eqref{solution} would be to act with the
vector operators $\vec{r},~(r\vec{\nabla}-i\vec{\Lambda}),~(r\vec{\nabla}+i\vec{\Lambda})$ on the standard angular functions $Y_{tm}(\theta,\varphi)$ (with $\vec{\Lambda}=\vec{r}\times\vec{\nabla}$).
The two methods are totally equivalent, but the one we choose here, following ref. 
\cite{Boulware:1976tv}, provides an easy way to understand the mechanism of rotating from the standard, cartesian frame to a ``radial" one.
Since equation (\ref{eea}) in its most general form admits mixing between particles of different charge,
in order to get the equation for the radial functions $S_t^n$ we have to write the order 1 fluctuations of the field around the classical solutions as a superposition of the functions describing 
a particle with definite charge $n$:
\be
\chi(\vec{r},x_0)_a=\sum_{n=-1,0,1}\mathrm{e}^{i\omega x_0}\frac{S_{t}^{n}(r)}{r}\phi_{ti}^{mn}(\theta,\varphi)_a
\ee
where the three functions $S_{t}^{n}(r)$ correspond to the three physical fluctuations with
charge $n=0,~\pm1$. We plug the above expansion for $\vec{\chi}(\vec{r})$
into Eq. (\ref{eea}); through this procedure we obtain the following system of equations for the radial functions
\begin{subequations}
\begin{align}
\nonumber\\
&S_{t}^{0''}(\xi)-\left(\frac{t(t+1)}{\xi^2}+2\frac{K(\xi)^2}{\xi^2}-\omega^2\right)S_{t}^{0}(\xi)-\lambda
\left(3\frac{H\left(\xi\right)^2}{\xi^2}-1\right)S_{t}^{0}(\xi)
\vspace{.3cm}
\nonumber\\
&+\frac{\sqrt{2t(t+1)}}{\xi^2}K(\xi)\left(S_{t}^{1}(\xi)
+S_{t}^{-1}(\xi)\right)=0
\\
\nonumber\\
&S_{t}^{1''}(\xi)-\left(\frac{t(t+1)-1}{\xi^2}+\frac{K(\xi)^2}{\xi^2}-\omega^2\right)S_{t}^{1}(\xi)
-\lambda\left(\frac{H\left(\xi\right)^2}{\xi^2}-1\right)S_{t}^{1}(\xi)
\vspace{.3cm}
\nonumber\\
&+\frac{\sqrt{2t(t+1)}}{\xi^2}K(\xi)S_{t}^{0}(\xi)=0
\\
\nonumber\\
&S_{t}^{-1''}(\xi)-\left(\frac{t(t+1)-1}{\xi^2}+\frac{K(\xi)^2}{\xi^2}-\omega^2\right)S_{t}^{-1}(\xi)
-\lambda\left(\frac{H\left(\xi\right)^2}{\xi^2}-1\right)S_{t}^{-1}(\xi)
\vspace{.3cm}
\nonumber\\
&+\frac{\sqrt{2t(t+1)}}{\xi^2}K(\xi)S_{t}^{0}(\xi)=0.
\\
\nonumber
\label{scalarsystem}
\end{align}
\label{scalarsystem}
\end{subequations}
where we have introduced the dimensionless variable $\xi=evr$. At the end of this section we will
discuss how to fix the scale and go to physical units.
As it is evident from the above system, a mixing occurs in the monopole core between different charges:
the term $\propto K(\xi)$ involves a mixing between charges that differ by one unit.
The above system of equations has been obtained in the most general case, for generic angular momentum $t$. Nevertheless, we have to keep in mind that there are some restrictions due to
the following requirement:
\be
\hat{r}\cdot\vec{T}=\hat{r}\cdot\vec{L}+\hat{r}\cdot\vec{I}=\hat{r}\cdot\vec{I}=n.
\ee
For this reason, in the case $t=0$ only the $n=0$ scalar fluctuation is allowed. 
The equation for this special case and its solution will be discussed in the following.

For $t>0$, the system of equations (\ref{scalarsystem}) is difficult to solve, due to the mixing
between the different radial functions. This mixing is due to the charge-exchange
reactions that can occur inside the monopole core. If the monopole
core is small (we have seen in Section 2.1 that lattice-based estimates for the monopole size
give $r_m\simeq0.15$ fm)
we can neglect the charge-exchange reactions. 
This corresponds to considering the above system of equations (\ref{scalarsystem}) in the limit
\be
K(\xi)\rightarrow 0~~~~~~~~~~~~~~~H(\xi)\rightarrow\xi.
\ee
In this 
approximation, it reduces to
\begin{subequations}
\begin{align}
&S_{t}^{0''}(\xi)-\left(\frac{t(t+1)}{\xi^2}-\omega^2+2\lambda\right)S_{t}^{0}(\xi)=0
\\
\nonumber\\
&S_{t}^{1''}(\xi)-\left(\frac{t(t+1)-1}{\xi^2}-\omega^2\right)S_{t}^{1}(\xi)=0
\\
\nonumber\\
&S_{t}^{-1''}(\xi)-\left(\frac{t(t+1)-1}{\xi^2}-\omega^2\right)S_{t}^{-1}(\xi)=0.
\end{align}
\label{scalarsystem2}
\end{subequations}
From the above system it is clear that we can identify the radial functions with 
spherical Bessel functions having index $t'$, which
is the positive root of
\be
t'(t'+1)=t(t+1)-n^2.
\ee
Namely, in the limit of small monopole core we have $S_{t}^{n}(r)\rightarrow j_{t'}(kr)$.
In general, $t'$ is not an integer number. The corresponding scattering phase will be
$\delta_{t'}=t'\pi/2$, independent of the energy of the incoming particle.

\begin{figure}[t]
\begin{center}
\includegraphics{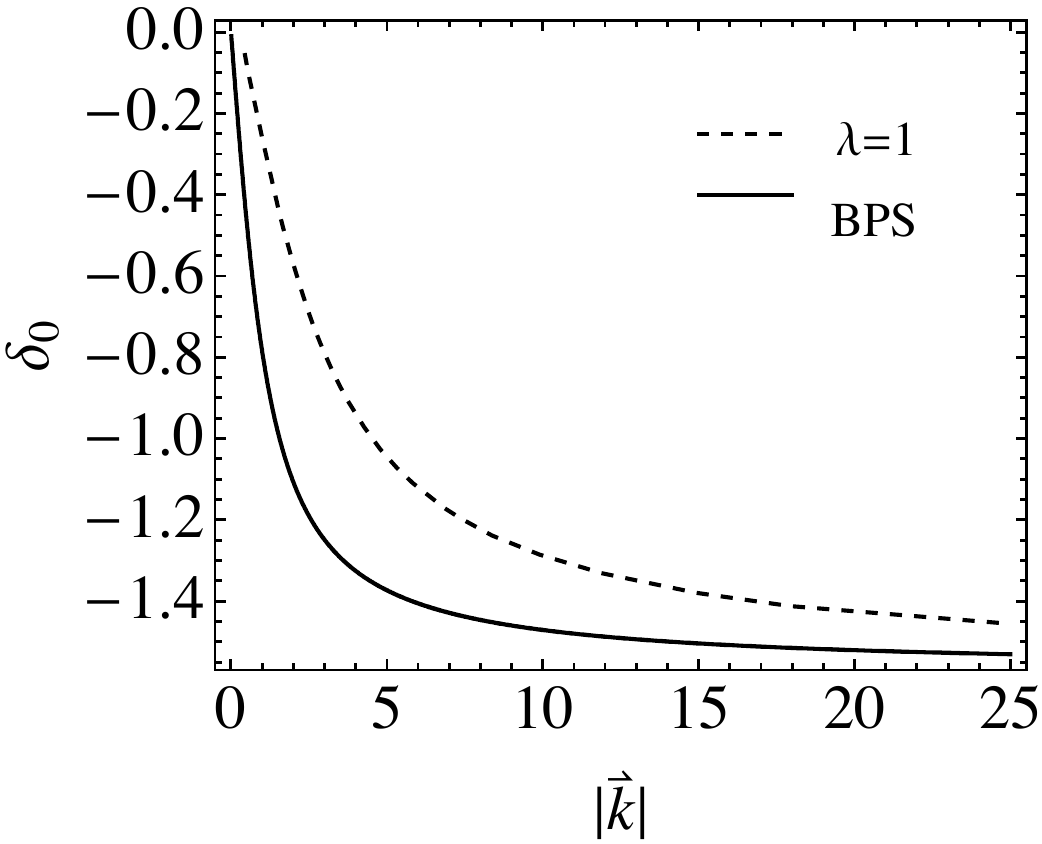}
\caption{Scattering phase $\delta_0$ as a function of $|\vec{k}|=\sqrt{\omega^2-2 \lambda}$. $\delta_0$
is obtained by solving Eq. (\ref{eq0}) with the boundary conditions (\ref{boundaryscalar}). The continuous
line corresponds to the BPS limit ($\lambda=0$), while the dashed line corresponds to $\lambda=1$.}
\label{pds0}
\end{center}
\end{figure}

\begin{figure}[t]
\begin{center}
\includegraphics{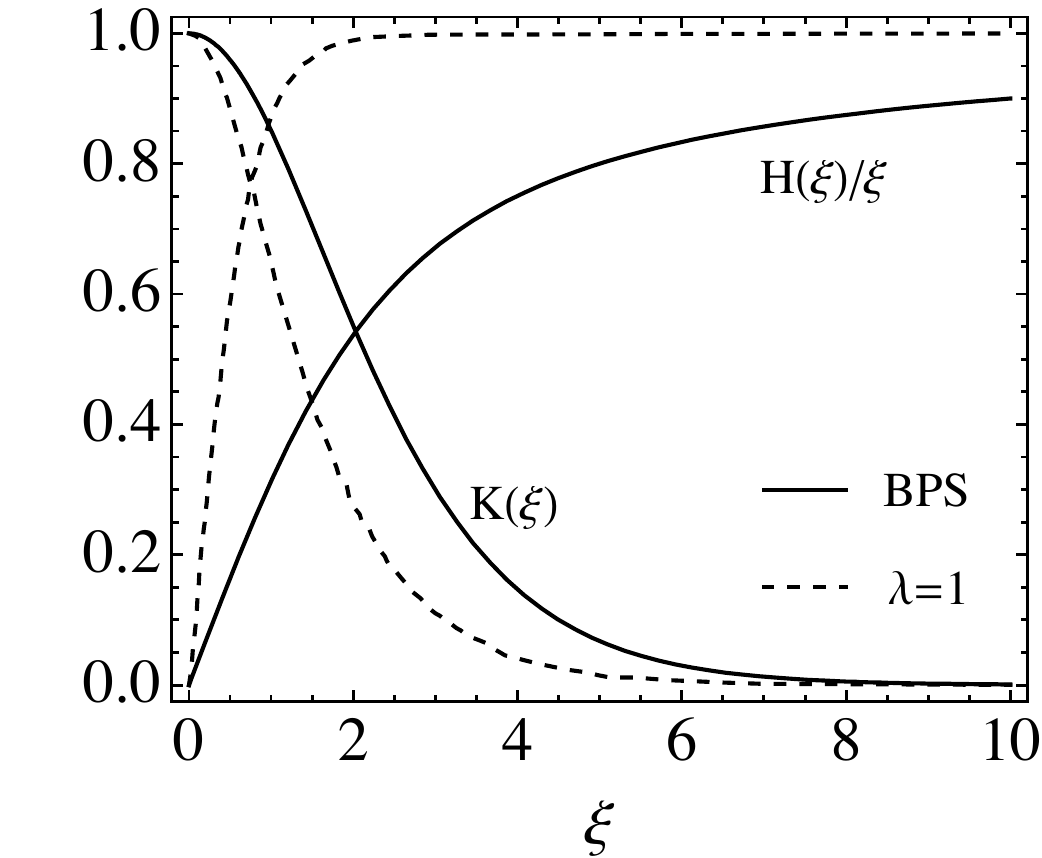}
\caption{Classical solutions $K(\xi)$ and $H(\xi)$ for the 't Hooft Polyakov monopole at $\lambda=0$ 
(BPS limit) and $\lambda=1$ (taken from Ref.~\cite{classic}).}
\label{classical}
\end{center}
\end{figure}

For $t=0$ we have only one fluctuation allowed, namely the one having zero-charge: $S_0^0(\xi)$.
It obeys the following
equation
\be
S_{0}^{0''}(\xi)-\left(2\frac{K(\xi)^2}{\xi^2}\right)S_{0}^{0}(\xi)-\lambda
\left(3\frac{H\left(\xi\right)^2}{\xi^2}-1\right)S_{0}^{0}(\xi)=
-\omega^2S_{0}^{0}\left(\xi\right).
\label{eq0}
\ee
In this case, there is no Coulomb potential of the form $1/\xi^2$, which is obvious since
a charge-neutral particle does not feel the Lorentz force. The scattering in this case is entirely
due to the monopole core.
We can solve the above equation numerically, thus obtaining the scattering phase as a function of the
energy of the incoming particle, by imposing the following boundary conditions
\ba
S_{0}^{0}(\xi)&=&\sin\left[\xi\sqrt{\omega^2-2 \lambda}+\delta_0\right]~~~~~~~~~~~~~~~~~~~~~~~~
\xi\rightarrow\infty
\nonumber\\
S_{0}^{0'}(\xi)&=&\sqrt{\omega^2-2 \lambda}\cos\left[\xi\sqrt{\omega^2-2 \lambda}+\delta_0\right]~~~~~~~~~~~~\xi\rightarrow\infty.
\label{boundaryscalar}
\ea
The scattering phase as a function of $|\vec{k}|=\sqrt{\omega^2-2 \lambda}$ is plotted
in Fig.~\ref{pds0} in the BPS (non-interacting) case, and also for a finite value of
the coupling $\lambda$ in the Higgs potential ($\lambda=1$).  Fig. \ref{classical} shows
the classical solutions
$H(\xi)$ and $K(\xi)$, both in the BPS limit and for $\lambda=1$.
At this point we need to fix the scale in our problem, and to estimate the
scattering length in physical units. In order to do it, we have to connect 
 the physical size of the core of the monopole, $r_m$
(\ref{eqn_size}), to the dimensionless units that we used in the Georgi-Glashow model.
We recall that we obtained $r_m$ through the width at half height of the ``non-Abelianicity",
which is defined as the square of the non-abelian components of the gauge potential. Therefore,
it is natural to 
fix the scale by imposing that, in the Georgi-Glashow model, $r_m$ coincides with $\xi_m$,
the width at half height of the function $K(\xi)^2$. By looking at 
Fig. \ref{classical}, we can see that, in the BPS limit, we have $\xi_m=1.49$, while for $\lambda=1$ 
we have $\xi_m=0.87$. Thus, $ev=9.94$ fm$^{-1}$ in the BPS limit, while $ev=5.8$ fm$^{-1}$
for $\lambda=1$.
The scattering length is defined as $a_{sl}\simeq -d\delta_{t'}/dk$. From Fig. \ref{pds0}, which shows the behavior of $\delta_0$ as a function of $|\vec{k}|$, we get $a_{sl}=1$ in the BPS limit and $a_{sl}=0.6$
for $\lambda=1$. Converting to physical units, we obtain $a_{sl}\simeq0.1$ fm for both values of
$\lambda$.

\section{Vector fluctuations}
\label{vectorfluctuations}
In the case of vector particles, the generalized angular momentum $\vec{J}$ is made up of three
components: the orbital angular momentum $\vec{L}$, the isotopic spin $\vec{I}$ and the spin 
$\vec{S}$:
\be
\vec{J}=\vec{L}+\vec{I}+\vec{S}=\vec{T}+\vec{S}.
\ee
There are generally two different ways of composing three vectors, depending on 
the two vectors to be added first.
The monopole vector spherical harmonics are eigenfunctions of $\vec{J}^{\,2}$ and $J_3$. Due
to the following relation
\be
\hat{r}\cdot\vec{J}=\hat{r}\cdot\vec{L}+\hat{r}\cdot\vec{I}+\hat{r}\cdot\vec{S}=\hat{r}\cdot\vec{I}
+\hat{r}\cdot\vec{S}=n+\sigma,
\label{constraint}
\ee
the allowed values of the total angular momentum quantum number $j$ are $|n|-1,~|n|,...$
except in the case of $n=0$, where $|n|-1$ is absent. 
There are different ways of building the monopole vector spherical harmonics; for example,
they can be constructed by making use of the standard Clebsch-Gordan technique
of addition of momenta.

Another possibility, which we will adopt here, is to build the vector harmonics with $j\geq n$
by applying vector operators to the scalar harmonics, as we will see
in the following \cite{Weinberg:1993sg}. By definition, these harmonics can be introduced as eigenfunctions of the
operator of the radial component of the spin, $\vec{S}\cdot\hat{r}$. We will show that this is
a very useful choice for the ``hedgehog" configuration we are working in:
in fact, there is a natural separation between radial and transverse vectors, making this choice
particularly useful for studying spherically symmetric problems.
The vector harmonics with the minimum allowed angular momentum $j=|n|-1$
cannot be constructed in this way and must be treated specially.
As already mentioned, there is more than one way to obtain a given value of $j$, and thus several
multiplets of harmonics with the same total angular momentum. In the following, we will
classify the multiplets by the eigenvalue of $\hat{r}\cdot\vec{S}=\sigma$. In general,
$\sigma=0,\pm1$, but it is further restricted by the requirement (\ref{constraint}), which
implies that $n+\sigma$ lies in the range $-j$ to $j$. This gives
\begin{itemize}
\item{for $j=0$:
\begin{itemize}
\item{$n=0$ and $\sigma=0$}
\item{$n=1$ and $\sigma=-1$}
\item{$n=-1$ and $\sigma=1$}
\end{itemize}}
\item{for $j=1$ all combinations are allowed, except:
\begin{itemize}
\item{$n=-1$ and $\sigma=-1$}
\item{$n=1$ and $\sigma=1$}
\end{itemize}
}
\end{itemize}
We denote the vector harmonics by $\Phi_{j,n}^{m,\sigma}(\theta,\varphi)_{ai}$. They obey the following eigenvalue equations
\be
 \left\{ 
 \begin{array}{l}  
 \vec{J}^{\,2} \\  J_3 \\  (\hat{r}\cdot\vec{I}) \\  (\hat{r}\cdot \vec{S} ) \end{array}
 \right\}  
 \Phi^{m,\sigma}_{j,n}(\theta,\varphi)_{ai}=
  \left\{  \begin{array}{l}  
 j(j+1) \\  m \\  n \\  \sigma \end{array}
 \right\}  
 \Phi^{m,\sigma}_{j,n}(\theta,\varphi)_{ai}.
 \label{eigenvalues3}
  \ee
 \subsection{$j\geq 2$}
 In this case, all the possible combinations of $n$ and $\sigma$ are allowed.
 We can therefore construct a set of vector spherical harmonics 
 $\Phi_{j,n}^{m,\sigma}(\theta,\varphi)_{ia}$ that will obey the eigenvalue equations 
 (\ref{eigenvalues3}) (in the following, $\vec{\Lambda}=\vec{r}\times\vec{\nabla}$):
 \ba
 \Phi_{j,0}^{m,1}(\theta,\varphi)_{ia}&=&\frac{1}{\sqrt{j(j+1)(2j+1)}}
 \left[\left(r\vec{\nabla}+i\vec{\Lambda}\right)_i\hat{r}_a
Y_{j}^{m}(\theta,\varphi)
\right.
\nonumber\\
&+&\left.\frac{1}{j(j+1)}\left(r\vec{\nabla}+i\vec{\Lambda}\right)_i\left(r\vec{\nabla}-i\vec{\Lambda}\right)_aY_{j}^{m}(\theta,\varphi)\right]
\nonumber\\
\nonumber\\
\nonumber\\
 \Phi_{j,0}^{m,-1}(\theta,\varphi)_{ia}&=&\frac{1}{\sqrt{j(j+1)(2j+1)}}\left[\left(r\vec{\nabla}-i\vec{\Lambda}\right)_i\hat{r}_a
Y_{j}^{m}(\theta,\varphi)\right.
\nonumber\\
&+&\left.\frac{1}{j(j+1)}\left(r\vec{\nabla}-i\vec{\Lambda}\right)_i\left(r\vec{\nabla}+i\vec{\Lambda}\right)_aY_{j}^{m}(\theta,\varphi)\right]
\nonumber
\ea
\ba
\Phi_{j,1}^{m,-1}(\theta,\varphi)_{ia}&=&\frac{1}{j(j+1)\sqrt{2(2j+1)}}\left[\left(r\vec{\nabla}-i\vec{\Lambda}\right)_i\left(r\vec{\nabla}+i\vec{\Lambda}\right)_aY_{j}^{m}(\theta,\varphi)\right]
\nonumber\\
\nonumber\\
\nonumber\\
\Phi_{j,-1}^{m,1}(\theta,\varphi)_{ia}&=&\frac{1}{j(j+1)\sqrt{2(2j+1)}}\left[\left(r\vec{\nabla}+i\vec{\Lambda}\right)_i\left(r\vec{\nabla}-i\vec{\Lambda}\right)_aY_{j}^{m}(\theta,\varphi)\right]
\nonumber\\
\nonumber\\
\nonumber\\
\Phi_{j,1}^{m,1}(\theta,\varphi)_{ia}&=&\sqrt{\frac{2}{(2j+1)(j-1)j(j+1)(j+2)}}\left[\left(r\vec{\nabla}+i\vec{\Lambda}\right)_i\hat{r}_a
Y_{j}^{m}(\theta,\varphi)
\right.
\nonumber\\
&+&
\left.\frac{1}{2}\left(r\vec{\nabla}+i\vec{\Lambda}\right)_i\left(r\vec{\nabla}+i\vec{\Lambda}\right)_aY_{j}^{m}(\theta,\varphi)\right.
\nonumber\\
&+&\left.\frac{1}{j(j+1)}\left(r\vec{\nabla}+i\vec{\Lambda}\right)_i\left(r\vec{\nabla}-i\vec{\Lambda}\right)_aY_{j}^{m}(\theta,\varphi)
\right]
\nonumber\\
\nonumber\\
\nonumber\\
\Phi_{j,-1}^{m,-1}(\theta,\varphi)_{ia}&=&\sqrt{\frac{2}{(2j+1)(j-1)j(j+1)(j+2)}}\left[\left(r\vec{\nabla}-i\vec{\Lambda}\right)_i\hat{r}_a
Y_{j}^{m}(\theta,\varphi)\right.
\nonumber\\
&+&
\left.\frac{1}{2}\left(r\vec{\nabla}-i\vec{\Lambda}\right)_i\left(r\vec{\nabla}-i\vec{\Lambda}\right)_aY_{j}^{m}(\theta,\varphi)
\right.
\nonumber\\
&+&\left.\frac{1}{j(j+1)}\left(r\vec{\nabla}-i\vec{\Lambda}\right)_i\left(r\vec{\nabla}+i\vec{\Lambda}\right)_aY_{j}^{m}(\theta,\varphi)
\right]
\nonumber
\ea
\ba
\Phi_{j,0}^{m,0}(\theta,\varphi)_{ia}&=&\sqrt{\frac{2}{2j+1}}\hat{r}_i\hat{r}_aY_{j}^{m}(\theta,\varphi)
\nonumber\\
\nonumber\\
\nonumber\\
\Phi_{j,1}^{m,0}(\theta,\varphi)_{ia}&=&\frac{1}{\sqrt{j(j+1)(2j+1)}}\hat{r}_i\left(r\vec{\nabla}+i\vec{\Lambda}\right)_a
Y_{j}^{m}(\theta,\varphi)
\nonumber\\
\nonumber\\
\nonumber\\
\Phi_{j,-1}^{m,0}(\theta,\varphi)_{ia}&=&\frac{1}{\sqrt{j(j+1)(2j+1)}}\hat{r}_i\left(r\vec{\nabla}-i\vec{\Lambda}\right)_a
Y_{j}^{m}(\theta,\varphi).
\label{sphericalharmonics}
\\
\nonumber
\ea
The above functions are normalized in the following way:
\be
\int_{0}^{2\pi} {\mathrm d}\varphi\int_{-1}^{1}{\mathrm d}\cos(\theta)(\Phi_{j,n}^{m,\sigma}(\theta,\varphi))^{\dagger}
\Phi_{j',n'}^{m',\sigma'}(\theta,\varphi)=\frac{2}{2j+1}\delta_{jj'}\delta_{nn'}\delta^{mm'}\delta^{\sigma\sigma'}.
\ee

We can therefore expand the vector fluctuations in terms of the above spherical harmonics basis: 
\ba
&&a(\vec{r},x_0)_{ia}=\frac{\mathrm{e}^{i\omega x_0}}{r}\left[T_{j1}(r)\Phi_{j,0}^{m,1}(\theta,\varphi)+T_{j2}(r)\Phi_{j,1}^{m,1}(\theta,\varphi)+
T_{j3}(r)\Phi_{j,-1}^{m,1}(\theta,\varphi)\right.
\nonumber\\
\nonumber\\
&&\left.+T_{j4}(r)\Phi_{j,0}^{m,-1}(\theta,\varphi)+
T_{j5}(r)\Phi_{j,1}^{m,-1}(\theta,\varphi)+T_{j6}(r)\Phi_{j,-1}^{m,-1}(\theta,\varphi)+\right.
\nonumber\\
\nonumber\\
&&\left.T_{j7}(r)\Phi_{j,0}^{m,0}(\theta,\varphi)+T_{j8}(r)\Phi_{j,1}^{m,0}(\theta,\varphi)+
T_{j9}(r)\Phi_{j,-1}^{m,0}(\theta,\varphi)\right]_{ia}.
\label{vectorfluctuation}
\ea
The coefficients
of the expansion are radial functions; by plugging eq. (\ref{vectorfluctuation}) into the corresponding field equation for the vector fluctuations (\ref{fieldeq}), we obtain a system of differential equations
for these coefficients. We find that the equations for $T_{j7}(r),~T_{j8}(r)$ and $T_{j9}(r)$ do not contain
a second derivative with respect to $\xi$. These amplitudes are closely related to
infinitesimal gauge transformations (see for example the discussion in Ref.~\cite{baacke}) and can
in fact be removed by such transformations.
Explicitly we have
\begin{subequations}
\begin{align}
&T_{j1}''(\xi)-\sqrt{\frac{j(j+1)}{2}}\frac{1}{\xi}T_{j7}'(\xi)+\frac{K\left(\xi\right)}{\xi}T_{j8}'(\xi)-\left[-\omega^2+\frac{j\left(j+1\right)+2K\left(\xi\right)^2}{2\xi^2}\right]T_{j1}(\xi)
\nonumber\\
&+\sqrt{\frac{(j-1)(j+2)}{2}}\frac{K\left(\xi\right)}{\xi^2}T_{j2}(\xi)
+\sqrt{\frac{j(j+1)}{2}}\frac{K\left(\xi\right)}{\xi^2}T_{j3}(\xi)
+\frac{j(j+1)}{2\xi^2}T_{j4}(\xi)
\nonumber\\
&
-\sqrt{2j(j+1)}\frac{K\left(\xi\right)}{\xi^2}T_{j5}(\xi)+\sqrt{\frac{j(j+1)}{2}}\frac{1}{\xi^2}T_{j7}(\xi)+
\frac{2K'(\xi)\xi-K(\xi)}{\xi^2}T_{j8}(\xi)=0
\tag{59a}
\nonumber\\
\nonumber
\end{align}
\end{subequations}
\begin{subequations}
\begin{align}
&T_{j2}''(\xi)-\sqrt{\frac{(j-1)(j+2)}{2}}\frac{1}{\xi}T_{j8}'\left(\xi\right)
-\left[-\omega^2+\frac{j\left(j+1\right)-2K\left(\xi\right)^2+2H\left(\xi\right)^2}{2\xi^2}\right]T_{j2}(\xi)
\nonumber\\
&+
\sqrt{\frac{(j-1)(j+2)}{2}}\frac{K\left(\xi\right)}{\xi^2}T_{j1}(\xi)
+\frac{\sqrt{(j-1)j(j+1)(j+2)}}{2\xi^2}T_{j5}(\xi)
\nonumber\\
&+\sqrt{\frac{(j-1)(j+2)}{2}}\frac{1}{\xi^2}T_{j8}\left(\xi\right)=0
\tag{59b}
\nonumber
\end{align}
\end{subequations}
\vspace{-.8cm}
\begin{subequations}
\begin{align}
&T_{j3}''(\xi)+\frac{K\left(\xi\right)}{\xi}T_{j7}'\left(\xi\right)-\sqrt{\frac{j(j+1)}{2}}\frac{1}{\xi}T_{j9}'\left(\xi\right)
\nonumber\\
&-\left[-\omega^2+\frac{j\left(j+1\right)-2+4K\left(\xi\right)^2+2H\left(\xi\right)^2}{2\xi^2}\right]T_{j3}(\xi)
+\sqrt{\frac{j(j+1)}{2}}\frac{K\left(\xi\right)}{\xi^2}T_{j1}(\xi)
\nonumber\\
&
-\sqrt{2j(j+1)}\frac{K\left(\xi\right)}{\xi^2}T_{j4}(\xi)+\frac{K\left(\xi\right)^2}{\xi^2}T_{j5}(\xi)
+\frac{\sqrt{(j-1)j(j+1)(j+2)}}{2\xi^2}T_{j6}(\xi)
\nonumber\\
&+\frac{2K'(\xi)\xi-K(\xi)}{\xi^2}T_{j7}(\xi)
+\sqrt{\frac{j(j+1)}{2}}\frac{1}{\xi^2}T_{j9}\left(\xi\right)
=0
\tag{59c}
\vspace{-.8cm}
\nonumber\\
\nonumber
\end{align}
\end{subequations}
\vspace{-1.3cm}\\
\begin{subequations}
\begin{align}
&T_{j4}''(\xi)-\sqrt{\frac{j(j+1)}{2}}\frac{1}{\xi}T_{j7}'\left(\xi\right)+\frac{K\left(\xi\right)}{\xi}T_{j9}'(\xi)
-\left[-\omega^2+\frac{j\left(j+1\right)+2K\left(\xi\right)^2}{2\xi^2}\right]T_{j4}(\xi)
\nonumber\\
&+\frac{j(j+1)}{2\xi^2}T_{j1}(\xi)-\sqrt{2j(j+1)}\frac{K\left(\xi\right)}{\xi^2}T_{j3}(\xi)
+\sqrt{\frac{j(j+1)}{2}}\frac{K\left(\xi\right)}{\xi^2}T_{j5}(\xi)
\nonumber\\
&+\sqrt{\frac{(j-1)(j+2)}{2}}\frac{K\left(\xi\right)}{\xi^2}T_{j6}(\xi)+
\sqrt{\frac{j(j+1)}{2}}\frac{1}{\xi^2}T_{j7}\left(\xi\right)+\frac{2K'(\xi)\xi-K(\xi)}{\xi^2}T_{j9}(\xi)
=0
\vspace{-.4cm}
\nonumber\\
\tag{59d}
\nonumber
\end{align}
\end{subequations}
\vspace{-1.3cm}\\
\begin{subequations}
\begin{align}
&T_{j5}''(\xi)+\frac{K\left(\xi\right)}{\xi}T_{j7}'(\xi)-\sqrt{\frac{j(j+1)}{2}}\frac{1}{\xi}T_{j8}'\left(\xi\right)
\nonumber\\
&
-\left[-\omega^2+\frac{j\left(j+1\right)-2+4K\left(\xi\right)^2+2H\left(\xi\right)^2}{2\xi^2}\right]T_{j5}(\xi)
-\sqrt{2j(j+1)}\frac{K\left(\xi\right)}{\xi^2}T_{j1}(\xi)
\nonumber\\
&
+\frac{\sqrt{(j-1)j(j+1)(j+2)}}{2\xi^2}T_{j2}(\xi)
+\frac{K\left(\xi\right)^2}{\xi^2}T_{j3}(\xi)+\sqrt{\frac{j(j+1)}{2}}\frac{K\left(\xi\right)}{\xi^2}T_{j4}(\xi)
\nonumber\\
&+\frac{2K'(\xi)\xi-K(\xi)}{\xi^2}T_{j7}(\xi)+
\sqrt{\frac{j(j+1)}{2}}\frac{1}{\xi^2}T_{j8}\left(\xi\right)=0
\vspace{-.4cm}
\nonumber\\
\tag{59e}
\nonumber
\end{align}
\end{subequations}
\vspace{-1.3cm}\\
\begin{subequations}
\begin{align}
&T_{j6}''(\xi)-\sqrt{\frac{(j-1)(j+2)}{2}}\frac{1}{\xi}T_{j9}'\left(\xi\right)
-\left[-\omega^2+\frac{j\left(j+1\right)-2K\left(\xi\right)^2+
2H\left(\xi\right)^2}{2\xi^2}\right]T_{j6}(\xi)
\nonumber\\
&+\frac{\sqrt{(j-1)j(j+1)(j+2)}}{2\xi^2}T_{j3}(\xi)
+
\sqrt{\frac{(j-1)(j+2)}{2}}\frac{K\left(\xi\right)}{\xi^2}T_{j4}(\xi)
\nonumber\\
&
+\sqrt{\frac{(j-1)(j+2)}{2}}\frac{1}{\xi^2}T_{j9}\left(\xi\right)
=0
\vspace{-.4cm}
\nonumber
\tag{59f}
\end{align}
\end{subequations}
\begin{subequations}
\begin{align}
&\frac{T_{j1}'\left(\xi\right)}{\xi}+\left(1-\frac{2K\left(\xi\right)^2}{j\left(j+1\right)}\right)\frac{T_{j4}'\left(\xi\right)}{\xi}
-\frac{\sqrt{2}K\left(\xi\right)}{\sqrt{j\left(j+1\right)}}\frac{T_{j5}'\left(\xi\right)}{\xi}
\nonumber\\
&+\frac{\sqrt{2(j-1)(j+2)}K\left(\xi\right)}{j(j+1)}\frac{T_{j6}'(\xi)}{\xi}
+\frac{\sqrt{2}K'\left(\xi\right)}{\sqrt{j\left(j+1\right)}}\frac{T_{j3}(\xi)}{\xi}+
\frac{2K\left(\xi\right)K'\left(\xi\right)}{j\left(j+1\right)}\frac{T_{j4}(\xi)}{\xi}
\nonumber\\
&+\frac{\sqrt{2}K'\left(\xi\right)}{\sqrt{j\left(j+1\right)}}\frac{T_{j5}(\xi)}{\xi}
+\frac{\sqrt{2}}{\sqrt{j\left(j+1\right)}}\left(\omega^2-\frac{j\left(j+1\right)}{\xi^2}\right)T_{j7}(\xi)
+\frac{2K\left(\xi\right)}{\xi^2}T_{j8}\left(\xi\right)
\nonumber\\
&-\frac{2 K\left(\xi\right) \left(-1+H\left(\xi\right)^2+K\left(\xi\right)^2-\omega^2 \xi^2\right)}
{j\left(j+1\right)\xi^2}T_{j9}\left(\xi\right)=0
\nonumber\\
\tag{59g}
\nonumber
\end{align}
\end{subequations}
\begin{subequations}
\begin{align}
&
\frac{T_{j3}'\left(\xi\right)}{\xi}-\frac{\sqrt{2} K\left(\xi\right)}{\sqrt{j (1+j)}}\frac{T_{j4}'\left(\xi\right)}\xi
+\sqrt{\frac{(j-1)(2+j)}{j(j+1)}} \frac{T_{j6}'\left(\xi\right)}{\xi}
+\frac{\sqrt{2} K'\left(\xi\right)}{\sqrt{j (1+j)}}\frac{T_{j4}(\xi)}{\xi}
\nonumber\\
&+\frac{2 K\left(\xi\right)}{\xi^2}T_{j7}\left(\xi\right)-\frac{\sqrt{2}\left(H\left(\xi\right)^2+j(j+1)-1+
K\left(\xi\right)^2-\omega^2 \xi^2\right)}{\sqrt{j (j+1)} \xi^2}T_{j9}\left(\xi\right)=0
\nonumber\\
\tag{59h}
\nonumber
\end{align}
\end{subequations}
\begin{subequations}
\begin{align}
&
\frac{T_{j2}'\left(\xi\right)}{\xi}+\frac{\sqrt{2}K\left(\xi\right)\left(j(j+1)-2K\left(\xi\right)^2\right)}{
j(j+1)\sqrt{(j-1)(j+2)}}\frac{T_{j4}'\left(\xi\right)}{\xi}+
\frac{\left(j(j+1)-2K\left(\xi\right)^2\right)}{
\sqrt{(j-1)j(j+1)(j+2)}}\frac{T_{j5}'\left(\xi\right)}{\xi}
\nonumber\\
&+\frac{2 K\left(\xi\right)^2}{j(j+1)}\frac{T_{j6}'\left(\xi\right)}{\xi}+
\frac{\sqrt{2}K'\left(\xi\right)}{\sqrt{(j-1)(j+2)}}\frac{T_{j1}\left(\xi\right)}{\xi}
+\frac{2K\left(\xi\right)K'\left(\xi\right)}{\sqrt{(j-1)j(j+1)(j+2)}}\frac{T_{j3}\left(\xi\right)}{\xi}
\nonumber\\
&
+\frac{2\sqrt{2}K\left(\xi\right)^2K'\left(\xi\right)}{
j(j+1)\sqrt{(j-1)(j+2)}}\frac{T_{j4}\left(\xi\right)}{\xi}+\frac{2K\left(\xi\right)K'\left(\xi\right)}{
\sqrt{(j-1)j(j+1)(j+2)}}\frac{T_{j5}\left(\xi\right)}{\xi}
\nonumber\\
&
-\frac{2\omega^2K\left(\xi\right)}{\sqrt{(j-1)j(j+1)(j+2)}}T_{j7}\left(\xi\right)
-\frac{\sqrt{2}\left(H\left(\xi\right)^2+j(j+1)-1-
K\left(\xi\right)^2-\omega^2 \xi^2\right)}{\sqrt{(j-1) (j+2)} \xi^2}T_{j8}\left(\xi\right)
\nonumber\\
&
-\frac{2\sqrt{2}K\left(\xi\right)^2\left(H\left(\xi\right)^2-1+
K\left(\xi\right)^2-\omega^2 \xi^2\right)}{
j(j+1)\sqrt{(j-1)(j+2)}}\frac{T_{j9}\left(\xi\right)}{\xi^2}=0
\nonumber
\tag{59i}
\end{align}
\end{subequations}

As we already did for the scalar channel, we consider the above system in the limit of small
monopole core.
This corresponds to the limit
\be
K(\xi)\rightarrow 0~~~~~~~~~~~~~~~H(\xi)\rightarrow\xi.
\setcounter{equation}{60}
\ee
Besides, we gauge away the unphysical longitudinal components of our fields
($T_{j7}(\xi)$, $T_{j8}(\xi),~T_{j9}(\xi)$), after which the
above system reduces to:
\begin{subequations}
\begin{align}
&T_{j1}''(\xi)-\left[-\omega^2+\frac{j\left(j+1\right)}{2\xi^2}\right]T_{j1}(\xi)
+\frac{j(j+1)}{2\xi^2}T_{j4}(\xi)=0
\\
\nonumber
\\
&T_{j4}''(\xi)-\left[-\omega^2+\frac{j\left(j+1\right)}{2\xi^2}\right]T_{j4}(\xi)
+\frac{j(j+1)}{2\xi^2}T_{j1}(\xi)=0
\\
\nonumber
\\
&
T_{j2}''(\xi)
-\left[-\omega^2+1+\frac{j\left(j+1\right)}{2\xi^2}\right]T_{j2}(\xi)
+\frac{\sqrt{(j-1)j(j+1)(j+2)}}{2\xi^2}T_{j5}(\xi)=0
\\
\nonumber
\\
&
T_{j5}''(\xi)
-\left[-\omega^2+1+\frac{j\left(j+1\right)-2}{2\xi^2}\right]T_{j5}(\xi)
+\frac{\sqrt{(j-1)j(j+1)(j+2)}}{2\xi^2}T_{j2}(\xi)
=0
\\
\nonumber
\\
&T_{j3}''(\xi)
-\left[-\omega^2+1+\frac{j\left(j+1\right)-2}{2\xi^2}\right]T_{j3}(\xi)
+\frac{\sqrt{(j-1)j(j+1)(j+2)}}{2\xi^2}T_{j6}(\xi)=0
\\
\nonumber
\\
&T_{j6}''(\xi)
-\left[-\omega^2+1+\frac{j\left(j+1\right)}{2\xi^2}\right]T_{j6}(\xi)
+\frac{\sqrt{(j-1)j(j+1)(j+2)}}{2\xi^2}T_{j3}(\xi)
=0
\\
\nonumber
\\
&T_{j1}'(\xi)=-T_{j4}'(\xi)
\\
\nonumber
\\
&T_{j2}'(\xi)=-\sqrt{\frac{j(j+1)}{(j-1)(j+2)}}T_{j5}'(\xi)
\\
\nonumber
\\
&T_{j3}'(\xi)=-\sqrt{\frac{(j-1)(j+2)}{j(j+1)}}T_{j6}'(\xi)
\end{align}
\end{subequations}
The last three equations in the above set are what remains of the equations for the longitudinal
components of the fields, namely for $T_{j7}(\xi),~T_{j8}(\xi),~T_{j9}(\xi)$, after we gauge these components away. They can be used as constraints for the other six $T_{ji}$s. Indeed, after taking into account these
constraints, we end up with the following set of (diagonal) equations for our radial functions:
\begin{subequations}
\begin{align}
&T_{j1}''(\xi)-\left[-\omega^2+\frac{j\left(j+1\right)}{\xi^2}\right]T_{j1}(\xi)=0
\\
\nonumber\\
&T_{j4}''(\xi)-\left[-\omega^2+\frac{j\left(j+1\right)}{\xi^2}\right]T_{j4}(\xi)=0
\\
\nonumber\\
&
T_{j2}''(\xi)
-\left[-\omega^2+1+\frac{j\left(j+1\right)-1}{\xi^2}\right]T_{j2}(\xi)=0
\\
\nonumber\\
&
T_{j5}''(\xi)
-\left[-\omega^2+1+\frac{j\left(j+1\right)-1}{\xi^2}\right]T_{j5}(\xi)
=0
\\
\nonumber\\
&T_{j3}''(\xi)
-\left[-\omega^2+1+\frac{j\left(j+1\right)-1}{\xi^2}\right]T_{j3}(\xi)=0
\\
\nonumber\\
&T_{j6}''(\xi)
-\left[-\omega^2+1+\frac{j\left(j+1\right)-1}{\xi^2}\right]T_{j6}(\xi)
=0.
\label{diag}
\end{align}
\end{subequations}
We observe that the above equations are those corresponding to Bessel functions with index
$j'=\frac12(-1+\sqrt{(2j+1)^2-4n^2})$. In analogy with the scalar fluctuations, also in this case we have
a constant scattering phase $\delta_{j'}=j'\pi/2$.
The same set of equations can be obtained, as a check, if we consider the general field equations 
for the fluctuations, in the limit of pointlike monopole core for the three fields $W_{n}~(n=0,\pm1)$:
\be
\left[\frac1\xi^2\frac{\partial}{\partial \xi}(\xi^2\frac{\partial}{\partial \xi})-\frac{\vec{T}^2-\left(\vec{I}\cdot\hat{r}
\right)^2+ 2n\left(\vec{S}\cdot\hat{r}\right)}{\xi^2}\right]W_n=0.
\label{eqshnir}
\ee
We remember that the operator $\vec{T}$ is given by
\be
\vec{T}=\vec{L}+\vec{I}=\vec{J}-\vec{S}
\ee
and therefore we have
\be
\vec{T}^2=\vec{J}^2+\vec{S}^2-2\vec{J}\cdot\vec{S}.
\ee
In the above equation, the only nondiagonal operator is $2\vec{J}\cdot\vec{S}$
and we therefore have to understand how it acts on our angular functions.
We obtain:
\ba
&&2\vec{J}\cdot\vec{S}\Phi_{j,n}^{0,1}(\theta,\varphi)=\sqrt{2(j-n)(j+1+n)}\Phi_{j,n}^{0,0}(\theta,\varphi)
+2(1+n)\Phi_{j,n}^{0,1}(\theta,\varphi)
\nonumber\\
&&\\
&&2\vec{J}\cdot\vec{S}\Phi_{j,n}^{0,-1}(\theta,\varphi)=\sqrt{2(j+n)(j+1-n)}\Phi_{j,n}^{0,0}(\theta,\varphi)
+2(1-n)\Phi_{j,n}^{0,-1}(\theta,\varphi).
\nonumber
\ea
Since the above equations involve mixing with the longitudinal components only (for example 
$\Phi_{j,n}^{0,1}(\theta,\varphi)$ only gets mixed with $\Phi_{j,n}^{0,0}(\theta,\varphi)$
and not with $\Phi_{j,n}^{0,-1}(\theta,\varphi)$, and since we have seen that the longitudinal
components are spurious and can be gauged away, we conclude that (\ref{eqshnir}) will
produce diagonal equations for the remaining radial functions.
In particular we get
\ba
&&\left[-\frac{\vec{J}^{~2}+\vec{S}^{~2}-2\vec{J}\cdot\vec{S}-\left(\vec{I}\cdot\hat{r}
\right)^2\pm 2\left(\vec{S}\cdot\hat{r}\right)}{\xi^2}\right]\Phi_{j,n}^{0,1}(\theta,\varphi)=
\nonumber\\
&&\left[-\frac{j(j+1)+s(s+1)-2(1+n)-n^2+2n}{\xi^2}\right]\Phi_{j,n}^{0,1}(\theta,\varphi)
\nonumber\\
&&\nonumber\\
&&\nonumber\\
&&\left[-\frac{\vec{J}^{~2}+\vec{S}^{~2}-2\vec{J}\cdot\vec{S}-\left(\vec{I}\cdot\hat{r}
\right)^2\pm 2\left(\vec{S}\cdot\hat{r}\right)}{\xi^2}\right]\Phi_{j,n}^{0,-1}(\theta,\varphi)=
\nonumber\\
&&\left[-\frac{j(j+1)+s(s+1)-2(1-n)-n^2+2n}{\xi^2}\right]\Phi_{j,n}^{0,-1}(\theta,\varphi).
\label{eqshnir2}
\ea
which results in the set of equations (62) for the radial functions.
\subsection{j=0}
\begin{figure}[t]
\begin{center}
\includegraphics{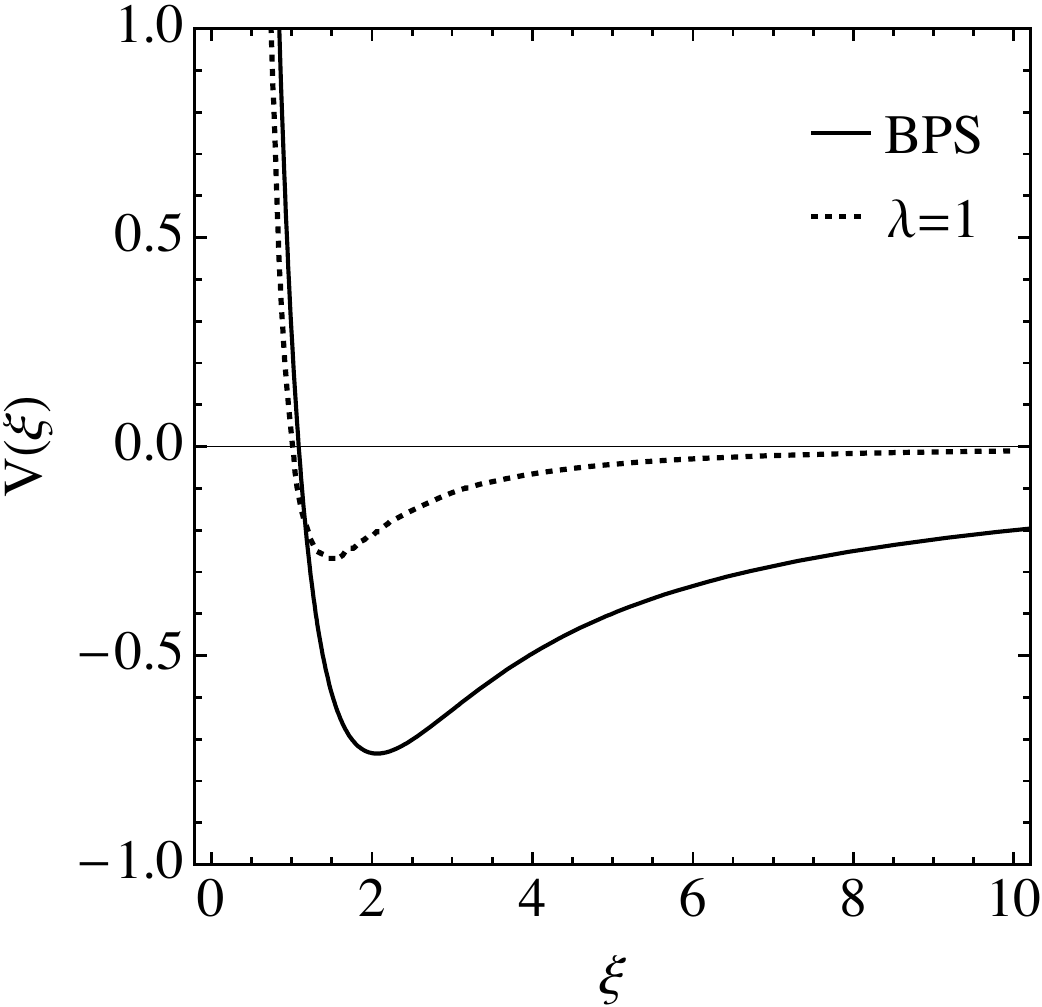}
\caption{Behavior of $V(\xi)$, the potential in the radial equations at $j=0$, in the
BPS limit (solid line) and for $\lambda=1$ (dashed line).}
\label{potential}
\end{center}
\end{figure}

\begin{figure}
\begin{minipage}{.48\textwidth}
\parbox{6cm}{
\scalebox{.65}{
\includegraphics{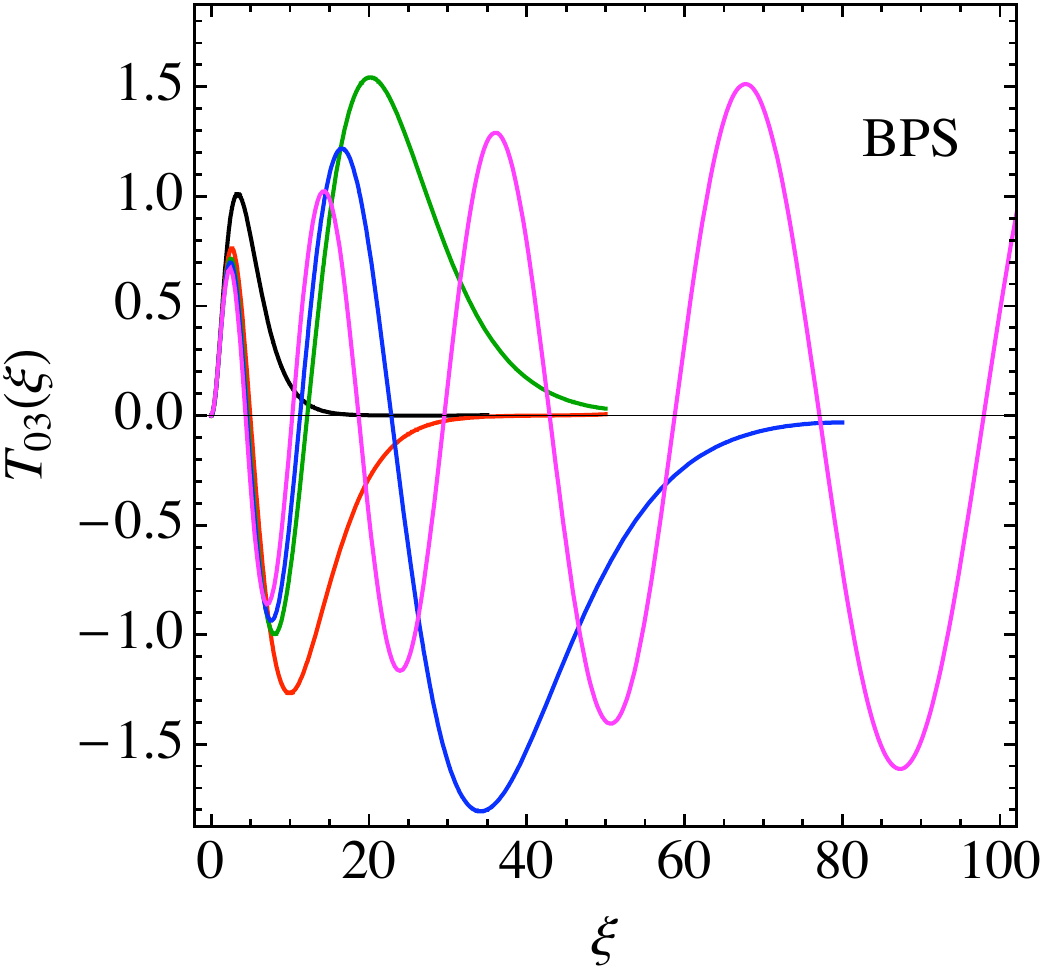}\\}}
\end{minipage}
\hspace{.4cm}
\begin{minipage}{.48\textwidth}
\parbox{6cm}{
\scalebox{.65}{
\includegraphics{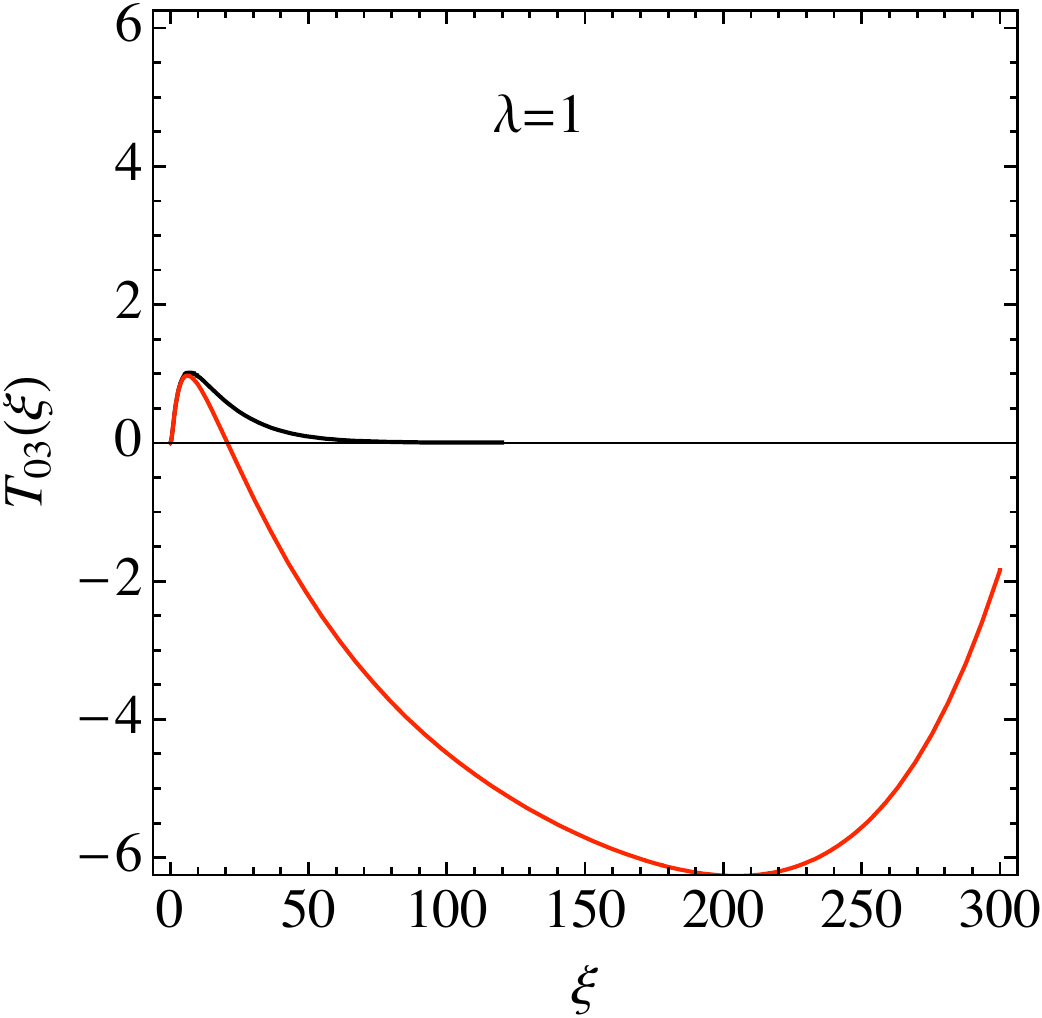}\\}}
\end{minipage}
\caption{Some radial wave functions
 in the BPS limit (left panel) and for $\lambda=1$ (right panel)
corresponding to bound states in the j=0 channel.}
\label{boundstates}
\end{figure}

As we have seen, for $j=0$, only three combinations of $n$ and $\sigma$ are allowed
\begin{itemize}
\item{$n=0$ and $\sigma=0$}
\item{$n=1$ and $\sigma=-1$}
\item{$n=-1$ and $\sigma=1$}.
\end{itemize}
We construct the corresponding set of vector spherical harmonics in the following way
\ba
(\Phi^{0,0}_{0,0}(\theta,\varphi))_{ia}&=&\sqrt{2}\hat{r}_i\hat{r}_aY_{0}^{0}(\theta,\varphi)
\nonumber\\
(\Phi^{0,-1}_{0,1}(\theta,\varphi))_{ia}&=&\frac{1}{\sqrt{2}}\left[r\vec{\nabla}-i\vec{\Lambda}\right]_i\hat{r}_a
Y_{0}^{0}(\theta,\varphi)
\nonumber\\
(\Phi^{0,1}_{0,-1}(\theta,\varphi))_{ia}&=&\frac{1}{\sqrt{2}}\left[r\vec{\nabla}+i\vec{\Lambda}\right]_i\hat{r}_a
Y_{0}^{0}(\theta,\varphi).
\ea
The above functions are normalized in the following way
\be
\int_{0}^{2\pi} {\mathrm d}\varphi\int_{-1}^{1}{\mathrm d}\cos(\theta)(\Phi_{0,n}^{0,\sigma}(\theta,\varphi))^{\dagger}
\Phi_{0,n'}^{0,\sigma'}(\theta,\varphi)=2\delta_{nn'}\delta^{\sigma\sigma'}.
\ee
We can write the vector fluctuations for $j=0$ as a superposition of the above vector spherical harmonics:
\be
a(\vec{r},x_0)_{ia}=\frac{\mathrm{e}^{i\omega x_0}}{r}\left(T_{07}(r)\Phi^{0,0}_{0,0}(\theta,\varphi)+
T_{03}(r)\Phi^{0,1}_{0,-1}(\theta,\varphi)+T_{05}(r)\Phi^{0,-1}_{0,1}(\theta,\varphi)\right)_{ia}.
\ee
The coefficients $T_{0i}(r)$ in the above expansion are radial functions which obey 
a set of coupled differential equations. The equation for the function $T_{07}(r)$ does not contain
a second derivative with respect to $\xi$. This amplitude is closely related to
an infinitesimal gauge transformation (see for example the discussion in Ref.~\cite{baacke}) and can
in fact be removed by such transformation.
The remaining two radial functions obey the following set of equations
\begin{subequations}
\begin{align}
&T_{03}''(\xi)-\left(\frac{H(\xi)^2+2K(\xi)^2-1}{\xi^2}-\omega^2\right)T_{03}(\xi)+\frac{K(\xi)^2}{\xi^2}T_{05}(\xi)=0
\\
\nonumber\\
&T_{05}''(\xi)-\left(\frac{H(\xi)^2+2K(\xi)^2-1}{\xi^2}-\omega^2\right)T_{05}(\xi)+\frac{K(\xi)^2}{\xi^2}T_{03}(\xi)=0
\\
\nonumber\\
&T_{03}'(\xi)+T_{05}'(\xi)-\frac{K'(\xi)}{K(\xi)}T_{03}(\xi)-\frac{K'(\xi)}{K(\xi)}T_{05}(\xi)=0
\label{vectorsystemj0}
 \end{align}
 \end{subequations}
 The last equation above can be used as a constraint on $T_{03}(\xi)$ and $T_{05}(\xi)$,
in order to obtain two decoupled equations for these functions:
\begin{subequations}
\begin{align}
&T_{03}''(\xi)-\left(\frac{H(\xi)^2+3K(\xi)^2-1}{\xi^2}-\omega^2\right)T_{03}(\xi)=0
\\
\nonumber\\
&T_{05}''(\xi)-\left(\frac{H(\xi)^2+3K(\xi)^2-1}{\xi^2}-\omega^2\right)T_{05}(\xi)=0
\label{vectorsystemj0diag}
 \end{align}
 \end{subequations}
We can solve the above equations both in the BPS limit and for a finite value of $\lambda$ (for example
we choose $\lambda=1$ in analogy with what we did in the scalar channel), by imposing
the following boundary conditions for $\xi\rightarrow\infty$:
\ba
T_{03}(\xi)&=&\sin\left[\xi\sqrt{\omega^2-1}+\delta_0\right]
\nonumber\\
\nonumber\\
T_{03}'(\xi)&=&\sqrt{\omega^2-1}\cos\left[\xi\sqrt{\omega^2-1}+\delta_0\right]
\nonumber\\
\nonumber\\
T_{05}(\xi)&=&\sin\left[\xi\sqrt{\omega^2-1}+\delta_0\right]
\nonumber\\
\nonumber\\
T_{05}'(\xi)&=&\sqrt{\omega^2-1}\cos\left[\xi\sqrt{\omega^2-1}+\delta_0\right].
\label{boundary}
\ea
We recall that the classical solutions for $K(\xi)$ and $H(\xi)$ that we plug in the equations are 
the ones that we show in Fig. \ref{classical}, both in the BPS limit and for $\lambda=1$. 
We plot the potential $V(\xi)=(H(\xi)^2+3K(\xi)^2-1)/\xi^2-1$ 
in Fig. \ref{potential} for these two values of $\lambda$ 
(we have subtracted the mass of the particle, 1 in dimensionless units); we see that it can
in principle admit bound states. In particular, in the BPS limit $V(\xi)\sim1/\xi$ at large $\xi$, namely
we have a Coulomb potential which admits infinitely many bound states.
For $\lambda=1$  the number of bound states is reduced: we still find two of them, as it can be seen 
from Fig. \ref{boundstates}, which shows the wave function $T_{03}(\xi)$ in the case of negative
(quantized) energy.

\begin{figure}
\begin{center}
\includegraphics{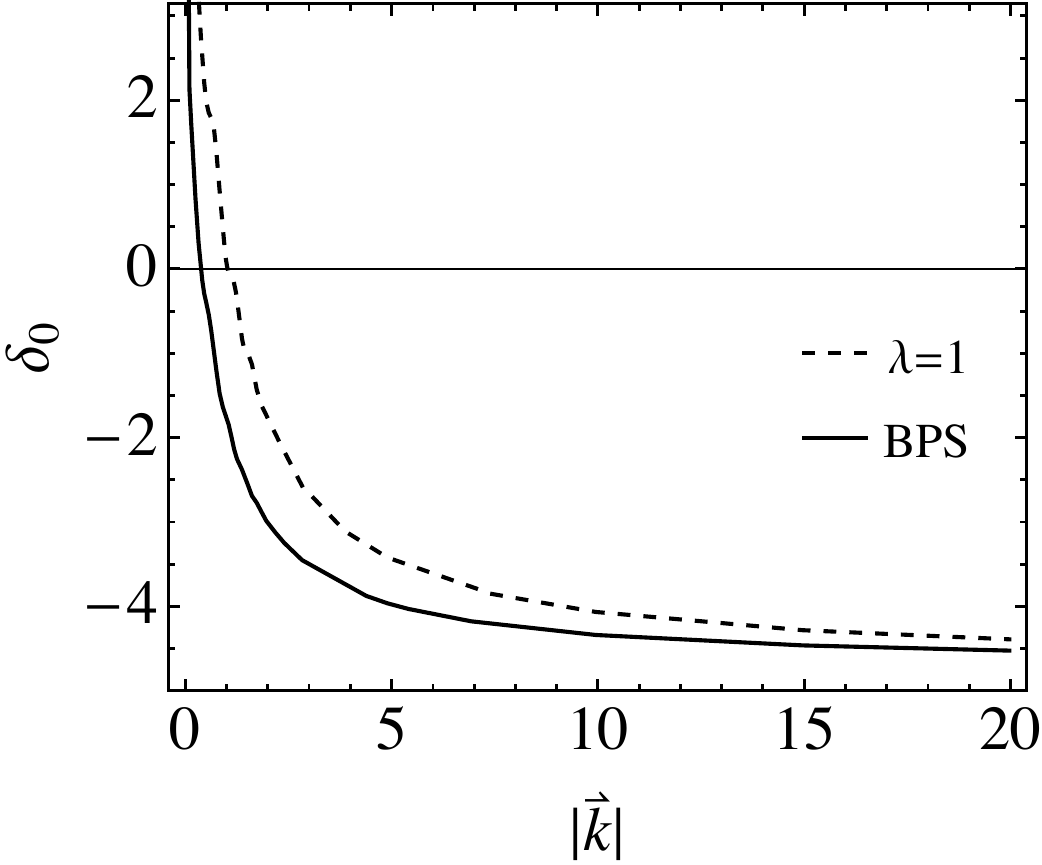}
\caption{Scattering phase $\delta_0$ as a function of $|\vec{k}|=\sqrt{\omega^2-1}$. The continuous
line corresponds to the BPS limit ($\lambda=0$), while the dashed line corresponds to $\lambda=1$.}
\label{deltavector}
\end{center}
\end{figure}

By solving the above equations, we obtain the resulting scattering phase $\delta_0$ as a function of the momentum $|\vec{k}|$ of the incoming particle: it is shown in Fig. \ref{deltavector} for the two values of $\lambda$ that we consider.
As we already did for the scalar case, by calculating $-d\delta_0/dk$ at $|\vec{k}|=0$ we can
obtain the value of the scattering length $a_{sl}$. In the BPS limit, since we have infinitely many
bound states, we also have an undetermined scattering length. For $\lambda=1$ we get 
 $d\delta_0/dk\simeq -10$ in dimensionless units, which translated to physical units gives a
 scattering length $a_{sl}\simeq 1.7$ fm. This rather large value can be explained by the presence
 of the two bound states: in particular, the second one is very weakly bound, therefore it
 shows a rather large effective radius.
\subsection{$j=1$}
 In this case, the set of vector spherical harmonics coincides with \eqref{sphericalharmonics}, but
 two of them are missing, namely $\Phi_{1,1}^{m,1}(\theta,\varphi)$ and $\Phi_{1,-1}^{m,-1}
 (\theta,\varphi)$. The corresponding equations for the radial functions are in this case:
 \begin{subequations}
\begin{align}
T_{11}''(\xi)&-\left[-\omega^2+\frac{1+K\left(\xi\right)^2}{\xi^2}\right]T_{11}(\xi)+
\frac{K\left(\xi\right)}{\xi^2}T_{13}(\xi)
\nonumber\\
&
+\frac{1}{\xi^2}T_{14}(\xi)
-2\frac{K\left(\xi\right)}{\xi^2}T_{15}(\xi)=0
\tag{75a}
\nonumber\\
\nonumber\\
T_{13}''(\xi)
&-\left[-\omega^2+\frac{2K\left(\xi\right)^2+H\left(\xi\right)^2}{\xi^2}\right]T_{13}(\xi)
\nonumber\\
&+\frac{K\left(\xi\right)}{\xi^2}T_{11}(\xi)
-2\frac{K\left(\xi\right)}{\xi^2}T_{14}(\xi)+\frac{K\left(\xi\right)^2}{\xi^2}T_{15}(\xi)=0
\tag{75b}
\end{align}
\end{subequations}
\begin{subequations}
\begin{align}
T_{14}''(\xi)&-\left[-\omega^2+\frac{1+K\left(\xi\right)^2}{\xi^2}\right]T_{14}(\xi)
+\frac{1}{\xi^2}T_{11}(\xi)
\nonumber\\
&-2\frac{K\left(\xi\right)}{\xi^2}T_{13}(\xi)
+\frac{K\left(\xi\right)}{\xi^2}T_{15}(\xi)
=0
\tag{75c}\\
\nonumber\\
T_{15}''(\xi)
&-\left[-\omega^2+\frac{2K\left(\xi\right)^2+H\left(\xi\right)^2}{\xi^2}\right]T_{15}(\xi)
-2\frac{K\left(\xi\right)}{\xi^2}T_{11}(\xi)
\nonumber\\
&+\frac{K\left(\xi\right)^2}{\xi^2}T_{13}(\xi)
+\frac{K\left(\xi\right)}{\xi^2}T_{14}(\xi)
=0.
\tag{75d}\\
\nonumber\\
\frac{T_{11}'\left(\xi\right)}{\xi}&+\left(1-K\left(\xi\right)^2\right)\frac{T_{14}'\left(\xi\right)}{\xi}
-K\left(\xi\right)\frac{T_{15}'\left(\xi\right)}{\xi}
\nonumber\\
&+K'\left(\xi\right)\frac{T_{13}(\xi)}{\xi}+
K\left(\xi\right)K'\left(\xi\right)\frac{T_{14}(\xi)}{\xi}
+K'\left(\xi\right)\frac{T_{15}(\xi)}{\xi}=0
\tag{75e}\\
\nonumber\\
\frac{T_{13}'\left(\xi\right)}{\xi}&-K\left(\xi\right)\frac{T_{14}'\left(\xi\right)}\xi
+K'\left(\xi\right)\frac{T_{14}(\xi)}{\xi}=0
\tag{75f}\\
\nonumber\\
K\left(\xi\right)&\left(1-K\left(\xi\right)^2\right)\frac{T_{14}'\left(\xi\right)}{\xi}+
\left(1-K\left(\xi\right)^2\right)\frac{T_{15}'\left(\xi\right)}{\xi}
+
K'\left(\xi\right)\frac{T_{11}\left(\xi\right)}{\xi}
\nonumber\\
\nonumber\\
+K\left(\xi\right)&K'\left(\xi\right)\frac{T_{13}\left(\xi\right)}{\xi}
+K\left(\xi\right)^2K'\left(\xi\right)\frac{T_{14}\left(\xi\right)}{\xi}+K\left(\xi\right)K'\left(\xi\right)
\frac{T_{15}\left(\xi\right)}{\xi}=0
\tag{75g}
\end{align}
\end{subequations}
The above system, in the limit of large distances reduces to
\begin{subequations}
\begin{align}
&T_{11}''(\xi)-\left[-\omega^2+\frac{1}{\xi^2}\right]T_{11}(\xi)
+\frac{1}{\xi^2}T_{14}(\xi)=0
\tag{76a}\\
&
\nonumber\\
&T_{13}''(\xi)
-\left[-\omega^2+1\right]T_{13}(\xi)=0
\tag{76b}\\
&
\nonumber\\
&T_{14}''(\xi)-\left[-\omega^2+\frac{1}{\xi^2}\right]T_{14}(\xi)
+\frac{1}{\xi^2}T_{11}(\xi)=0
\tag{76c}
\end{align}
\end{subequations}
\begin{subequations}
\begin{align}
&T_{15}''(\xi)
-\left[-\omega^2+1\right]T_{15}(\xi)=0
\tag{76d}\\
\nonumber\\
&\frac{T_{11}'(\xi)}{\xi}+\frac{T_{14}'(\xi)}{\xi}=0
\tag{76e}\\
\nonumber\\
&T_{13}'(\xi)=0
\tag{76f}\\
\nonumber\\
&T_{15}'(\xi)=0
\tag{76g}
\end{align}
\end{subequations}
Due to the constraint given by the last two equations, we find that, in the case of $j=1$,
the two radial functions $T_{13}(\xi)$ and $T_{15}(\xi)$ are actually constant.
This means that, for this value of the angular momentum, only $T_{11}(\xi)$ and
$T_{14}(\xi)$ will appear in the expansion of the general solution in terms of
partial waves.
\section{Applications}
\subsection{Density of states and thermodynamics}
Among the applications of the results obtained above,
we start with thermodynamic quantities.
Indeed, with a small admixture of monopoles 
in the electric plasma, one can use the standard virial expansion
for these quantities. The
 Beth-Uhlenbeck-like formula from the theory
of nonideal gases
gives well known corrections to the partition function, induced by
the scattering on monopoles 
\be 
\delta M_m=-{T\over \pi} \sum_j (2j+1) \int dk {d\delta_j\over dk} f(k,T)
\label{eqn_BU}
\ee
where $\delta M_m$ is the modification of the monopole mass,
$\delta_j(k)$ 
is the scattering phase and $f(k,T)$ is the Bose (Fermi)
distribution for gluons (quarks).
These partial waves 
imply the decomposition of scattering amplitudes
into the usual basis of angular polynomials
\be  
2ik f(\theta,k) =  
\sum_j (2j+1)(e^{2i\delta_j(k)}-1)P_j(cos\theta) \ee  
rather than the Wigner d-functions we have
to use in the current problem. One may
think that a complicated decomposition of the amplitude should be
needed. This however can be avoided in general because of the 
following identity
\be  \sum_j (2j+1) {d\delta_j\over dk}={d \over dk}Re(f(\theta=0))+{ik^2\over
  4\pi}\int d\Omega \left(f {\partial f^* \over \partial k} -f^* {\partial
  f \over \partial k}  \right)   \ee
expressing the relevant sum in terms of the amplitude itself.

  It has been shown above, that the scattering problem involves
   (a) many partial waves with $j>0$ and (b) 
the exceptional $j=0$ channels. In the former, the
scattering is dominated by the Coulomb magnetic field of the monopole,
while in the latter the Abelian magnetic field has
no effect and the scattering happens on the monopole core. 
Let us discuss the $j>0$ case first, returning
to the  exceptional $j=0$ channel at the end of the section.

 Since the Coulombic magnetic field possesses no dimensional scale,
 one finds that the dimensionless combination
$ 2ik f(\theta,k)$ has $no$ dependence on momentum, being 
a function of the
angle only.  The scattering phases are just numbers:
they don't have any logarithmic dependence on $k$ familiar
from electric Coulomb problem.
Therefore, in this approximation
the whole Beth-Uhlenbeck correction to the statistical sum 
$vanishes$ identically!

The origin of this somewhat unexpected result can be traced to
the fact that the Beth-Uhlenbeck expression was derived from a
semiclassical counting of  the
density of states in a large (spherical) box containing the monopole.
However the semiclassical density of states, related to classical
phase space, is insensitive to magnetic fields because
the corresponding integral
\be\Omega_{cl}(E)= \int {d^3p d^3x \over (2\pi)^3} \delta\left(E-H(p,x)\right)\ee
for an electric particle in any magnetic
field $H=(\vec p-e \vec A(x))^2/2m$ does not depend 
on the field at all (in order to see that this explanation is correct, consider
for example a scattering on a dyon, which was also solved
in the same papers 
\cite{Boulware:1976tv,Schwinger:1976fr}. In this case
there is a Coulomb phase $\delta_j \sim (e_1g_1+e_2g_2)\log(k)$
and thus a nonzero change in the density of states).

Let us now turn to the
contribution of the exceptional $j=0$ channels. The corresponding
scattering phase has a non-trivial $k$-dependence,
and thus a nonzero contribution to thermodynamics
appears. 

Before we estimate the effect of this scattering on the
thermal monopole mass,
let us briefly clarify the issue of ``mean gluon momentum''.
When one calculates the typical momentum of black-body
radiation, related to integrals like $\int d^3k k f(k/T)$,
one finds that those are $k\sim 3T$. However such
estimate is not adequate for the integrals which appear
in both the thermal mass (considered here) and the scattering rates
(considered in the next section). Indeed, if
in the expression (\ref{eqn_BU}) one can replace
$d\delta_j/dk\approx a_{sl}$ where $ a_{sl}$ is the scattering length,
assuming that $ka_{sl}$ is a small parameter (to be justified in the following),
then the integral is of the 1d type  $\int dk f(k/T)$ and thus
the dominant $k\sim 1T$. The same argument applies to the
calculation of the scattering rates, since the integral is
similar
\be \dot w \sim \int dn\sigma\sim  \int d^3k f(k/T)/k^2 \ee
with the cross section being $\sigma\sim 1/k^2$ (for non-exceptional
channels). These differences between bulk and single-particle
properties are basically enforced by the dimension.

  Let us now look at the absolute magnitude of the scattering
lengths.
Our results for $\lambda=0,1$ for the exceptional channels show that the gluon-monopole 
scattering length is large
 $ a^{gm}_{sl}\approx 10 a$ 
where $a\approx .15$ \, fm=.76 \, GeV$^{-1}$
is the monopole core size. 
 This happens due to shallow 
 gluon-monopole
bound states: we remind that we found two of them (one near-zero)
for $\lambda=1$ and infinitely many in the BPS limit.

However we don't expect to have these states in physical QCD
because the effective Higgs selfcoupling (and thus mass)
are expected to be large, as seen from near-Coulomb
monopole correlations \cite{D'Alessandro:2007su}.
We already noted that currently the monopole structure
is too closed to UV lattice cutoff to get a realistic
value of the effective Higgs selfcoupling $\lambda$. As soon as
it is determined, one may
 solve the classical equations
for $K(\xi)$ and $H(\xi)$ at the corresponding
 value of $\lambda$ and determine the scattering phase.
It seems highly probable that this coupling  is large enough
to prevent  formation of the bound states.
One  expects that, without bound states the scattering length is small
$ a^{gm}_{sl}\approx 0.1 - 0.2$ fm, analogous to that in the scalar
channel.

Thus indeed $ka<1$ for temperatures of the
order of $T_c$ (.19 GeV in QCD and .27 GeV in pure gauge $SU(3)$ theory)
and the contribution to the thermal monopole mass is small
\be \delta M_m\approx {a_{sl} T\over \pi}\int {dk \over
  exp(\sqrt{k^2+m^2}/T) -1 }\approx {a_{sl} T^2\over \pi}\log({m\over T}) \ee 
where the latter integral has only weak (logarithmic) dependence on the
gluon mass $m$, provided it is small. Since this
was obtained under the assumption that monopoles
have a small size, 
$a_{sl} T\ll 1$, this contribution to the monopole mass is
relatively small. 

At asymptotically high $T$ this assumption gets violated and 
monopoles size would be at the magnetic scale,
so that this parameter gets large $ka\sim 1/e^2(T)\sim \log(T)\gg 1$.
So to say, hard thermal gluons now resolve the monopole
structure. The answer in this regime then is the perturbative one-loop
correction to the classical monopole mass, already mentioned
in the introduction.

\subsection{Transport coefficients: the benchmarks}

 In the gas approximation, all transport coefficients
are inversely proportional to the so-called 
transport cross section, normally defined as
\be \sigma_t=\int  (1-\cos\theta) d\sigma\ee
where $\theta$ is the scattering angle.
While the factor in brackets vanishes at
small angles, the Rutherford singularity in the cross section,
for any charged particle,
leads to its logarithmic divergence. 
Since we will be comparing the gluon scattering on monopoles with that
on gluons, let us first introduce those benchmarks, namely
 the well-known (lowest 
order) QCD processes, the $gg$ and $\bar q q$ scatterings:
\be {d\sigma_{\bar q q} \over dt}={e^4  \over  36 \pi} 
\left( { s^4+t^4+u^4 \over s^2 t^2 u^2 } -{8 \over 3 t u  }
\right)  \ee
\be   
{d\sigma_{gg} \over dt}={9 e^4  \over  128 \pi} 
{(s^4+t^4+u^4)(s^2+t^2+u^2) \over s^4 t^2 u^2} 
\ee
where (we remind) the electric coupling is related to
$\alpha_s$ as usual: $e^2/4\pi=\alpha_s$.

\begin{figure}[t]
\begin{center}
\includegraphics[width=8cm]{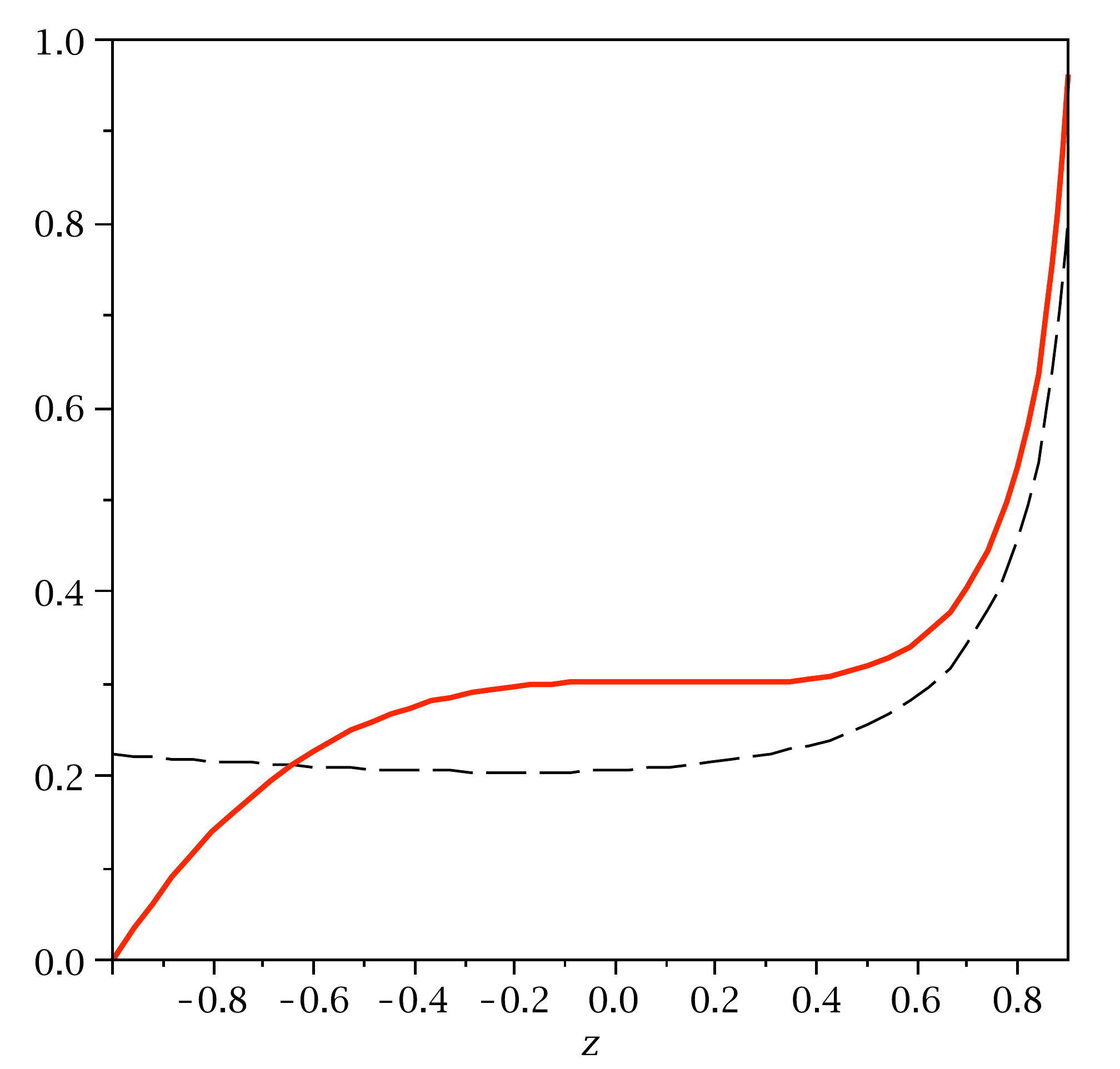}
\end{center}
\caption{Angular dependence of the integrand of the
transport cross sections vs  $z=\cos{\theta}$, where $\theta$ is the 
scattering angle. The (red) solid
line shows the integrand for the $gg$ scattering,
$d\sigma_{gg}/dt*s*t*(1+z)/2
/e^4$ , while the (black) dashed
line  shows  that 
for the $\bar q q$ scattering scaled up by a factor 4:
$4*d\sigma_{\bar q q}*dt*s*t/e^4$.
}
\label{fig_angular_gg_barqq}
\end{figure}

While for non-identical particles the transport
cross section is simply given by the cross section weighted 
by momentum transport $t\sim (1-z)$, for identical ones such as $gg$
one needs to introduce the additional factor $(1+z)/2$ 
in order to suppress
backward scattering as well. 
The integrands of the transport cross section  for these two
processes
are compared in Fig. \ref{fig_angular_gg_barqq}. The
integrated transport cross sections themselves are given by
\be \sigma^t_{gg}   = {3 e^4\over 320\pi s}\left(105 \log(3)-16
+30 \log{\left(4 \over \theta_{min}^2\right)} \right)  
\label{sigmatgg}
\ee
\be \sigma^t_{\bar q q} = {e^4 \over  54\pi s}
\left( 4+7 \log(3)+ 3 \log{\left(4 \over \theta_{min}^2 \right)} \right) \ee
where the smallest scattering angle can be related to the
(electric) screening mass by $\theta_{min}^2=2*M_D^2/s$.
Note that the forward scattering log in the $gg$ case 
has a coefficient which is roughly four times larger, 
as a consequence of the gluon color
being roughly twice that of a fundamental quark.
Note also that the $gg$ scattering is significantly larger at large
angles, as compared to the $\bar q q$ scattering.

\subsection{Transport cross section on monopoles}
\begin{figure}[t]
\begin{center}
\includegraphics[width=6cm]{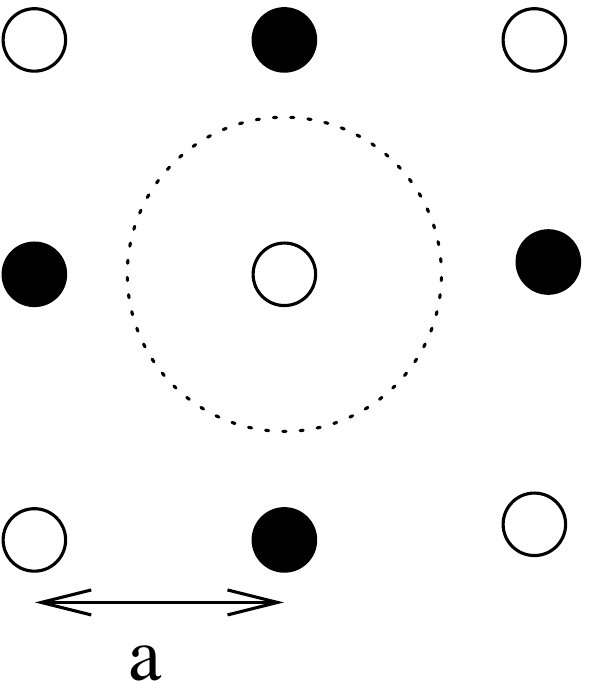}
\caption{A charge scattering on a 2-dimensional array of 
correlated monopoles (open points) and antimonopoles (closed points).
The dotted circle indicates a region of impact parameters for
which scattering on a single monopole is a reasonable approximation.}
\label{fig_monos_2d}
\end{center}
\end{figure}

We explained (already  in the introduction) that the
charge-monopole scattering is Rutherford-like 
 at small angles: this comes 
 from harmonics with
large  angular momenta (large impact parameters).
However, in matter there is a finite density of
 monopoles, so the issue of the scattering should be
reconsidered.
 A sketch of the setting,
assuming strong correlation of monopoles into a crystal-like
structure, is shown
 in Fig. \ref{fig_monos_2d}.  A
 ``sphere of influence of one monopole''(the dotted circle)
gives the maximal impact parameter
to be used. 

As a result,
 the impact parameter is limited from above by some $b_{max}$, which implies that
 only a finite number of partial waves should be included.
The
range of partial waves to be included in the scattering amplitude
can be estimated as follows
\be 
j_{max}=\langle p_x\rangle n_{m}^{-1/3}/2\sim aT\sim  1/e^2(T) \sim \log(T).
\ee
Since at asymptotically high $T$ the monopole
 density $n_{m}\sim (e^2 T)^3$ is small compared to
the density of quarks and gluons $\sim T^3$,
 $j_{max}$ asymptotically grows  logarithmically with $T$. So,
only in the academic limit
$T\rightarrow \infty$ one gets $j_{max}\rightarrow \infty$ 
and the usual free-space
scattering amplitudes calculated in \cite{Boulware:1976tv} 
where all partial
waves are recovered.
However, in reality
 we have to recalculate the scattering,
retaining only several lowest partial waves from the sum.
As we will see, this dramatically changes the angular distribution,
by strongly  depleting scattering at small angles and enhancing
scattering backwards. 
\subsubsection{Scalar particle}
By making use of the partial-wave solutions to the wave equations that we
obtained in Sec.~\ref{scalarfluctuations}, we can proceed to construct the
scattering solutions. At large distance from the origin, they should behave as 
a plane wave plus an outgoing spherical wave.

We consider a particle entering from $z=-\infty$; the $z$ component
of its total angular momentum is well defined: $J_{3}=T_3=
-(\vec{T}\cdot\hat{r})=-n$. Therefore, the appropriate angular functions
that we must use to expand our solution are $\phi_{ji}^{-nn}(\theta,\varphi)$.
In the following, we consider only $n\geq0$; a negative $n$ would
just produce an overall phase factor.
We consider the following partial-wave sum
\ba
\phi^{(+)}(\vec{r})&=&e^{-i\pi n}\sum_{j=|n|}^{j_{max}}(2j+1)e^{i\pi j}e^{-i\pi j'/2}j_{j'}(kr)
\left[U\left(-\varphi,\theta,\varphi\right)\chi_i^n\right]\mathcal{D}^{(j)}_{n,-n}
(-\varphi,\theta,\varphi)
\nonumber\\
&=&\left[U\left(-\varphi,\theta,\varphi\right)\chi_i^n\right]
\psi^{(+)}(\vec{r}).
\ea
We recall that the index $j'$ in the above formula is
the positive root of
\be
j'(j'+1)=j(j+1)-n^2.
\ee
The square brackets in the above formula contain a vector which
is the charge eigenstate. Since in our approximation we
don't have any charge-flip reactions (they would be possible
inside the monopole core, which we consider pointlike), this factor 
will remain the same for the incoming and outgoing waves. 
The physically significant wave function is therefore given by
 \ba
\psi^{(+)}(\vec{r})&=&e^{-i\pi n}\sum_{j=|n|}^{j_{max}}(2j+1)e^{i\pi j}e^{-i\pi j'/2}j_{j'}(kr)
\mathcal{D}^{(j)}_{n,-n}(-\varphi,\theta,\varphi)
\nonumber\\
&=&e^{-i\pi n}\sum_{j=|n|}^{j_{max}}(2j+1)e^{i\pi j}e^{-i\pi j'/2}j_{j'}(kr)
e^{-2in\varphi}d^{(j)}_{n,-n}(\theta).
\ea
The above expression  
contains the functions $d^{(j)}_{n,-n}(\theta)$ which can be
written via the generalized Rodriguez formula as $(z=\cos{\theta})$
\be d^{(j)}_{n,-n}(z)={(-1)^{j-n} \over 2^j (j-n)!} (1-z)^{-n}
\left({d\over dz} \right)^{j-n} \left[(1-z)^{j+n} (1+z)^{j-n} \right]\ee   
So, the first term with $j=n$ is simply $\sim (1-\cos{\theta})^n$, a function
strongly peaked backwards at large $n$.  

Obviously, a tendency to scatter backwards is very important for
transport properties and equilibration of electric plasma.

The asymptotic value of the wave function $\psi^{(+)}(\vec{r})$ is then

\be
\psi^{(+)}(\vec{r})\sim e^{-2in\varphi}\left[e^{ikz}+f(\theta)\frac{e^{ikr}}{r}
\right]
\ee
with the scattering amplitude $f(\theta)$ given by
\be
2ik f(\theta)=\sum_{j=|n|}^{j_{max}} (2j+1)e^{i\pi (j'-j)} d^{(j)}_{n,-n}(\theta).
\label{scatteringamplitude}
\ee

The above sum would be badly divergent if we would sum
over $j$ up to $\infty$. It is possible to improve its
convergence by separating the most singular behavior
(see for example the discussion in \cite{Boulware:1976tv}).
We do not have such convergence problems, since we
have an ultraviolet cutoff on $j$ due to 
the finite density of monopoles. Therefore we calculate
the scattering amplitude by directly
using Eq. (\ref{scatteringamplitude}).

\begin{figure}
\begin{minipage}{.48\textwidth}
\parbox{6cm}{
\scalebox{.65}{
\includegraphics{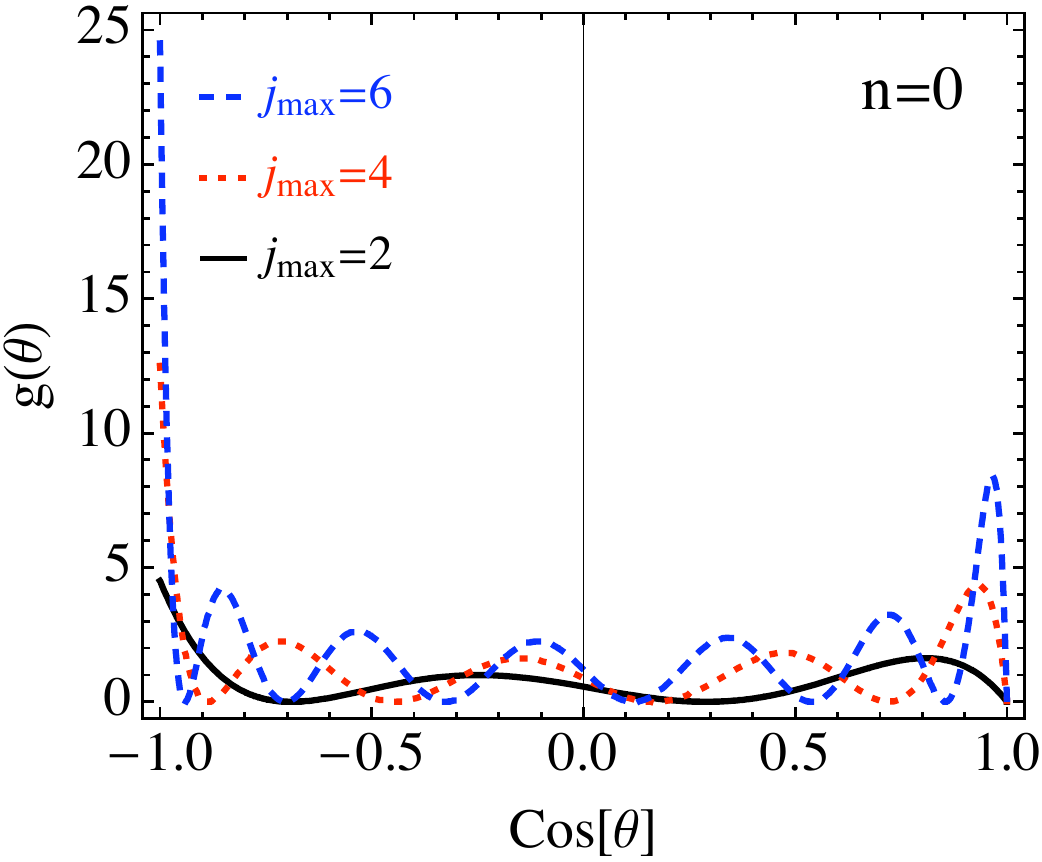}\\}}
\end{minipage}
\hspace{.4cm}
\begin{minipage}{.48\textwidth}
\parbox{6cm}{
\scalebox{.65}{
\includegraphics{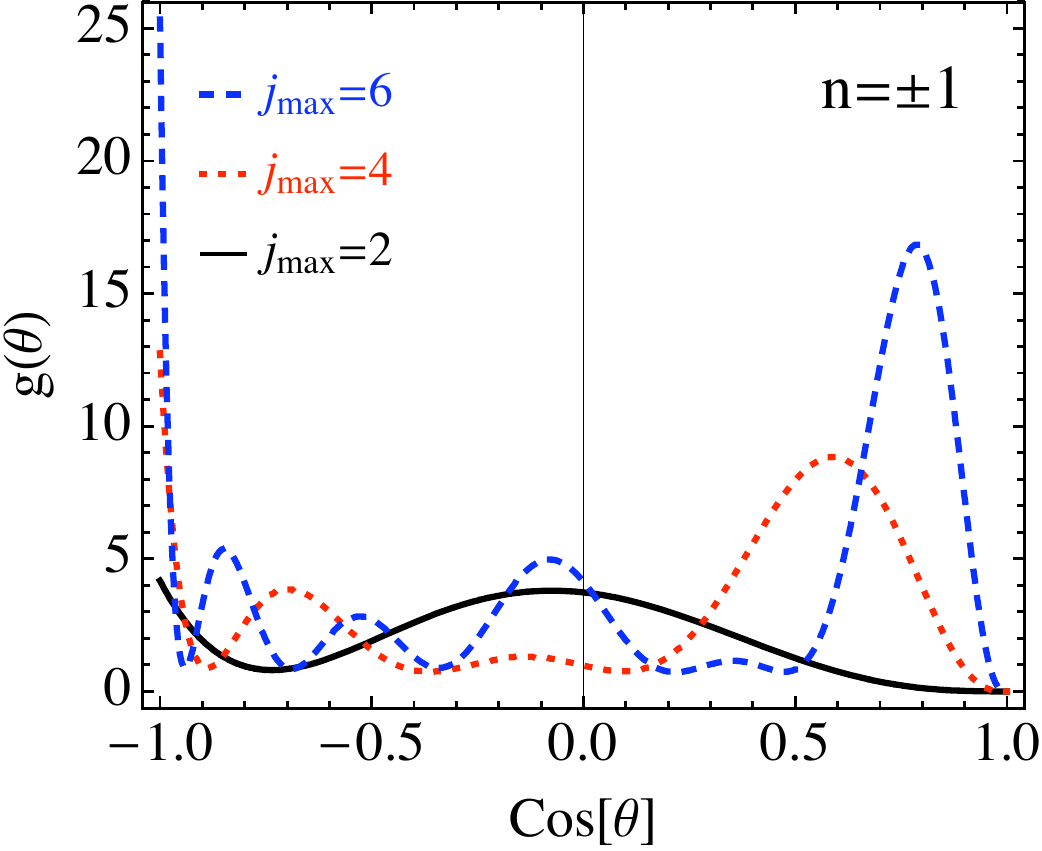}\\}}
\end{minipage}
\caption{Scaled Integrand of the trasport cross section 
$g(\theta)=k^2(1-\cos{\theta})|f(\theta)|^2$
with only 2, 4 and 6 lowest partial waves
included, for a scalar particle with $n=0$ (left) and
$n=\pm1$ (right). The curves can be easily recognized by higher $j_{max}$
having more oscillations. }
\label{fig_246}
\end{figure}

\begin{figure}
\begin{center}
\includegraphics[width=9cm]{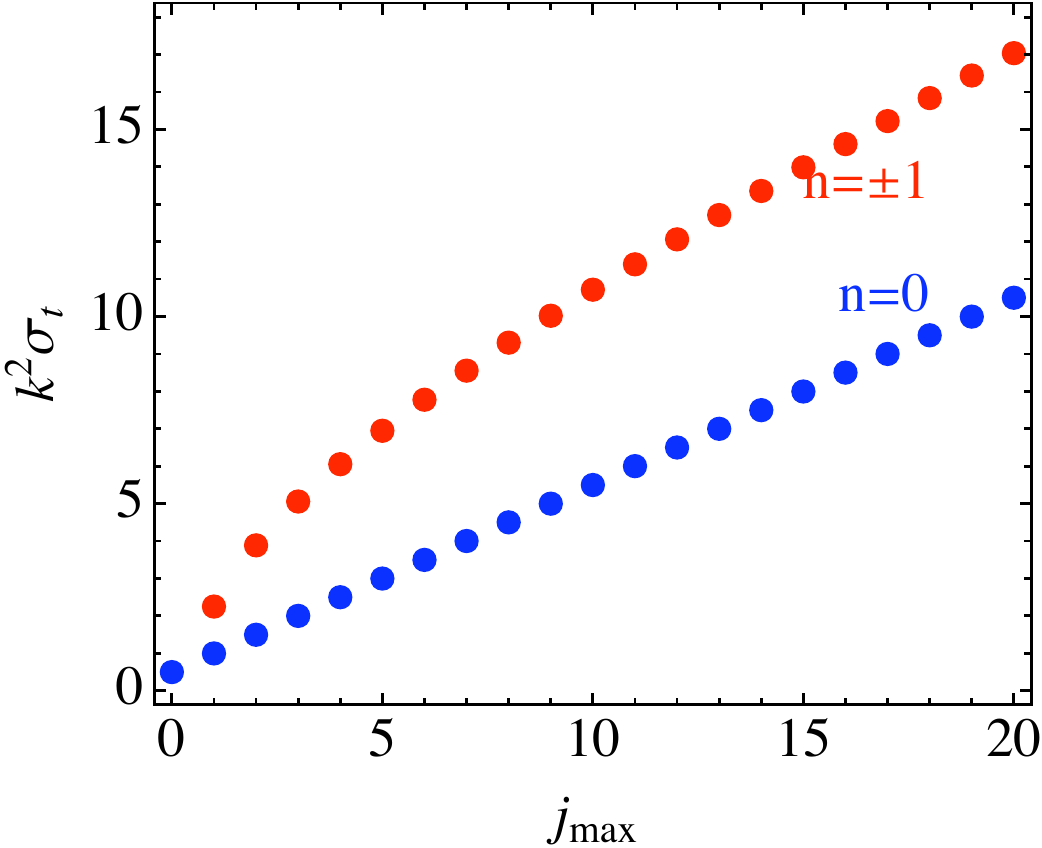}
\caption{Normalized transport cross section as a function of cutoff
in maximal harmonics retained, for $n=0,1$.}
\label{fig_sigma_trans}
\end{center}
\end{figure}

The
integrands of the transport cross section
$(1-\cos{\theta})|f(\theta)|^2$ are shown in Fig. \ref{fig_246} for
$n=0,~j_{max}=2,4,6$ (left panel),
$n=\pm1,~j_{max}=2,4,6$ (right panel).
One can see how much their angular distribution 
is distorted. Strong oscillations of this function occur because
we use a sharp cutoff for the higher harmonics, which represents
diffraction of a sharp edge. This edge in reality does not exist
and can be removed by any smooth edge prescription (for example a gaussian weight).
However, we further found that the transport cross section itself
is rather insensitive to these oscillations, and thus there is no
need in smoothening the scattering amplitude. The transport cross section
as a function of $j_{max}$ is shown in Fig. \ref{fig_sigma_trans}:
it is large and smoothly rising with the cutoff.  

\subsubsection{Vector particle}
\begin{figure}
\begin{center}
\parbox{6cm}{
\scalebox{.8}{
\includegraphics{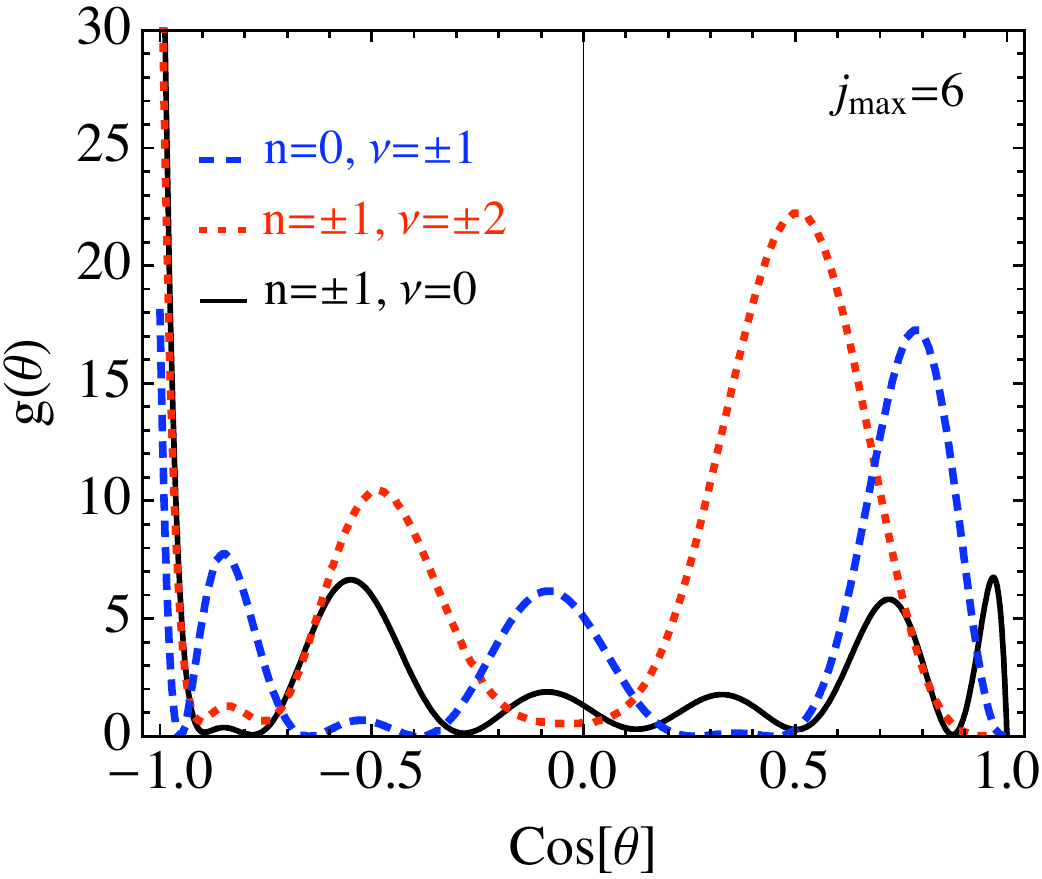}\\}}
\caption{Integrand of the transport cross section 
$g(\theta)=(1-\cos(\theta))|f(\theta)|^2$
with only 6 lowest partial waves
included, for a vector particle with $n=0,~\nu=\pm1$,
$n=\pm1,~\nu=0$ and $n=\pm1,~\nu=\pm2$. }
\label{fig_vector}
\end{center}
\end{figure}

\begin{figure}
\begin{center}
\includegraphics[width=9cm]{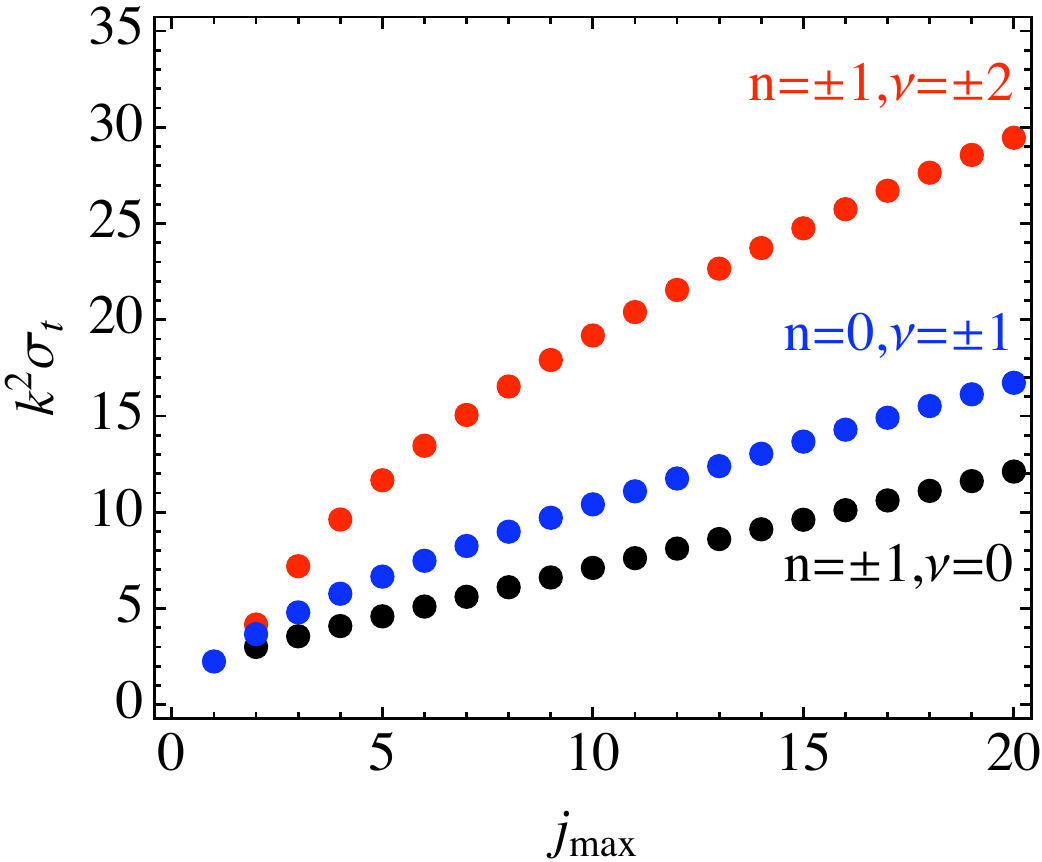}
\caption{Normalized transport cross section as a function of cutoff
in maximal harmonics retained, for $n=0,\pm1$ and the corresponding
polarizations.}
\label{fig_sigma_vector}
\end{center}
\end{figure}

We consider a vector particle entering from $z=-\infty$. As we already saw in the case
of the scalar particle, the
$z$ component of its total angular momentum is
well-defined: we have seen that, if we consider
angular functions that are simultaneous eigenfunctions of
$(\vec{T}\cdot\hat{r})$ and $(\vec{S}\cdot\hat{r})$,
the equations for the corresponding radial
functions are diagonal. This means that,
once we neglect the monopole core,
we can consider an incoming particle with definite charge and
radial polarization, and these quantities will be conserved in
the final state, after the scattering.
The third component of the angular momentum will then be
\be
J_3=-\left[\left(\vec{T}\cdot\hat{r}\right)+\left(\vec{S}\cdot\hat{r}\right)\right]=
-\left[n+\sigma\right]=-\nu.
\ee
We have seen that, after gauge fixing, for each particle of given charge
we have two possible spin polarizations $\sigma=\pm1$. This means that, for
a vector particle with charge $n=1(-1)$, we can have $\nu=0,2(0,-2)$ respectively,
while for the vector particle with $n=0$ (the photon) we have
$\nu=\pm1$.
We consider the following partial-wave sum
\ba
\Phi^{(+)}\left(\vec{r}\right)&=&
e^{-i\pi \nu}\sum_{j=|\nu|}^{j_{max}}\left(2j+1\right)
e^{i\pi j}e^{-i\pi j'/2}j_{j'}\left(kr\right)\left[\Xi\left(\theta,\varphi\right)_{n}^{\sigma}\right]
\mathcal{D}^{(j)}_{\nu,-\nu}\left(-\varphi,\theta,\varphi\right)
\nonumber\\
&=&\left[\Xi\left(\theta,\varphi\right)_{n}^{\sigma}\right]\Psi^{(+)}\left(\vec{r}\right)
\ea
where the factor $\left[\Xi\left(\theta,\varphi\right)_{n}^{\sigma}\right]$
is a two-index tensor representing the charge and $(\vec{S}\cdot\hat{r})$ 
eigenstate, which is not altered in the peripheral collision that we are considering.
As we already did for the scalar channel, we consider the physically significant
wave function
\ba
\Psi^{(+)}(\vec{r})&=&e^{-i\pi \nu}\sum_{j=|\nu|}^{j_{max}}(2j+1)e^{i\pi j}e^{-i\pi j'/2}j_{j'}(kr)
\mathcal{D}^{(j)}_{\nu,-\nu}(-\varphi,\theta,\varphi)
\nonumber\\
&=&e^{-i\pi \nu}\sum_{j=|\nu|}^{j_{max}}(2j+1)e^{i\pi j}e^{-i\pi j'/2}j_{j'}(kr)
e^{-2i\nu\varphi}d^{(j)}_{\nu,-\nu}(\theta).
\ea
for which we can write the asymptotic behavior
\be
\Psi^{(+)}(\vec{r})\sim e^{-2i\nu\varphi}\left[e^{ikz}+f(\theta)\frac{e^{ikr}}{r}
\right]
\ee
with the scattering amplitude $f(\theta)$ given by
\be
2ik f(\theta)_{n,\nu}=\sum_{j=|\nu|}^{j_{max}} (2j+1)e^{i\pi (j'-j)} d^{(j)}_{\nu,-\nu}(\theta).
\label{scatteringamplitudev}
\ee
Actually, there are some exceptions to the above expression for $f(\theta)$: in
Sec. \ref{vectorfluctuations} we have seen that the cases $j=0$ and $j=1$ are somehow
exceptional. For $j=0$ we have only two modes, with charge $\pm1$.
The corresponding radial function obeys
a differential equation which is not Bessel-like.Yet, this solution
 still has a sinusoidal behavior at large distances, with 
a momentum-dependent scattering phase. We have also seen that, for $j=1$,
the only surviving modes are the two transverse 
polarizations of the particle with charge $n=0$.
Therefore, the above general formula for the scattering amplitude \eqref{scatteringamplitudev}
must be separately written, for particles with different charge.
For the neutral particle (with $\nu=\pm1$) we have
\be
2ik f(\theta)_{0,\pm1}=
\sum_{j=1}^{j_{max}} (2j+1)d^{(j)}_{1,-1}(\theta),
\label{scatteringamplitudev01}
\ee
where the factor $e^{i\pi (j'-j)}$ is absent in this case, since for a neutral particle we have $j'=j$.
We have seen that, for the positive (negative) charge particle, 
we have two possible values of $\nu=0,~2~(-2)$.
Therefore we have:
\ba
2ik f(\theta)_{\pm1,\pm2}&=&\sum_{j=2}^{j_{max}} (2j+1)e^{i\pi (j'-j)} d^{(j)}_{2,-2}(\theta),
\\
2ik f(\theta)_{\pm1,0}&=&e^{2 i \delta_{0}(k)}d^{(0)}_{0,0}(\theta)+
\sum_{j=2}^{j_{max}} (2j+1)e^{i\pi (j'-j)} d^{(j)}_{0,0}(\theta),
\nonumber
\label{scatteringamplitudev02}
\ea
where $\delta_0(k)$ is the scattering phase plotted in Fig.~\ref{deltavector}.
The
integrands of the transport cross section
$(1-\cos(\theta))|f(\theta)_{n,\nu}|^2$ are shown in Fig.~\ref{fig_vector} for
$j_{max}=6,~n=0,\pm1$ and the corresponding values of $\nu$.
One can see how much their angular distribution 
is distorted. The resulting transport cross section,
\be
(\sigma_t)_{n,\nu}=\int_{-1}^{1}d\cos\theta(1-\cos\theta)|f(\theta)_{n,\nu}|^2
\ee
is shown in Fig.~\ref{fig_sigma_vector} as a function of $j_{max}$ for
the different possible combinations of charge and polarization. As we can
notice, it is large and smoothly rising with the cutoff, in spite of the large oscillations
in the integrand, due to our choice of a sharp cutoff.
In the following we will use $j_{max}=6$.
Now we proceed to evaluate the scattering rate of gluons on monopoles.
We recall that it is defined as
\be
\frac{\dot{w}_{gm}}{T}=\frac{\langle n_m (\sigma_t)_{gm}\rangle}{T}
\ee
where the $\langle...\rangle$ indicates an average over the incoming gluon.
Following Ref. \cite{Meisinger:2003id}, we take into account the interaction
of transverse gluons with a temporal background gauge field $\mathcal{A}_0$ which
is related to the Polyakov loop through
\be
\Phi=\frac{1}{N_c}tr\left[\mathcal{P}\exp{\left(i\int_{0}^{\beta}\mathcal{A}_0d\tau\right)}\right]
=\frac13tr\exp\left[\frac{i\mathcal{A}_{0}^{a}\lambda_a}{T}\right]
\ee
(the above definition of the gauge field actually includes the electric coupling $e$).
If we select the abelian gauge for $\mathcal{A}_0$, for example along $\lambda_3$, we get:
\be
\Phi=\frac13\left[2\cos{\frac{\mathcal{A}_{0}^{3}}{T}}+1\right].
\ee
The presence of this background gauge field breaks color symmetry. The eight gluons
interact with $\mathcal{A}_{0}^{3}$ in different ways, depending on their color charge. For example,
since we chose $\mathcal{A}_{0}$ along the abelian direction, the two abelian gluons
don't interact with $\mathcal{A}_0$ at all.
In particular, the gluon density has the following form
\ba
\!\!\!\!\!\!n_g(T)\!\!\!&=&\!\!\!\frac{8\pi}{(2\pi)^3}\int k^2dk\left[\frac{2}{\exp{(\beta\epsilon_k)}-1}
+\frac{2}{\exp{(\beta\epsilon_k)}\exp{(i\beta \mathcal{A}_{0}^{3})}-1}
\right.
\nonumber\\
\!\!\!&+&\!\!\!\left.\frac{2}{\exp{(\beta\epsilon_k)}\exp{(-i\beta \mathcal{A}_{0}^{3})}-1}+
\frac{1}{\exp{(\beta\epsilon_k)}\exp{(2i\beta \mathcal{A}_{0}^{3})}-1}
\right.
\nonumber\\
\!\!\!&+&\!\!\!\left.\frac{1}{\exp{(\beta\epsilon_k)}\exp{(-2i\beta \mathcal{A}_{0}^{3})}-1}\right]
=\frac{4\pi}{(2\pi)^3}\int k^2dk\rho_g(k,T)
\label{gluondensity}
\ea
where we use for $\mathcal{A}_{0}^{3}$ the expectation value which
fits the available lattice results for the Polyakov loop in pure-gauge
$SU(3)$ gauge theory (see Fig. \ref{fig_m1m2}) and we define
$\epsilon_k=\sqrt{k^2+m_g(T)^2}$, with
an effective gluon mass $m_g(T)=e(T)T/\sqrt{2}$. For $\alpha_s(T)$
we use the estimate given in Ref. \cite{Liao:2008jg}. 
Notice that $\rho_g$ contains a factor 2 to account for gluon spin polarization.
The coupling of gluons to the Polyakov
loop gives a suppression of these degrees of freedom approaching $T_c$ from above.
This is evident from Fig. \ref{fig_density}, where we plot $n_g(T)/T^3$ as defined in 
eq. \eqref{gluondensity} as a function of the temperature (continuous line).

In order to obtain the formula for the (scaled) scattering rate, we need to average $\sigma_{t}(k)$
over the momentum of the incoming gluon: we use the distribution
taken from eq. \eqref{gluondensity}; in taking the integral over momenta, we have to recall
that the transport cross section, and the gluon screening due to $\mathcal{A}_0$, both depend on the charge of
gluons: for this reason, we have to split $\rho_g(k,T)$ into a ``charge-neutral" part, which will be used
as a weight for $(\sigma_t(k))_{0,1}$, and a ``charged part", which will be used to
weight $(\sigma_t(k))_{1,0}$ and $(\sigma_t(k))_{1,2}$:
\ba
\frac{\dot{w}_{gm}}{T}&=&\frac{n_m(T)}{n_g(T)T}
\frac{4\pi}{(2\pi)^3}\int k^2dk\left[\frac{4(\sigma_t(k))_{0,1}}
{\exp{(\beta\epsilon_k)}-1}
\right.
\nonumber\\
&+&\left.\left((\sigma_t(k))_{1,0}+(\sigma_t(k))_{1,2}\right)
\left(\frac{2}{\exp{(\beta\epsilon_k)}\exp{(i\beta \mathcal{A}_{0}^{3})}-1}
\right.\right.
\nonumber\\
\!\!\!&+&\!\!\!\left.\left.\frac{2}{\exp{(\beta\epsilon_k)}\exp{(-i\beta \mathcal{A}_{0}^{3})}-1}+
\frac{1}{\exp{(\beta\epsilon_k)}\exp{(2i\beta\mathcal{A}_{0}^{3})}-1}
\right.\right.
\nonumber\\
\!\!\!&+&\!\!\!\left.\left.\frac{1}{\exp{(\beta\epsilon_k)}\exp{(-2i\beta \mathcal{A}_{0}^{3})}-1}\right)\right].
\ea
We recall that the momentum dependence of the transport cross section is trivially $\sim1/k^2$,
except for the case $(\sigma^{t}(k))_{1,0}$ for which we have the exceptional case
$j=0$, with a momentum-dependent scattering phase $\delta_0(k)$.
In the above equation, $n_m(T)$ is the monopole density as a function of the temperature. 
We take this information from the avaliable lattice results for this quantity, plotted in Fig. \ref{fig_density}.
We show $\dot{w}_{gm}/T$ in Fig. \ref{fig_viscosity} (the red, continuous line in the left panel).
Also shown is the same quantity for the $gg$ scattering process (black, dotted line),
obtained through the following equation
\be
\frac{\dot{w}_{gg}}{T}=\frac{1}{n_{g1}}\int\frac{4\pi k_{1}^{2}dk_1}{(2\pi)^3}\int\frac{2\pi k_{2}^{2}dk_2}{(2\pi)^3}\int_{-1}^{1}
d\cos{\theta}\sigma_{gg}^{t}(k_1,k_2,\cos{\theta})\rho_g(k_1,T)\rho_g(k_2,T)
\label{wgg}
\ee
where $\sigma_{gg}^{t}(k_1,k_2,\cos{\theta})$ was defined in eq. \eqref{sigmatgg},
with
\be
s=2k_1k_2(1-\cos{\theta})+2m_g(T)^2.
\ee
\section{Conclusions and discussion}

In this paper we studied the role of magnetic monopoles
in the ``electric'' 
QGP phase. In distinction with papers by Liao and Shuryak, we have 
not focused on the near-$T_c$ region, in which monopoles
seem to be dominant over gluons in number, and
may even expel electric fields into flux tubes as they do
in the confined phase.
Instead, we considered the more traditional QGP away from $T_c$,
 dominated by
the usual ``electric'' constituents -- quarks and gluons,
the magnetic monopoles being subleading in number.
  
More specifically, we focused on scalar-monopole and gluon-monopole
 scattering. Classical molecular dynamics, studied in  refs. \cite{Liao:2006ry},
 suggested a very interesting mechanism of` ``mutual trapping'' by electric/magnetic quasiparticles.
 Based on results of 
quantum scattering on monopoles
 \cite{Boulware:1976tv,Schwinger:1976fr,Kazama:1976fm} for scalar-monopole
 and spinor-monopole scatterings, in this paper we studied this effect further. Indeed, we found significant
 large angle and even backward scattering for small impact parameters, complemented with a
 Rutherford-like
forward scattering peak. 

Our main result is a detailed derivation of the
gluon-monopole scattering amplitude, especially in the lowest partial
waves. For gauge particles, the existence of spin and isospin
leads to complications which so far precluded this problem from
being solved.

Our first important finding is that the gluon-monopole
scattering  has little effect on thermodynamics, because scattering
phases -- albeit large -- are mostly {\em energy independent},
except the ``exceptional channel'' with $j=0$.  
This explains why
 lattice studies which focused on thermodynamic observables
were unable to see such effects.

Our second (and the main) finding is that 
the contribution of gluon-monopole scattering
is very important for transport properties. While
the monopole density may be small, the $gm$ scattering
amplitudes have $e^2g^2\sim O(1)$ coupling instead of small $e^4\ll1$.
 Furthermore,
in our setting (with a
limited number of partial waves $j<j_{max}$  included)
 there is an additional enhancement for large angle (or 
even backward) scattering.

\begin{figure}
\begin{minipage}{.48\textwidth}
\parbox{6cm}{
\scalebox{.65}{
\includegraphics{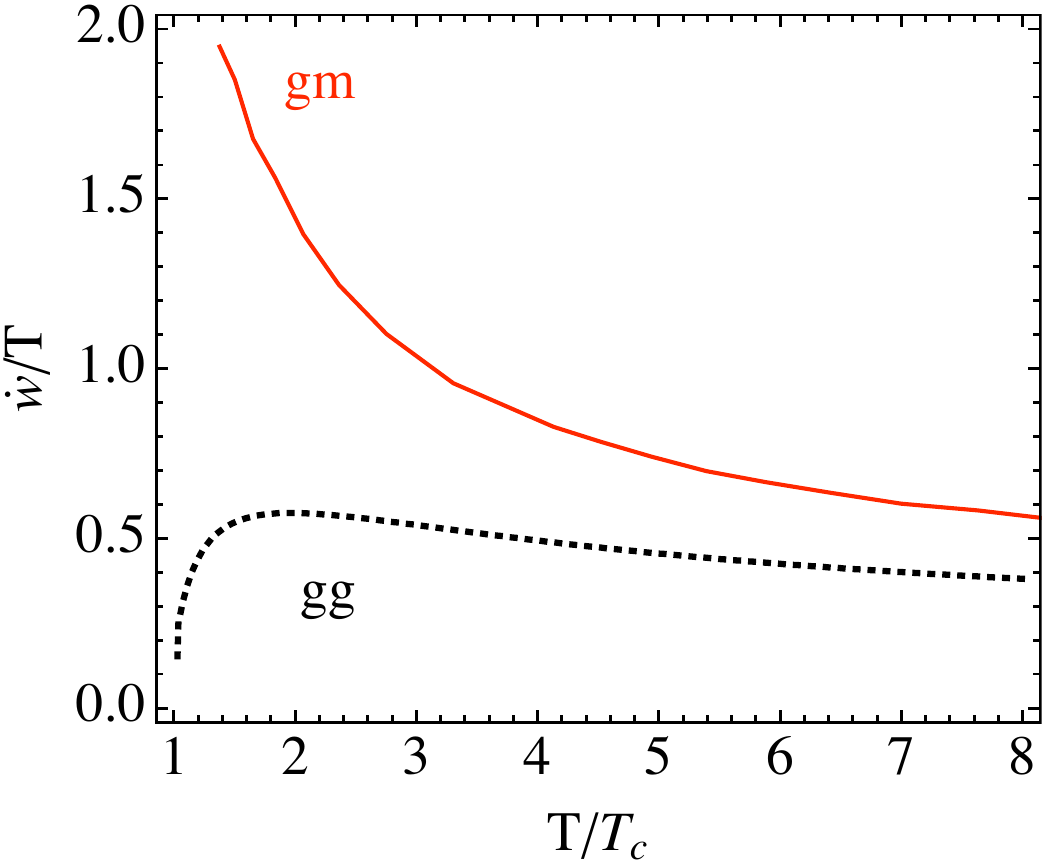}\\}}
\end{minipage}
\hspace{.4cm}
\begin{minipage}{.48\textwidth}
\parbox{6cm}{
\scalebox{.65}{
\includegraphics{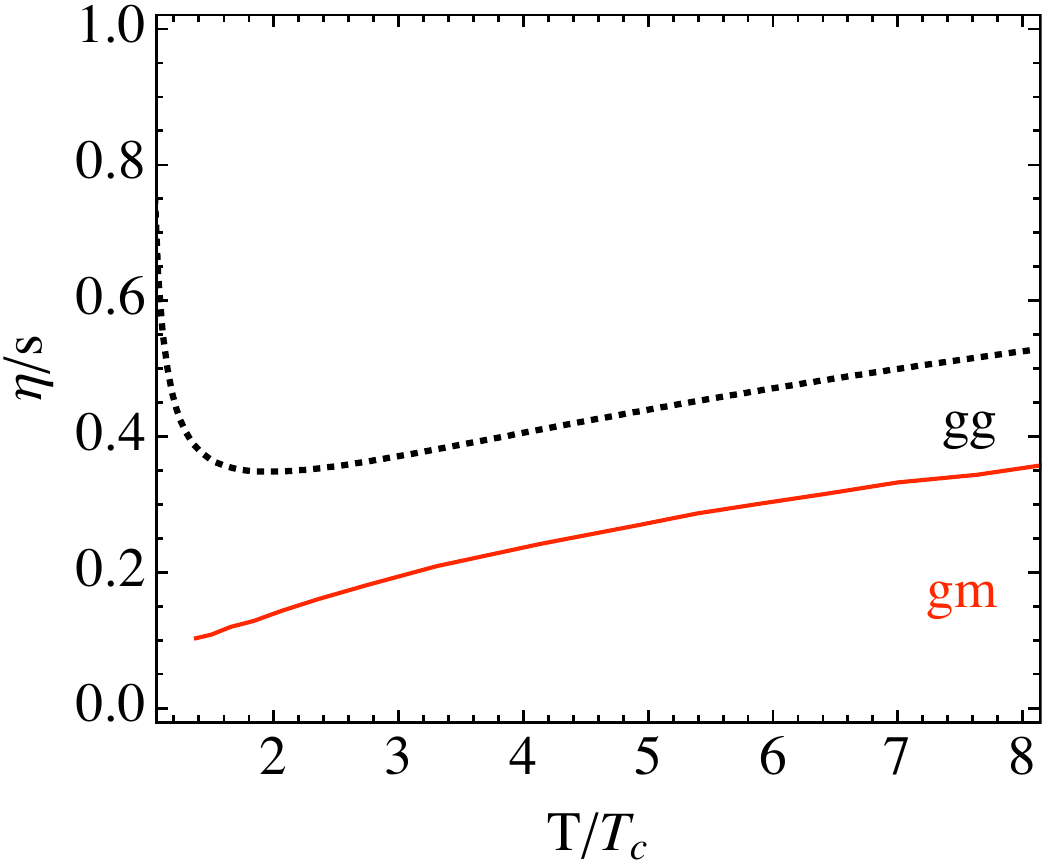}\\}}
\end{minipage}
\caption{Left panel: gluon-monopole and gluon-gluon scattering rate. Right panel: gluon-monopole
and gluon-gluon viscosity over entropy ratio, $\eta/s$.}
\label{fig_viscosity}
\end{figure}

Convoluting the cross sections found with the monopole density 
and gluon momentum distribution,
we plot  the scattering rates $n\sigma_t$ per 
gluon vs $T$ in Fig.\ref{fig_viscosity}.

It follows from this comparison of the gluon-monopole
curve with the gluon-gluon one that the former remains the leading
effect
till very high $T$, although asymptotically it is expected to get subleading.
This maximal $T$  expected  at LHC does not exceed 4$T_c$, where
$\eta/s\sim .2$. This value is well in the region which would ensure
hydrodynamical radial and elliptic flows, although deviations from
ideal hydro would be larger than at RHIC (and measurable!).

The approximate relation of these rates to viscosity/entropy
ratio is 
\be 
{\eta \over s}\approx {T \over 5 \dot w};
\ee
We plot $\eta/s$ in the right panel of Fig. \ref{fig_viscosity}.

Let us now comment on some details of the calculations, which were not
yet taken into account and should be included later.  
One important fact is that 
the lattice data for the monopole density we used actually include all magnetically charged
particles, including dyons . Their electric charge would lead to additional Rutherford-like
scattering which we have ignored.

In the calculation above we have not included Bose enhancement for scattered gluons. This
effect adds the factor  $(1+f(p'))$ for $gm$ and 
$(1+f(p_1'))(1+f(p_2'))$ for $gg$, with the prime marking the
$secondary$ gluons. If the gluon mass is small (high $T$)
 those corrections
are small: their magnitude is $\langle f\rangle\sim (T/m)\sim 1/e(T)\gg 1$.
In the experimentally relevant region, when $m/T=O(1)$,  
the effect is not enhanced and is additionally suppressed
by the expectation values of the  Polyakov lines\footnote{
We do not agree with Hidaka and Pisarski \cite{Hidaka:2008dr}
in their conclusion that $\langle L\rangle<1$ makes viscosity $\eta$ smaller
by  $\langle L\rangle^2$.
}
 $\sim\langle L\rangle^2$.

As we have already mentioned in the introduction,
Xu, C. Greiner and St\"ocker \cite{Xu:2007jv} have suggested an
alternative explanation  for small
QGP viscosity, namely
the next-order radiative processes,
 $gg\leftrightarrow ggg$. Using perturbative matrix elements
and $\alpha_s=0.3..0.6$, they found
$\eta/s$ several times smaller than for
 the $gg\leftrightarrow gg$ process,
 close to what we get from the $gm$ scattering.
Obviously, both mechanisms, albeit having such different origin, 
would thus be sufficient to explain the
 well-known hydrodynamic
results for radial and elliptic flow at RHIC. 

It will require much more work to see how $both$ results
will change, when further  refinements are performed.
We have discussed those for monopoles above:
let us now mention a few questions for $gg\leftrightarrow ggg$ :\\
 (i) Xu et al used
near-massless perturbative gluons: while in RHIC-LHC range
the lattice
quasiparticle masses are instead much larger than $T$, about $3T$
or so. This would suppress emission of extra gluons.\\
(ii) in RHIC-LHC range one should
 include the suppression by the Polyakov VEV $\langle L\rangle$ for any gluon
effects (see Fig. \ref{fig_density}\\
(iii) Inclusion of
higher order corrections in badly divergent perturbative series
needs further studies. As shown
 years ago in \cite{Xiong:1992cu}, similarly treated 
 processes $gg \rightarrow ng$ 
with larger $n=4,...$  lead to even larger rates! The development
of convergent  series for $\eta/s$ itself still remains to be
an open challanging problem.

{\bf Acknowledgements}
  We thank 
 Jinfeng Liao and Alfred Goldhaber for multiple useful discussions, as well as Ernst-Michael
 Ilgenfritz and
Massimo D'Elia, who provided us with
 unpublished lattice results. The
 work is partially
supported by the US-DOE grants DE-FG02-88ER40388 and
DE-FG03-97ER4014.

\section*{Appendix}
\appendix
\renewcommand{\theequation}{A-\arabic{equation}}
  \setcounter{equation}{0}  
\section{The Georgi-Glashow model}
The Lagrangian of the Georgi-Glashow model describes coupled gauge and Higgs fields and reads
\ba
\mathcal{L}&=&
-\frac12\mathrm{Tr}F_{\mu\nu}F^{\mu\nu}+\mathrm{Tr}\left(D^\mu\phi\right)\left(D_\mu\phi\right)
-V\left(\phi\right)
\nonumber\\
&=&-\frac14F^{a}_{\mu\nu}F^{a\mu\nu}+\frac12\left(D^\mu\phi^a\right)\left(D_\mu\phi^a\right)
-V\left(\phi\right),
\label{gglagrangian}
\ea
where $F_{\mu\nu}=F^{a}_{\mu\nu}I^a,~\phi=\phi^aI^a$ and 
Tr$(I^aI^b)=\frac12\delta_{ab},~a,b=1,2,3$. For the gauge group $SU(2)$, 
the matrices $I_a$ satisfy the algebra
\be
\left[I^a,I^b\right]=i\epsilon_{abc}I^c.
\ee
The covariant derivative is defined as
\be
D_\mu=\partial_\mu+ieA_\mu
\ee
which gives, for the Higgs field,
\be
D_\mu\phi=\partial_\mu\phi+ie\left[A_\mu,\phi\right]~~~~~~\mathrm{or}~~~~~~
D_\mu\phi^a=\partial_\mu\phi^a-e\epsilon_{abc}A^{b}_{\mu}\phi^c
\ee
and the potential for the scalar field reads
\be
V(\phi)=\frac{\lambda}{4}\left(\phi^a\phi^a-v^2\right)^2.
\ee
In the above formulas, $e$ and $\lambda$ are the gauge and scalar coupling constants, respectively.
The field strength tensor is
\be
F^{a}_{\mu\nu}=\partial_{\mu}A^{a}_{\nu}-\partial_{\nu}A^{a}_{\mu}-e\epsilon_{abc}A^{b}_{\mu}
A^{c}_{\nu}
\ee
or, in matrix form,
\be
F_{\mu\nu}=\partial_{\mu}A_{\nu}-\partial_{\nu}A_{\mu}+ie\left[A_\mu,A_\nu\right]=\frac{1}{ie}
\left[D_\mu,D_\nu\right].
\ee
The field equations corresponding to the Lagrangian (\ref{gglagrangian}) can be written as
\be
D_{\nu}F^{a\mu\nu}=-e\epsilon_{abc}\phi^bD^\mu\phi^c,~~~~~~~~~D_\mu D^\mu\phi^a
=-\lambda\phi^a\left(\phi^b\phi^b-v^2\right).
\ee
One looks for a classical solution that is invariant under a combined space and isospin rotation
generated by the $\vec{J}$ spin operator $\vec{J}=\vec{T}+\vec{S}$, with
$\vec{T}=\vec{L}+\vec{I}$, and $\vec{S}$ is the spin of the particle. An ansatz with this generalized ``spherical
symmetry" is~\cite{thooft,polyakov}
\be
\Phi^a=\frac{r^a}{er^2}H\left(\xi\right),~~~~~~\mathcal{A}^{a}_{n}=\epsilon_{amn}\frac{r^m}{er^2}
\left[1-K\left(\xi\right)\right],~~~~~~\mathcal{A}^{a}_{0}=0,
\label{hedgehog}
\ee
were $H(\xi)$ and $K(\xi)$ are functions of the dimensionless variable $\xi=ver$. These functions
are solutions of the following system of coupled equations
\be
\xi^2\frac{d^2K}{d\xi^2}=KH^2+K\left(K^2-1\right),~~~~~~~~~\xi^2\frac{d^2H}{d\xi^2}=2K^2H
+\frac{\lambda}{e^2}H\left(H^2-\xi^2\right)
\label{KHequations}
\ee
with boundary conditions
\ba
&&K\left(\xi\right)\rightarrow 1,~~~~~~~~~H\left(\xi\right)\rightarrow 0~~~~~~\mathrm{as}~~~~~~
\xi\rightarrow 0
\nonumber\\
&&K\left(\xi\right)\rightarrow 0,~~~~~~~~~H\left(\xi\right)\rightarrow \xi~~~~~~\mathrm{as}~~~~~~
\xi\rightarrow
\infty.
\ea
In the Bogomol'nyi-Prasad-Sommerfield (BPS) limit, namely for $\lambda=0$, 
the equations (\ref{KHequations}) have an analytic solution in terms of elementary functions
\be
K\left(\xi\right)=\frac{\xi}{\mathrm{sinh}\xi},~~~~~~~~~H\left(\xi\right)=\xi\mathrm{coth}\xi-1.
\ee
We are interested in fluctuations around the classical solutions (\ref{hedgehog}) in the general
case in which $\lambda\neq0$. We therefore write the fields in the Lagrangian (\ref{gglagrangian})
as the sum of the classical solution, considered to be of order 0, and a fluctuation field which is
considered to be of order 1:
\be
A^{a}_{\mu}=\mathcal{A}^{a}_{\mu}+a^{a}_{\mu},~~~~~~~~~\phi^a=\Phi^a+\chi^a
\ee
The Lagrangian is then expanded up to order 2 and is therefore quadratic in the
fluctuations $a^{a}_{\mu}$ and $\chi^a$: 
\ba
\mathcal{L}^{(2)}&=&-\frac14\left\{\partial_{\mu}a^{a}_{\nu}\partial_{\mu}a^{a}_{\nu}-\partial_{\mu}a^{a}_{\nu}
\partial_{\nu}a^{a}_{\mu}-\partial_{\nu}a^{a}_{\mu}\partial_{\mu}a^{a}_{\nu}+\partial_{\nu}a^{a}_{\mu}
\partial_{\nu}a^{a}_{\mu}\right.
\nonumber\\
&-&\left.2e\epsilon_{abc}\left[\left(\partial_{\mu}a^{a}_{\nu}-\partial_{\nu}a^{a}_{\mu}
\right)\left(\A^{b}_{\mu}a^{c}_{\nu}+a^{b}_{\mu}\A^{c}_{\nu}\right)+
\left(\partial_{\mu}\A^{a}_{\nu}-\partial_{\nu}\A^{a}_{\mu}\right)a^{b}_{\mu}a^{c}_{\nu}\right]\right.
\nonumber\\
&+&\left.
e^2\left[4\A^{b}_{\mu}\A^{c}_{\nu}a^{b}_{\mu}a^{c}_{\nu}-2\A^{b}_{\mu}\A^{b}_{\nu}a^{c}_{\mu}
a^{c}_{\nu}+2\A^{b}_{\mu}\A^{b}_{\mu}a^{c}_{\nu}a^{c}_{\nu}-2\A^{b}_{\mu}\A^{c}_{\nu}
a^{c}_{\mu}a^{b}_{\nu}-2\A^{b}_{\mu}\A^{c}_{\mu}a^{b}_{\nu}a^{c}_{\nu}\right]\right\}
\nonumber\\
&+&\frac12\partial_\mu\chi^a\partial_\mu\chi^a-e\epsilon_{abc}\left[
\left(\partial_\mu\Phi^a\right)a_{\mu}^b\chi^c+\left(\partial_\mu\chi^a\right)a_\mu^b\Phi^c
+\left(\partial_\mu\chi^a\right)\A_\mu^b\chi^c\right]
\nonumber\\
&+&e^2\left[2\A_\mu^b\Phi^ca_\mu^b\chi^c-\A_\mu^b\Phi^ca_\mu^c\chi^b
-\A_\mu^b\Phi^ba_\mu^c\chi^c
+\frac12a^{b}_{\mu}a^{b}_{\mu}\Phi^c\Phi^c-\frac12a^{b}_{\mu}a^{c}_{\mu}\Phi^b\Phi^c
\right.
\nonumber\\
&+&\left.\frac12\A_\mu^b\A_\mu^b\chi^c\chi^c-\frac12\A_\mu^b\A_\mu^c\chi^b\chi^c\right]
-\frac\lambda 4\left[2\chi^a\chi^a\left(\Phi^b\Phi^b-v^2\right)+4\left(\Phi^a\chi^a\right)^2\right]
\ea
so that their equations of motion are linear.

If we only allow fluctuations in the scalar and vector channel separately, we get the following
equations
\ba
\!\!\!\!\!\!\!\!\!\!&&\!\!\!\!\!\!\!\!\!\!\!\!\!\!\!\!\!\!\!\!\partial_\mu\partial^\mu\chi^a-2e\epsilon_{abc}\A_\mu^b\left(\partial_\mu\chi^c\right)
+e^2\left[\A_\mu^a\A_\mu^b\chi^b-\A_\mu^b\A_\mu^b\chi^a\right]
-\lambda\left[2\Phi^a\Phi^b\chi^b+\chi^a\left(\Phi^b\Phi^b-v^2\right)\right]=0
\nonumber\\
&&
\nonumber\\
&&\!\!\!\!\!\!\!\!\!\!\!\!\!\!\!\!\!\!\!\!\partial_\nu\partial_\mu a_\nu^a-\partial_\nu\partial_\nu a_\mu^a+e\epsilon_{abc}
\left[2\left(\partial_\nu a_\mu^c\right)\A_\nu^b+\left(\partial_\nu a_\nu^b\right)\A_\mu^c
+2a_\nu^b\partial^\nu\A_\mu^c-\A_\nu^b\left(\partial_\mu a_\nu^c\right)
-a_\nu^b\partial_\mu\A_\nu^c\right]
\nonumber\\
&&\!\!\!\!\!\!\!\!\!\!\!\!\!\!\!\!\!\!\!\!+e^2\left[2\A_\mu^a\A_\nu^b a_\nu^b-\A_\mu^b\A_\nu^b a_\nu^a+\A_\nu^b\A_\nu^ba_\mu^a
-\A_\mu^b\A_\nu^aa_\nu^b-\A_\nu^a\A_\nu^b a_\mu^b-a_\mu^a\Phi^b\Phi^b+a_\mu^b\Phi^b\Phi^a
\right]=0.
\nonumber\\
\label{fieldeq}
\ea
The equation for the scalar particles can be rewritten in polar coordinates by making use of the following
definition
\be
\vec{L}=-i\vec{r}\times\vec{\nabla};~~~~~~L_k=-i\epsilon_{kmn}r_m\partial_n
\ee
and by remembering that
\be
\partial_\mu\partial^\mu=\frac{\partial^2}{\partial r^2}-\frac{2}{r}\frac{\partial}{\partial r}-\frac{L^2}{r^2}-
\partial_0^2.
\ee
Therefore, we can write
\ba
\left[\frac{\partial^2}{\partial r^2}-\frac{2}{r}\frac{\partial}{\partial r}-\frac{L^2}{r^2}-
\partial_0^2\right]\chi^a&-&2e\epsilon_{abc}\A_\mu^b\left(\partial_\mu\chi^c\right)
+e^2\left[\A_\mu^a\A_\mu^b\chi^b-\A_\mu^b\A_\mu^b\chi^a\right]
\nonumber\\
&-&\lambda\left[2\Phi^a\Phi^b\chi^b+\chi^a\left(\Phi^b\Phi^b-v^2\right)\right]=0.
\ea
Now we remember the classical solution for the monopole (\ref{hedgehog}) so that we can write
\ba
-2e\epsilon_{abc}\A_\mu^b\left(\partial_\mu\chi^c\right)&=&
-2i\frac{\epsilon_{ack}L_k}{r^2}\chi^c
\left[1-K\left(\xi\right)\right],
\nonumber\\
e^2\left[\A_\mu^a\A_\mu^b\chi^b-\A_\mu^b\A_\mu^b\chi^a\right]
&=&-\left(\frac{\delta_{ab}}{r^2}+\frac{r^ar^b}{r^4}\right)\left[1-K\left(\xi\right)\right]^2
\ea
and by recalling that the isospin operator is $(I^a)_{bc}=-i\epsilon_{abc}$ we get
\ba
&&\left[\frac{\partial^2}{\partial r^2}-\frac{2}{r}\frac{\partial}{\partial r}-\frac{\left(\vec{L}+\vec{I}\left[1-
K\left(\xi\right)\right]\right)^2}{r^2}-
\partial_0^2\right]\chi+\frac{\left(\hat{r}\cdot\vec{I}\left[1-
K\left(\xi\right)\right]\right)^2}{r^2}\chi
\nonumber\\
&&-\lambda\left[2\frac{r^ar^b}{e^2r^4}H(\xi)^2\chi^b+\chi^a\left(
\frac{H(\xi)^2}{e^2r^2}-v^2\right)\right]=0.
\ea
The above equation can be rewritten as
\ba
&&\left[\frac{\partial^2}{\partial r^2}-\frac{2}{r}\frac{\partial}{\partial r}-\frac{\left(\vec{T}^2
-\left(\hat{r}\cdot\vec{I}\right)^2\right)}{r^2}-
\partial_0^2\right]\chi+\frac{2K(\xi)\left(\vec{I}\cdot\vec{T}-\left(\hat{r}\cdot\vec{I}\right)^2
\right)}{r^2}
\nonumber\\
&-&\frac{K(\xi)^2\left[\vec{I}^2-\left(\hat{r}\cdot\vec{I}\right)^2\right]}{r^2}
-\lambda\left[2\frac{r^ar^b}{e^2r^4}H(\xi)^2\chi^b+\chi^a\left(
\frac{H(\xi)^2}{e^2r^2}-v^2\right)\right]=0.
\nonumber\\
\label{ea}
\ea
The term $\propto K(\xi)$ in the above equation induces charge-exchange reactions.
The term proportional to $K(\xi)^2$, modifies the deep-scattering amplitude for the 
charge-preserving processes.

\section{SU(3) 't Hooft-Polyakov monopoles}
\renewcommand{\theequation}{B-\arabic{equation}}
  \setcounter{equation}{0}  

  Although in the text above we deal with monopoles
in a simplified way, we briefly review 
the elements  of the $SU(3)$ Lie algebra and monopoles. It is given by
a set of traceless Hermitian $3\times 3$ matrices 
\be
T^a=\lambda^a/2~~~~~~~~~~~~~~~~~~a=1,2...8
\ee
where $\lambda^a$ are the standard Gell-Mann matrices. They are normalized as
$Tr T^aT^b=\delta^{ab}/2$. The structure constants of the Lie algebra are
$f^{abc}=\frac14Tr[\lambda^a,\lambda^b]\lambda^c$ and in the adjoint 
representation we have $(T^a)_{bc}=f^{abc}$.
We are interested in the diagonal, or Cartan, subalgebra of $SU(3)$. It is 
given by two generators:
\be
H_1=T^3=\frac12\left(
\begin{tabular}{ccc}
1&0&0\\
0&-1&0\\
0&0&0
\end{tabular}
\right)~~~~~~~~~~~~~~~~~~~~~
H_2=T^8=\frac{1}{2\sqrt{3}}\left(
\begin{tabular}{ccc}
1&0&0\\
0&1&0\\
0&0&-2
\end{tabular}
\right)
\label{acca}
\ee
which are composed into the vector $\vec{H}=(H_1,H_2)$. The number of positive
roots is 3, since the dimension of the group is 8. We can take the basis of 
simple roots as
\be
\vec{\beta}_1=(1,0),~~~~~~~~\vec{\beta}_2=(1/2,-\sqrt{3}/2).
\label{betai}
\ee
The third positive root is given by the composition of the first two roots
$\vec{\beta}_3=\vec{\beta}_1-\vec{\beta}_2=(1/2,\sqrt{3}/2)$. Since all these
roots have unit length, our choice corresponds to the self-dual basis
$\vec{\beta}_{i}^{*}=\vec{\beta}_i$.

Let us now remind the usual construction of 
a 't Hooft-Polyakov monopole in QCD,
once we assume that the role of the Higgs is played by the expectation value
of the field $A_0$, here considered to be real in Minkowski time.
The QCD Lagrangian 
\ba
\mathcal{L}=&&-\frac14\left[F_{\mu\nu}^{a}\left[\mathcal{A}\right]+D_{\mu}^{ac}Q_{\nu c}
-D_{\nu}^{ab}Q_{\mu b}-gf^{abc}Q_{\mu}^{b}Q_{\nu}^{c}\right]
\nonumber\\
&&\times
\left[F^{\mu\nu}_{a}\left[\mathcal{A}\right]+D^{\mu}_{ae}Q^{\nu e}
-D^{\nu}_{ad}Q^{\mu d}-gf_{ade}Q^{\mu}_{d}Q^{\nu}_{e}\right]-V[\mathcal{A}_{0}^{a}],
\ea
where we have written the gauge field as the sum of a background field 
and fluctuations
\ba
A_{\mu}^a&=&\mathcal{A}_{\mu}^a+Q_{\mu}^a
\nonumber\\
\mathcal{A}_{\mu}^a&\rightarrow&\mathrm{background~field}
\nonumber\\
Q_{\mu}^a&\rightarrow&\mathrm{fluctuations}
\ea
and the potential $V[A_{0}^{a}]$ will be defined later.
In our case, the background field will have the following form:
\be
\mathcal{A}_{\mu}^a=\delta_{\mu 0}\delta^{a3}\mathcal{A}_{0}^{3}.
\ee
The field $\mathcal{A}_{0}^{a}$ will play the role of a Higgs field.
Since we suppose that our background field is diagonal in color space, we can 
rewrite it as follows:
\be
\mathcal{A}_0=\mathcal{A}_0^3\vec{h}\cdot\vec{H}
\ee
where $\vec{H}$ was defined after Eq.~(\ref{acca}) and clearly $\vec{h}=(1,0)$.
If the monopole solution obeys the Bogomol'nyi equations, in the direction
chosen to define $\mathcal{A}_0$, the asymptotic magnetic field of a BPS monopole is
of the form
\be
B_n=\vec{g}\cdot\vec{H}\frac{r_n}{r^3}.
\ee
Here, the magnetic charge $g=\vec{g}\cdot\vec{H}$ is defined as a vector in the
root space. The charge quantization condition may be obtained from the 
requirement of topological stability, namely the phase factor must be 
restricted by
\be
\exp\left\{ie\vec{g}\cdot\vec{H}\right\}=1.
\ee
A general solution of this equation is given by
\be
\vec{g}=\frac{4\pi}{e}\sum_{i=1}^{r}n_i\vec{\beta}_{i}^{*}
=\frac{4\pi}{e}\left(n_1\vec{\beta}_{1}^{*}+n_2\vec{\beta}_{2}^{*}\right)
=g_1\vec{\beta}_{1}^{*}+g_2\vec{\beta}_{2}^{*}
\ee
where $n_1$ and $n_2$ are non-negative integers and $\vec{g}_1,~\vec{g}_2$ are
the magnetic charges associated with the corresponding simple roots. Our basis of simple
roots is self-dual: $\vec{\beta}_{1}^{*}=\vec{\beta}_1;~\vec{\beta}_{2}^{*}=\vec{\beta}_2$.
Thus we get:
\ba
g=\vec{g}\cdot\vec{H}=\frac{4\pi}{e}\left[\left(n_1+\frac{n_2}{2}\right)H_1-\frac{\sqrt{3}}{2}
n_2H_2\right].
\ea
The magnetic charge is not so trivially quantized as in the $SU(2)$ model.

Since our Higgs vector $\vec{h}$ is not orthogonal to any of the simple roots 
$\vec{\beta}_i$ (\ref{betai}), there is a unique set of simple roots with 
positive inner product with $\vec{h}$, Thus, the symmetry is maximally broken
on the maximal abelian torus $U(1)\times U(1)$. In this case, both the numbers
$n_1$ and $n_2$ have the meaning of topological charges. The topological charge
of a non-Abelian monopole is given by the integral over the surface of sphere
$S^2$
\be
G=\frac{1}{\mathcal{A}^{3}_{0}}\int dS_n Tr\left(B_n\mathcal{A}_0\right)=\vec{g}\cdot\vec{h}=g_1+\frac12g_2.
\ee
Therefore, if $\vec{h}$ is orthogonal to one of the roots $\vec{\beta}_i$,
only one component of $\vec{g}$ may be associated with the topological charge.
In our case, there are two topological integers that are associated with a 
monopole. 

\begin{center}
\begin{figure}
\hspace{1.5cm}
\parbox{6cm}{
\scalebox{.95}{
\includegraphics{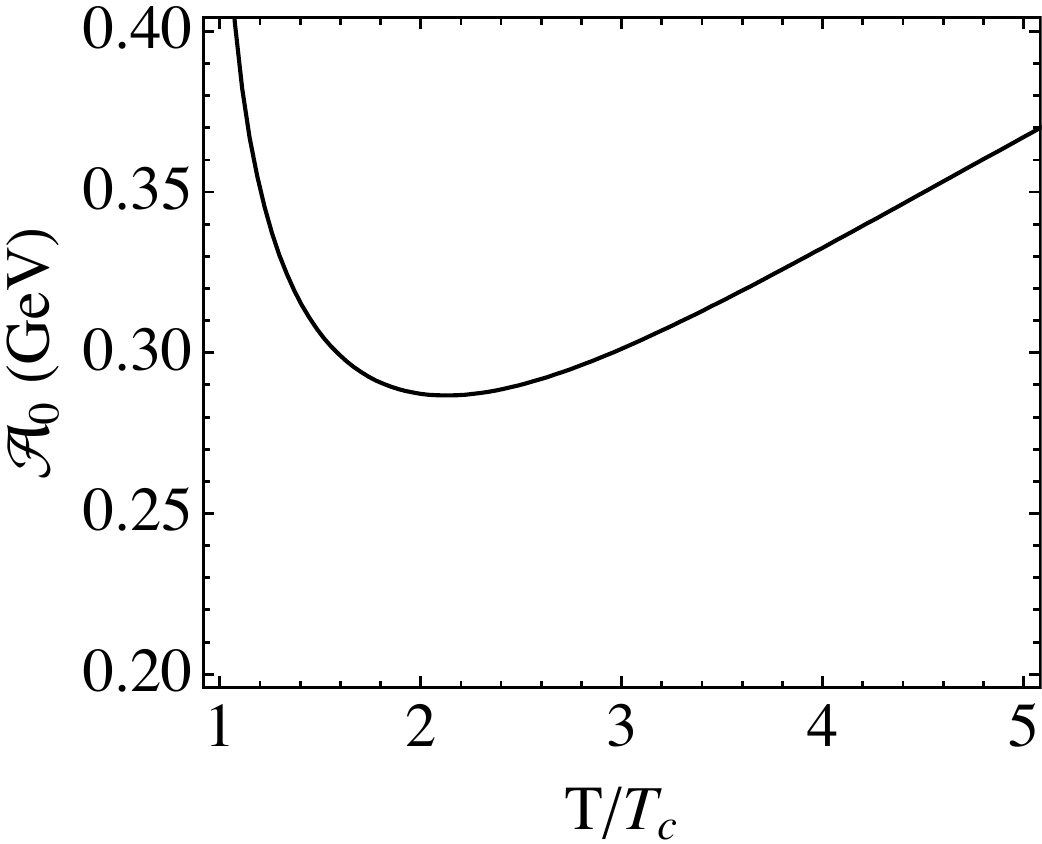}\\}}
\caption{$A_0$ as a function of $T/T_c$.}
\label{fig_m1m2}
\end{figure}
\end{center}

The above definition of topological charge can be used in order to generalize 
the
Bogomol'nyi bound for the $SU(N)$ monopoles. If we do not consider degrees of 
freedom
 that are related with electric charges of the configuration, it becomes
 \be
 M=\mathcal{A}^{3}_{0}|G|=\frac{4\pi \mathcal{A}^{3}_{0}}{e}\sum_{i=1}^{r}n_i\left(\vec{h}
\cdot\vec{\beta}_i \right) =\sum_{i=1}^{r}n_iM_i=\frac{4\pi \mathcal{A}^{3}_{0}}{e}
\left(n_1+\frac{n_2}{2}\right)
 \ee
 where $M_i=\frac{4\pi \mathcal{A}^{3}_{0}}{e}\vec{h}\cdot\vec{\beta}_i$ and we suppose 
that the
 orientation of the Higgs field uniquely determines a set of simple roots that 
satisfies
 the condition $\vec{h}\cdot\vec{\beta}_i\geq0$ for all $i$. Thus, it looks 
like there are $r$ individual monopoles of masses $M_i$.
In Figure \ref{fig_m1m2} we show the $\mathcal{A}_0$ that we obtain through
a fit of the lattice results for the Polyakov loop, as a function of the 
temperature.


\end{document}